\documentclass[%
 twocolumn,longbibliography,preprintnumbers,
nofootinbib,
 amsmath,amssymb,
 aps,
 pre,
]{revtex4-1}
\pdfoutput=1
\usepackage{graphicx}
\usepackage[utf8]{inputenc}
\usepackage{flushend}
\usepackage{dcolumn}
\usepackage{bm}
\usepackage{balance}
\usepackage{mathrsfs,epigraph,bm}
\usepackage{appendix}

\usepackage{tikz}
\usetikzlibrary{mindmap,trees}

\usepackage{tcolorbox}
\usepackage{tabularx}
\usepackage{array}
\usepackage{colortbl}
\tcbuselibrary{skins}

\usepackage{tcolorbox}
\definecolor{mycolor}{rgb}{0.122, 0.435, 0.698}
\makeatletter
\newcommand{\mybox}[1]{%
  \setbox0=\hbox{#1}%
  \setlength{\@tempdima}{\dimexpr\wd0+13pt}%
  \begin{tcolorbox}[colframe=mycolor,boxrule=0.5pt,arc=4pt,
      left=6pt,right=6pt,top=6pt,bottom=6pt,boxsep=0pt,width=0.98\linewidth]
    #1
  \end{tcolorbox}
}
\makeatother

\newtcbox{\boxbello}{on line,
  colframe=orange,colback=yellow!20!white,
  boxrule=0.5pt,arc=4pt,boxsep=0pt,left=6pt,right=6pt,top=6pt,bottom=6pt,width=0.98\linewidth}
\newcolumntype{Y}{>{\raggedleft\arraybackslash}X}

\tcbset{tab1/.style={fonttitle=\bfseries\large,fontupper=\normalsize\sffamily,
colback=yellow!10!white,colframe=red!75!black,colbacktitle=purple!40!white,
coltitle=black,center title,freelance,frame code={
\foreach \n in {north east,north west,south east,south west}
{\path [fill=purple!75!black] (interior.\n) circle (3mm); };},}}

\tcbset{tab2/.style={enhanced,fonttitle=\bfseries,fontupper=\normalsize\sffamily,
colback=yellow!10!white,colframe=red!50!black,colbacktitle=purple!40!white,
coltitle=black,center title}}
\usepackage[normalem]{ulem}

\usepackage[colorlinks = true,
            linkcolor = blue,
            urlcolor  = blue,
            citecolor =green,
            anchorcolor = blue]{hyperref}
\usepackage{verbatim}
\usepackage{color,ulem}
\usepackage[english]{babel}

\usepackage[utf8]{inputenc}
\input Starburst.fd
\newcommand*\initfamily{\usefont{U}{Starburst}{xl}{n}}\initfamily

\newcommand{\beq}{\begin{eqnarray}}
\newcommand{\eeq}{\end{eqnarray}}
\usepackage{amsmath}
\usepackage{tikz}
\usetikzlibrary{decorations.pathmorphing}
\usetikzlibrary{shapes.misc}
\tikzset{cross/.style={cross out, draw=black, minimum size=8*(#1-\pgflinewidth), inner sep=0pt, outer sep=0pt},
cross/.default={1pt}}
\usetikzlibrary{patterns,math}
\begin{document}

\title{Solution to the two-body Smoluchowski equation with shear flow for charge-stabilized colloids at low to moderate Péclet numbers}

\author{\textbf{Simone Riva}$^{1}$}%
 \email{simone.riva5@studenti.unimi.it}
 \author{\textbf{Luca Banetta}$^{2}$}
 \email{luca.banetta@polito}
\author{\textbf{Alessio Zaccone}$^{1,3}$}%
 \email{alessio.zaccone@unimi.it}
 
 \vspace{1cm}
 
\affiliation{$^{1}$Department of Physics ``A. Pontremoli'', University of Milan, via Celoria 16,
20133 Milan, Italy.}
\affiliation{$^{2}$Department of Applied Science and Technology (DISAT), Politecnico of Turin, Italy.}
\affiliation{$^{3}$Cavendish Laboratory, University of Cambridge, JJ Thomson
Avenue, CB30HE Cambridge, U.K.}

\begin{abstract}
We developed an analytical theoretical method to determine the microscopical structure of weakly to moderately sheared colloidal suspensions in dilute conditions. The microstructure is described by the static structure factor, obtained by solving the stationary two-body Smoluchowski advection-diffusion equation. The singularly perturbed PDE problem is solved by performing an angular averaging over the extensional and compressing sectors and by the rigorous application of boundary-layer theory (intermediate asymptotics). This allows us to expand the solution to a higher order in Péclet with respect to previous methods. The scheme is independent of the type of interaction potential. We apply it to the example of charge-stabilized colloidal particles interacting via the repulsive Yukawa potential and study the distortion of the structure factor. It is predicted that the distortion is larger at small wavevectors $k$ and its dependence on $Pe$ is a simple power law. At increasing $Pe$, the main peak of the structure factor displays a broadening and shift towards lower $k$ in the extensional sectors, which indicates shear-induced spreading out of particle correlations and neighbor particles locally being dragged away from the reference one.
In the compressing sectors, instead, a narrowing and shift towards high $k$ is predicted, reflecting shear-induced ordering near contact and concomitant depletion in the medium-range.
An overall narrowing of the peak is also predicted for the structure factor averaged over the whole solid angle. Calculations are also performed for hard spheres, showing good overall agreement with experimental data. It is also shown that the shear-induced structure factor distortion is orders of magnitude larger for the Yukawa repulsion than for the hard spheres.
\end{abstract}

\maketitle
\section{Introduction}

The determination of the microstructure of colloidal dispersions under shear is an important problem for both its relevance in their theoretical study and many applications from atmospheric science \cite{Falkovich_2002}, chemistry \cite{Likos}, preparation of emulsions \cite{Meleson_2004, Mason_2006} and industrial products \cite{Preziosi_2013}, and plasmas \cite{Nosenko}. The information about the spatial arrangement of a suspension of particles in a continuous liquid medium is encoded in the pair correlation function $g(\bm{r},t)$ or equivalently, via a Fourier transform, in the structure factor $S(\bm{k},t)$. Scattering with light in the visible spectrum, which typically corresponds to a wavelength of the order of the colloids, is sensible to the level of mesoscopic structure, thus providing a way of experimentally measuring their structure factor \cite{Dhont_book, Hunter_book}.

Johnson, de Kruif et al. \cite{Johnson,deKruif}, through small angle neutron scattering experiments, studied the structure factor distortion of concentrated hard-sphere dispersions under steady flow. Thanks to their experimental setup they were able to study the flow-induced distortion along the flow and vorticity axes, as well as along the extensional and compressing axes. Difficulties arise when trying to develop a theory in the same conditions of the experiments, because the latter are often difficult to be conducted in dilute conditions, which are, on the other hand, required in most theories, as they allow to neglect hydrodynamic interactions and many-body effects. However, some of the results from theoretical works \cite{Dhont_1989,Ronis} have also been found qualitatively to agree with experiments in \cite{Johnson,deKruif}, even though a quantitative comparison was not possible. In particular, when increasing the shear rate, a decrease and shift towards low $k$ of the peak of the structure factor along the flow direction was observed, suggesting that the particles concentration decreases in that sector. A decrease and shift towards large $k$ was found in the vorticity direction, suggesting the possible formation of transient accumulations of particles near contact \cite{Johnson}. As for the extensional and compressing directions, which are particularly relevant in our study, they found that, in the former, the peak is lowered and shifted towards large $k$ (even though experimental uncertainties were relevant), while the structure factor at small wavevector increases; in the latter case, the peak is increased and shifted towards large $k$, while the structure factor at small $k$ decreases \cite{deKruif}.

The Smoluchowski equation with shear \cite{Dhont_book,Fuchs} is a partial differential equation (PDE) which provides a means to determine the pair correlation function (pcf) of colloidal suspensions. It is the equation of motion for the time-dependent probability density function $P(\bm{r}_1, ..., \bm{r}_N, t)$ of the positions $\bm{r}_1$, ..., $\bm{r}_N$ of $N$ Brownian interacting particles, which are $3N$ stochastic variables \cite{Dhont_book}. $P(\bm{r}_1, ..., \bm{r}_N, t)$ represents the probability that each $i$-th particle is found within an infinitesimal volume $d^3r_i$ around the point $\bm{r}_i$ at a given time $t$. The Smoluchowski equation is valid on the Brownian or diffusive time scale, that is, for a large enough time resolution that the solvent coordinates are relaxed to thermal equilibrium and so are the momentum coordinates of the Brownian particles with the heat bath of the solvent; for this reason the statistical properties of the solution only depend on the position coordinates of the colloid particles.

The Smoluchowski equation with shear for $N$ interacting particles reads \cite{Dhont_book}:
\begin{equation} \label{eq:smol0}
\begin{split}
    \frac{\partial}{\partial t} P(\bm{r}_1, ..., \bm{r}_N, t) & =  \sum_{i,j=1}^N \nabla_{r_i} \cdot \bm{D}_{ij} \cdot \big\{ \beta \left[ \nabla_{r_j} \Phi (\bm{r}_1, ..., \bm{r}_N) \right] \\
    & P(\bm{r}_1, ..., \bm{r}_N, t) + \nabla_{r_j} P(\bm{r}_1, ..., \bm{r}_N, t) \big\} - \\
    & - \sum_{j=1}^N \nabla_{r_j} \cdot \left[ \bm{v} (\bm{r}_j) P(\bm{r}_1, ..., \bm{r}_N, t) \right] .
\end{split}
\end{equation}
Here $\Phi$ is the total potential energy of interaction of a given configuration. Each $\bm{D}_{ij}$, $i,j=1,...,N$, is a $3 \times 3$ matrix, called microscopic diffusion matrix, and describes the influence of the flow caused by the motion of the $j$-th particle on the force acting on the $i$-th particle; finally $\bm{v} (\bm{r})$ is the flow field velocity.

The Smoluchowski equation contains terms describing the two main contributions to the dynamics of the system, namely the Brownian diffusion and inter-particle interaction (which are collected under the name of Brownian-induced effects) and the effects of the flow field (shear-induced effects). The relative importance between these two contributions is expressed by the Péclet number,
\begin{equation} \label{eq:Pe}
    Pe = \frac{\dot{\gamma} a^2}{D_0},
\end{equation}
where $\dot{\gamma}$ is the shear rate, $D_0 = (6 \pi \eta_0 a \beta)^{-1}$ is the Stokes-Einstein diffusion coefficient for a single spherical particle, and $a$ is the particle radius. The higher the Péclet number, the more important the shear-induced distortion of the equilibrium (i. e. zero-shear) structure will be. However, $Pe$ being low does not guarantee that the Brownian motion dominates over the hydrodynamic motion over the whole space.

The Smoluchowski equation therefore provides a way to study theoretically the microstructure of real fluids where both Brownian and shear-induced contributions can be relevant, as well as having many applications in the study of colloidal dispersions and their coagulation rate \cite{Zaccone_2009}, phase transitions of complex fluids under shear and rheological properties of suspensions \cite{Brady_1997,Fuchs,Nazockdast_2012}.

Several solutions were proposed with the aim of determining the pair correlation function under the competing effects of Brownian-induced and shear-induced contributions. Many works were focused on hard-sphere systems under strong shear flow, i.e. with high Péclet numbers. Batchelor and Green \cite{Batchelor} derived a solution of the two-body equation in the limit of $Pe=\infty$. Brady and Morris \cite{Brady_1997} solved the Smoluchowski equation with analytical methods to obtain the pair correlation function of hard-sphere suspensions under high shear ($Pe \gg 1$), which featured the presence of a boundary layer of thickness $\mathcal{O}(Pe^{-1})$. The value of $g(\bm{r})$ at contact was found to be $\mathcal{O}(Pe^{0.78})$ if the particles are only subject to hydrodynamic interactions and $\mathcal{O}(Pe)$ when a hard-core repulsive potential is added. They also asserted that the same features are expected for particles interacting via a repulsive short-range force, with a boundary layer whose thickness depends on $Pe$, but they did not carry out the calculations for this case.

In another work by Nazockdast and Morris \cite{Nazockdast_2012}, the Smoluchowski equation was solved in a perturbative paradigm for a wide range of Péclet numbers and the volume fraction $\phi$, including the near-equilibrium limit $Pe \ll 1$. Many-body effects, which are significant at high volume fraction, were also accounted for. However, the solution was performed numerically rather than analytically. For weakly sheared suspensions the profile of $g(\bm{r})$ at the surface of the reference particle was found as a function of the two angular coordinates. In the shear plane, the pair correlation function was shown to be a sinusoidal function of the in-plane angle $\theta$ for $Pe=0.10$ independently of the concentration.

Recently Banetta and Zaccone \cite{Banetta_2019} have developed an analytical method based on intermediate asymptotics for solving the Smoluchowski equation and determining the pair correlation function of sheared suspensions. This method can be applied to systems displaying different types of inter-particle interactions. It has been successfully used to determine the radial distribution function of a hard-sphere fluid, for which previous numerical results were recovered, and of an attractive Lennard-Jones fluid, where new features were found.

Later it was also applied to particles interacting with the repulsive Yukawa (or Debye-H{\"u}ckel) potential, representing colloidal particles with charge stabilization, in conditions of high Péclet numbers \cite{Banetta_2020}. It was found that, at $Pe=1000$, the behaviour of the pair correlation function is completely dominated by the shear-induced contribution and only a small role is played by the interaction potential, which means that the flow-field convection is the main responsible for the spatial arrangement of the particles. On the contrary, at intermediate Péclet numbers ($Pe=10$), the effect of the interaction potential on the microstructure is significant. The extensional quadrants, where the particles are moving away from the reference one, feature a depletion layer near contact where the pcf is identically null. On the other hand, the compressing quadrants, where the particles approach the reference one, are characterized by a maximum near contact, followed by a lower peak, due to the interplay between advection and electrostatic repulsion, for sufficiently high values of the Debye length.

This work was addressed to strongly sheared systems and in this case the shear effects are dominant over the diffusion, or at least produce significant distortions in the microstructure of the system. The opposite limit of weakly or moderately sheared systems (i. e. Péclet number smaller than unity or about unity) has not been investigated with the same approach yet. The main difficulty of the problem is its singularly perturbed structure \cite{Dhont_1989}, meaning that the solution is not an analytic function of the perturbative parameter for large $\bm{r}$, hence the necessity to treat this region by means of boundary-layer theory \cite{Bender_1978, Dhont_1989}.

A similar boundary-layer method to the one described above was adopted by Dhont \cite{Dhont_1989}, who carried out a rigorous perturbative calculation to solve the Smoluchowski equation for colloidal systems at low Péclet. Dhont applied it to a hard-sphere suspension at $Pe = 0.05$ and calculated the leading terms of the asymptotic solution. He found that the distortion of the static structure factor is of order $10^{-3}$ at this shear rate. He also showed that the distortion is larger for charged particles interacting via the Yukawa potential than for hard spheres \cite{Dhont_1987}. The Smoluchowski equation was solved by Dhont in Fourier or momentum space, where the singularity appears at small wavevector $k$, thus making the perturbative problem a standard, non-singular one, and analytically solvable.

Brady and Vicic \cite{Brady_1995} proposed an alternative solution, calculated directly in the real space, for the case of hard-sphere particles. However, their approach is based on some crucial assumptions which simplify the whole derivation. The first one is that the inter-particle forces must balance the equilibrium value of the Brownian forces. While this is certainly true at equilibrium, it seems quite an important approximation when the system is out of equilibrium. This allows Brady and Vicic to write a simplified Smoluchowski equation for the distortion of the pair correlation function; its solution is treated as a regular expansion in integer powers of $Pe$. This approximation, while certainly useful, however does not fully account for the singularly perturbed nature of the problem.

Ronis \cite{Ronis} also calculated the static structure factor of colloidal dispersions under shear flow in a theoretical way but with a different approach. At small shear rates, the linear response theory for the correlation functions out of equilibrium was used to derive the structure factor in the linear regime, while in the non-linear regime a phenomenological fluctuating-diffusive equation was used, which applies to large shear rates and reduces to the previous one for small shear rates. The results for small shear rates at the quadratic order are similar to those of the present work, even though Ronis adopted a phenomenological approach and did not treat the singularity with perturbative methods, but rather expanded the solution in powers of the shear rate. As a consequence, he did not derive the boundary-layer part of the solution.

Lionberger and Russel \cite{Lionberger} derived the structure of colloids interacting with a hard-core Yukawa potential by solving the Smoluchowski equation considering also many-body contributions to the inter-particle force acting on each colloid. These are accounted for by introducing different thermodynamic closure approximations, which are analyzed to determine which one produces the best agreement with simulations. However, their solution for the pair correlation function is based on a regular perturbative expansion consisting in a linear correction in the Péclet number. This method does not take into account the singularly perturbed character of the problem. In particular, this regular-perturbation model is expected to fail to predict the structure oscillations at large separation (outer layer), where the perturbation is singular. In this respect, a more complete treatment from the point of view of perturbation theory is desirable.

Finally, classical density functional theory (DFT) methods \cite{Roth_2010} also may provide a route to determining the solution to the Smoluchowski equation for colloids in shear flow, although at present this has been possible only using the approximate ansatz of \cite{Brady_1995} for the dilute pcf, which, as explained above, does not fully account for the boundary-layer structure of the problem.

\section{Aim and structure of work}
Our aim in this paper is to develop a more rigorous analytic method to address the problem which can be applied to any physicochemical interaction potential and implement in particular the screened-Coulomb Yukawa potential, which is a more realistic interaction to describe charge-stabilized colloids than the hard-sphere potential used in most cases \cite{Dhont_1989,Brady_1997,Nazockdast_2012}, and relevant also for plasmas and dusty plasmas \cite{Nosenko}. Solving the Smoluchowski equation for sheared colloidal systems and describing the distortion caused by weak-to-moderate flows is the object of the present work, where a very similar approach to \cite{Dhont_1989, Dhont_1987} is used in the limit of low Péclet numbers and combined with important improvements from \cite{Banetta_2019, Banetta_2020}, in particular the introduction of an averaging over the angular components of the wavevector $\bm{k}$. This turns the PDE problem into an effectively one-dimensional ordinary differential equation (ODE), which considerably simplifies the mathematical treatment compared to \cite{Dhont_1989}. The separate angular averaging over extensional and compressing quadrants has a price to pay in terms of losing angular resolution but allows us to gain one more order in the perturbative solution (see below). This geometric structure of the solution, however, has also experimental relevance, as shown by de Kruif et al. \cite{deKruif}, who designed a flow cell to collect scattering data in $20^\circ$ sectors centered about the compressing and extensional axes.

The angular averaging allows us to analytically derive one additional order (the quadratic order) in $Pe$ in the asymptotic solution, which was not calculated by Dhont, and has several consequences. Some of these are the possibility to extend the theory to higher values of $Pe$ and the prediction that the distortion does not vanish if averaged on all directions, which would not be possible in a linear-in-$Pe$ theory. The effects of the Péclet number in determining the microstructure of the system is studied and discussed, and a comparison between hard-sphere and Yukawa interaction potentials is presented.

In Section \ref{theory} we present the proposed framework, based on the two-body Smoluchowski equation and we apply the singular-perturbation theory to derive its solution. In Section \ref{results} we implement the obtained analytical solutions to particles interacting via the Yukawa potential and show the effects of the Péclet number on the structure factor and its distortion. We derive a scaling law for the distortion as a function of $Pe$ and predict a broadening and narrowing of the structure factor peak in the extensional and compressing angular sectors, respectively, as well as a decrease of its maximum. These features were not observed in previous theoretical works, to our knowledge. We discuss the results in the final Section \ref{conclusions}.

\section{Theory} \label{theory}

\subsection{The model}

We operate in the limit of dilute solutions and neglect hydrodynamic interactions (HI), namely the effects that the flow generated by the presence of one particle produces on other particles, which have a limited relevance in such conditions \cite{Nazockdast_2012}. The assumption of low $Pe$ further justifies this assumption. In the absence of HI, the diffusion matrices take the form:
\begin{equation} \label{eq:diffusion_matrix}
    \bm{D}_{ij} = D_0 \delta _{ij} \bm{I}.
\end{equation}

\begin{figure}
    \centering
    \includegraphics[width=0.45\textwidth, keepaspectratio]{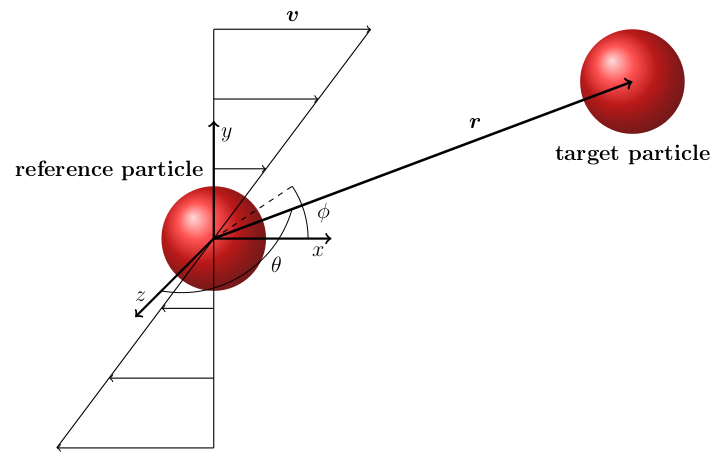}
    \caption{Schematic illustration of two interacting particles in a simple shear flow. The reference system is centered in the reference particle, so that the position of the target particle equals their relative distance $\bm{r}$. The flow velocity is null on the reference particles and is proportional to the $y$ coordinate.}
    \label{fig:two_particles}
\end{figure}

The suspension is considered dilute enough that interactions between three or more particles are extremely unlikely to take place and one can focus on two particles by choosing $N=2$ in Eq. (\ref{eq:smol0}). This corresponds to studying how a reference particle in a dilute system influences the surrounding space and where a second target particle is more likely to be found. Moreover, we consider a steady simple shear flow: steadiness makes the time derivative in Eq. (\ref{eq:smol0}) vanish as the solution is time independent. The flow is directed along the $x$ axis and its gradient along the $y$ axis, as illustrated in Fig. \ref{fig:two_particles}.
The velocity field of such flow is $\bm{v}(\bm{r}) = \bm{\Gamma} \cdot \bm{r}$, where the flow velocity gradient matrix $\bm{\Gamma}$ takes the form
\begin{center}
$\quad \bm{\Gamma}= \dot{\gamma}$
$ \begin{pmatrix}
    0 & 1 & 0 \\
    0 & 0 & 0 \\
    0 & 0 & 0
\end{pmatrix} $,\\ 
\end{center}
$\dot{\gamma}$ being the shear rate.

The described system is invariant under rigid translations as the relative flow velocity between the positions of the two particles does not vary, even though the absolute velocities change. For this reason it is possible to write the pcf as a function of the relative position of the two particles, which we refer to as $\bm{r}=\bm{r}_1-\bm{r}_2$:
\begin{equation} \label{eq:homogen}
    g(\bm{r}_1,\bm{r}_2) = g(\bm{r}_1-\bm{r}_2) = g(\bm{r}).
\end{equation}
In this form, $g(\bm{r})$ expresses the probability of finding the target particle at a relative centre-to-centre position $\bm{r}$ from the reference one.

Under these assumptions, the two-particle Smoluchowski equation with shear flow in terms of the relative coordinate becomes
\begin{equation} \label{eq:smol}
    2 D_0 \nabla \cdot \{ \beta \left[ \nabla V(r) \right] g(\bm{r}) + \nabla g(\bm{r}) \} - \nabla \cdot \left[ \bm{\Gamma} \cdot \bm{r} g(\bm{r}) \right] = 0,
\end{equation}
in which $V(r)$ is the two-particle interaction potential, assuming that it is spherically symmetric. The subscript $r$ after the operator $\nabla$ is implied.

Equation (\ref{eq:smol}) can be found in  \cite{Dhont_book}, p. 363, where its origin from the N-body equation is discussed. It contains all the features of the investigated system. The two terms inside the curly brackets describe the Brownian-induced effects. They account respectively for the inter-particle interactions and the Brownian forces. The remaining term describes the shear-induced effects in terms of the relative flow velocity, $\bm{v}(\bm{r}) = \bm{\Gamma} \cdot \bm{r}$. With the given definitions of $Pe$, Eq. (\ref{eq:Pe}), Eq. (\ref{eq:smol}) can be written as
\begin{equation} \label{eq:smol_small_Pe}
    2 a^2 \nabla \cdot \{ \beta \left[ \nabla V(r) \right] g(\bm{r}) + \nabla g(\bm{r}) \} - Pe \, y \frac{\partial}{\partial x}g(\bm{r}) = 0.
\end{equation}

\subsection{Derivation}

Eq. (\ref{eq:smol_small_Pe}) is to be solved perturbatively in the parameter $Pe$, which multiplies the advection term associated with the flow disturbance to the equilibrium structure. One would wish to be able to expand the solution as a regular power series of $Pe$:
\begin{equation} \label{eq:regular_expansion}
    g(\bm{r}) = g_0(\bm{r}) + Pe \, g_1(\bm{r}) + Pe^2 \, g_2(\bm{r}) + ...
\end{equation}
The main obstacle in doing so is the fact that the perturbation is singular \cite{Dhont_1989}. Indeed, for large enough separation in the $y$ direction ($\mathcal{O}(Pe^{-1/2}$)), the Brownian term does not prevail over the perturbation given by the shear term. As a consequence, the solution does not reduce to the equilibrium solution in the limit of small $Pe$, meaning that it is not an analytical function of $Pe$. Hence the impossibility of writing an expansion like Eq. (\ref{eq:regular_expansion}). The problem is said to have a boundary-layer behaviour at sufficiently large inter-particle separations and one has to resort to boundary-layer theory \cite{Bender_1978} to solve it.

First, we have to bring Eq. (\ref{eq:smol_small_Pe}) to the standard form of a singular perturbation problem, that is, with the perturbative parameter multiplying the highest-order derivative, leading to the presence of a boundary layer at some distance from the origin of the reference system. To do so we apply a Fourier transform, so that the singularity appears at small $\bm{k}$ in reciprocal space, which corresponds to large $\bm{r}$ in real space. If one were to stay in real space, the advective term would lead to a divergence at large $\bm{r}$.
This is the procedure also followed by Dhont \cite{Dhont_1989} and the Fourier transform of the Smoluchowski equation, Eq. (\ref{eq:smol}), is the same as in \cite{Dhont_1989}:
\begin{equation} \label{eq:smol_k}
\begin{split}
    \Gamma_{ij} k_i \frac{ \partial } {\partial k_j} S(\bm{k}) = 2 D_0 k^2 \left[ S(\bm{k}) -1 + \rho \beta V(k) \right] + \\
    + \frac{2 D_0}{(2 \pi)^3} \beta \bm{k} \cdot \int d^3k' \bm{k'} V(k') \left[ S(\bm{k-k'}) - 1 \right].
\end{split}
\end{equation}
Here $V(k)$ indicates the Fourier transform of the pairwise interaction potential between particles $V(r)$, whose form we do not specify yet, and $\rho$ is the number density of particles. This is the PDE for the static structure factor $S(\bm{k})$ that we are going to solve perturbatively.

\subsubsection*{Angular averaging}

We define the velocity in $\bm{k}$-space by analogy with $\bm{v}(\bm{r}) = \bm{\Gamma} \cdot\bm{r}$ as
\begin{equation} \label{eq:velocity_k}
    \bm{v}(\bm{k}) = \bm{k} \cdot \bm{\Gamma},
\end{equation}
which is the factor appearing on the left-hand side of Eq. (\ref{eq:smol_k}), and express it in spherical coordinates, finding
\begin{equation} \nonumber
    \left \{ \begin{array}{rl}
    & v_k = \dot \gamma k \sin^2{\theta} \sin{\phi} \cos{\phi} \\
    & v_\theta = \dot \gamma k \sin{\theta} \cos{\theta} \sin{\phi} \cos{\phi} \\
    & v_\phi = \dot \gamma k \sin{\theta} \cos^2{\phi}
    \end{array}
    \right. 
\end{equation}

If we go back to real ($\bm{r}$) space for a moment, we can see from immediate inspection of Eq. (\ref{eq:smol}) for zero shear rate ($\bm{\Gamma}=0$), that the equilibrium solution for a spherically symmetric pairwise potential is trivially \cite{Dhont_1989, Lionberger}
\begin{equation} \label{eq:g_eq}
    g_{eq}(r) = e^{-\beta V(r)},
\end{equation}
since it makes the expression in the curly brackets vanish. According to equilibrium statistical mechanics, this is in general a good approximation in dilute systems. Thus we obtain the equilibrium structure factor
\begin{equation} \label{eq:Se}
    S_{eq}(k) = 1 + \rho \int d^3r \left[ e^{- \beta V(r)} - 1 \right] e^{i \bm{k} \cdot \bm{r}},
\end{equation}
which solves the zero-shear Smoluchowski equation in reciprocal ($\bm{k}$) space.

We now subtract the equation for the system without shear from the original one, Eq. (\ref{eq:smol_k}):
\begin{equation} \label{eq:smol_deltaS}
\begin{split}
     & \bm{v}(\bm{k}) \cdot \nabla_{\bm{k}} S(\bm{k}) = 2 D_0 k^2 \left[ S(\bm{k}) - S_{eq}(k) \right] + \\
     & + \frac{2 D_0}{(2 \pi)^3} \beta \bm{k} \cdot \int d^3k' \bm{k'} V(k') \left[ S(\bm{k-k'}) - S_{eq}(| \bm{k-k'} |) \right].
\end{split}
\end{equation}

As pointed out by Dhont (\cite{Dhont_book}), the integral on the right-hand side of Eq. (\ref{eq:smol_deltaS}) is negligible because the integrand is almost anti-symmetric while the integration is carried out on the whole reciprocal space:
\begin{equation} \label{eq:smol_deltaS_integ}
\begin{split}
     \bm{v}(\bm{k}) \cdot \nabla_{\bm{k}} S(\bm{k}) = 2 D_0 k^2 \left[ S(\bm{k}) - S_{eq}(k) \right] .
\end{split}
\end{equation}

Following \cite{Banetta_2019,Banetta_2020}, we now apply an approximation, originally introduced in \cite{Zaccone_2009} (see also \cite{Nazockdast_2012}), to transform the three-dimensional PDE into an effective one-dimensional ODE which is analytically solvable. It is based on applying an angular average to Eq. (\ref{eq:smol_deltaS_integ}), namely applying $ \langle ... \rangle = \frac{1}{4 \pi} \int d \theta \sin{\theta} \int d \phi (...)$. We define the angular-averaged structure factor $\langle S(\bm{k}) \rangle = S(k)$, which only depends on the magnitude of the vector $\bm{k}$.

We also make use of a decoupling approximation: supposing that $\bm{v}$ and $\nabla S$ are weakly correlated, we can write
\begin{equation} \nonumber
    \langle \bm{v}(\bm{k}) \cdot \nabla_{\bm{k}} S(\bm{k}) \rangle = \langle \bm{v}(\bm{k}) \rangle \cdot \langle \nabla_{\bm{k}} S(\bm{k}) \rangle.
\end{equation}
This approximation was shown to be reasonable in comparison with numerical data in Refs. \cite{Zaccone_2009,Banetta_2019,Banetta_2020}.
Considering the dependence on the magnitude of $\bm{k}$ only, this expression becomes
\begin{equation} \nonumber
    \langle \bm{v}(\bm{k}) \cdot \nabla_{\bm{k}} S(\bm{k}) \rangle = \langle v_k \rangle \frac{d}{dk} S(k).
\end{equation}
With this procedure, the PDE, Eq. (\ref{eq:smol_deltaS_integ}), is reduced to the following ODE:
\begin{equation} \label{eq:smol_average}
    \langle v_k \rangle \frac{d}{dk} S(k) = 2 D_0 k^2 \left[ S(k) - S_{eq}(k) \right].
\end{equation}

It is possible to compute $\langle v_k \rangle$ in the compressing and extensional quadrants of the $\bm{k}$ space. These terms refer to the regions where the velocity flow field is directed radially towards the origin of the reference system and radially away from it, respectively. In real space, given the geometry of the system (see Fig. \ref{fig:two_particles}), the partition clearly corresponds to the four quadrants of the $x-y$ plane. Being $\phi$ the usual in-plane angular coordinate, the quadrants with $0 < \phi < \pi/2$ and $\pi < \phi < 3\pi/2$ have a flow that drives the target particle away from the origin, where the reference particle is, and are therefore extensional, while the ones with $\pi/2 < \phi < \pi$ and $3\pi/2 < \phi < 2\pi$ are compressing.

\begin{figure} [tbh]

\centering
\includegraphics[width=0.45\textwidth, keepaspectratio]{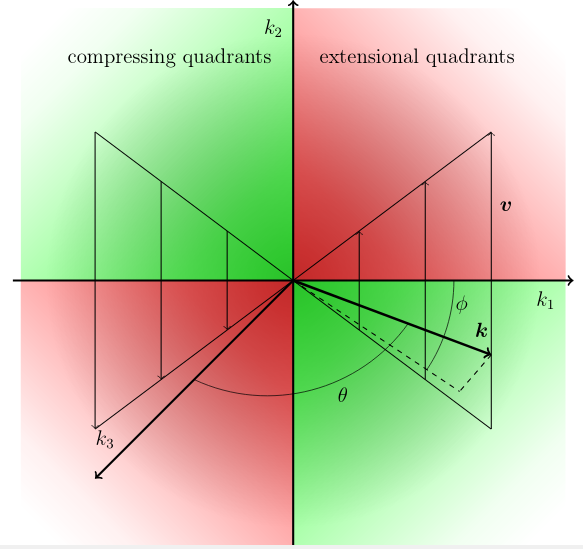}

\caption{
Representation of the flow velocity field in reciprocal space as defined by Eq. (\ref{eq:velocity_k}) (vertical arrows in the $k_1$-$k_2$ plane). The consequent partition into compressing and extensional quadrants comes from the radial component of the velocity being positive in the first and third quadrants of the $k_1$-$k_2$ plane and negative in the second and fourth. All the quadrants extend indefinitely in the $k_3$ direction.
} \label{fig:quadrants}
\end{figure}

We are now working in reciprocal space, where we cannot visualize particle as moving towards or away from each other in order to define the compressing and extensional quadrants. Therefore, we have to rely on the definition of the radial flow velocity being negative in the first ones and positive in the second ones. Surprisingly, even if it is not obvious \emph{a priori}, the partition is the same as the one described in real space, where $\phi$ is now the angle in the $k_1$-$k_2$ plane. This can be seen from the previous expression of the radial component of velocity $v_k$, requiring that it is positive in the extensional quadrants and negative in the compressing ones. Figure \ref{fig:quadrants} shows the distinction between the quadrants according to the direction of the velocity $\bm{v}(\bm{k})$.

We first consider the extensional case. With a simple calculation we find
\begin{equation} \nonumber
\begin{split}
    \langle v_k \rangle = \dot \gamma k \frac{1}{4 \pi} \int_0^\pi d \theta \sin{\theta} \sin^2{\theta} \left[ \int_0^{\pi/2} d \phi \sin{\phi} \cos{\phi} + \right. \\
    \left. + \int_\pi^{3\pi/2} d \phi \sin{\phi} \cos{\phi} \right] = \alpha_e \dot \gamma k,
\end{split}
\end{equation}
with $\alpha_e = \frac{1}{3\pi}$ \cite{Zaccone_2009}. Following the same procedure, integrating $\phi$ over the complementary intervals of the solid angle, one finds an analogous expression in the compressing sectors, with $\alpha_c = - \frac{1}{3\pi}$.

The two coefficients, $\alpha_e =  \frac{1}{3\pi}$ and $\alpha_c = - \frac{1}{3\pi}$, are important in the theory: the sign of $\alpha$ is what mainly distinguishes the extensional and the compressing sectors, i.e. the two separate regions of solid angle in which we are going to determine the structure factor of the colloidal suspension.

It is interesting to notice that $\langle v_k (\bm{k}) \rangle$ is formally identical to the corresponding expression for $\langle v_r (\bm{r}) \rangle$ found in direct space \cite{Banetta_2019,Banetta_2020}, which is, neglecting HIs, $\langle v_r \rangle = \alpha \dot{\gamma} r$. This is the same expression as the one we have obtained except for $k$ being replaced by $r$. The coefficient $\alpha$ also takes the same value of $\frac{1}{3\pi}$ and $-\frac{1}{3\pi}$ in the different quadrants.

Substituting $\langle v_k \rangle$ into Eq. (\ref{eq:smol_average}), we can synthetically write
\begin{equation} \label{eq:smol_deltaS_ave}
\begin{split}
     \alpha \dot \gamma \frac{d}{dk} S(k) = 2 D_0 k \left[ S(k) - S_{eq}(k) \right] ,
\end{split}
\end{equation}
where $\alpha$ stands either for $\alpha_e$ or $\alpha_c$.

This procedure of angular averaging is of remarkable importance because it simplifies considerably the form of the Smoluchowski equation, allowing us to solve it analytically to a higher order in $Pe$ in the perturbative expansion compared to the earlier work by Dhont \cite{Dhont_1989}, who, while retaining all the components of $\bm{k}$, obtained the solution only to leading order in $Pe$. Our gain of a higher order term in $Pe$ in the solution obviously comes at the expense of losing angular resolution and only being able to calculate the average of the structure factor on the extensional and compressing quadrants, which however is relevant for experimental scattering setups \cite{Johnson,deKruif}.

\subsubsection*{Perturbative solution}

We introduce the dimensionless wavevector $\tilde{k} = k a$, so that $\frac{d}{d\tilde{k}} = \frac{1}{a} \frac{d}{dk}$. The Smoluchowski equation in reduced units becomes
\begin{equation} \label{eq:smol_deltaS_adim}
     Pe \frac{d}{d\tilde{k}} S(\tilde{k}) = \frac{2\tilde{k}}{\alpha} \left[ S(\tilde{k}) - S_{eq}(\tilde{k}) \right].
\end{equation}

Within boundary-layer theory \cite{Bender_1978}, the solution of this equation must be evaluated separately in the boundary layer (or inner layer), which is the inner region around $\bm{k}=0$ where the perturbation is singular, and in the outer layer, with two separate asymptotic expansions in powers of $Pe$.

\subsubsection*{Outer solution}

In the outer region the equation is regular, thus we can expand the solution of Eq. (\ref{eq:smol_deltaS_adim}) in power series of the small parameter $Pe$:
\begin{equation} \label{eq:Sout_expansion}
    S_{out} (\tilde{k}) = S_{0,out} (\tilde{k}) + Pe \, S_{1,out} (\tilde{k}) + Pe^2 \, S_{2,out} (\tilde{k}) + ... .
\end{equation}
By substituting Eq. (\ref{eq:Sout_expansion}) into Eq. (\ref{eq:smol_deltaS_adim}) and comparing the coefficients of each power of $Pe$ we obtain a series of differential equations for the functions $S_{n,out} (\tilde{k})$, $n=0,1,2,...$. We calculate the first three terms (up to $n=2$).

\begin{itemize}
    \item Leading order ($Pe ^0$):
    \begin{equation} \label{eq:S0out}
    S_{0,out} (\tilde{k}) = S_{eq} (\tilde{k}).
    \end{equation}
    This is trivially obtained by substituting $Pe = 0$ and it is the correct asymptotic limit of the solution for vanishing $Pe$.
    
    \item First order ($Pe ^1$):
    \begin{equation}
    \frac{d}{d\tilde{k}} S_{0,out} (\tilde{k}) = \frac{2\tilde{k}}{\alpha} S_{1,out} (\tilde{k}).\nonumber
    \end{equation}
    With the previous result for $S_{0,out}$, the solution is
    \begin{equation} \label{eq:S1out}
    S_{1,out} (\tilde{k}) = \frac{\alpha}{2\tilde{k}} \frac{d}{d\tilde{k}} S_{eq} (\tilde{k}).
    \end{equation}
    
    \item Second order ($Pe ^2$): 
    \begin{equation}
    \frac{d}{d\tilde{k}} S_{1,out} (\tilde{k}) = \frac{2\tilde{k}}{\alpha} S_{2,out} (\tilde{k}).\nonumber
    \end{equation}
    
    The solution is
    \begin{equation} \label{eq:S2out}
    \begin{split}
        \quad S_{2,out} (\tilde{k}) & = \frac{\alpha}{2\tilde{k}} \frac{d}{d\tilde{k}} S_{1,out} (\tilde{k}) = \\
        & = \frac{\alpha^2}{4\tilde{k}^2} \left[ \frac{d^2}{d\tilde{k}^2} S_{eq} (\tilde{k}) - \frac{1}{\tilde{k}} \frac{d}{d\tilde{k}} S_{eq} (\tilde{k}) \right].
        \end{split}
    \end{equation}
\end{itemize}

To evaluate the outer solution up to 2nd-order in $Pe$, $S_{eq} (\tilde{k})$ must be differentiated twice. In the end, the solution takes the form
\begin{equation} \label{eq:Sout_approx}
\begin{split}
    S_{out} (\tilde{k}) = & S_{eq} (\tilde{k}) + Pe \frac{\alpha}{2\tilde{k}} \frac{d}{d\tilde{k}} S_{eq} (\tilde{k}) + \\
    & + Pe^2 \frac{\alpha^2}{4\tilde{k}^2} \left[ \frac{d^2}{d\tilde{k}^2} S_{eq} (\tilde{k}) - \frac{1}{\tilde{k}} \frac{d}{d\tilde{k}} S_{eq} (\tilde{k}) \right].
\end{split}
\end{equation}
$S_{eq} (\tilde{k})$ is calculated from Eq. (\ref{eq:Se}) once the pairwise potential $V(r)$ is given.

Additional perturbative corrections can be obtained by iterating the same scheme. The $n$-th term will contain a combination of derivatives of $S_{eq} (\tilde{k})$ up to the $n$-th, thus for the third order correction one has to calculate $\frac{d^3}{d\tilde{k}^3} S_{eq} (\tilde{k})$ in addition to $\frac{d^2}{d\tilde{k}^2} S_{eq} (\tilde{k})$ and $\frac{d}{d\tilde{k}} S_{eq} (\tilde{k})$, which are already necessary for the second order.

Notice that, starting from the first order, the solving scheme is recursive, i. e. for the $n$-th order with $n \geq 1$ it has the form,
\begin{equation}
    \frac{d}{d\tilde{k}} S_{n-1,out}(\tilde{k}) = \frac{2 \tilde{k}}{\alpha} S_{n,out}(\tilde{k}),
\end{equation}
and its solution is given in terms of the derivative of the previous step:
\begin{equation}
    S_{n,out}(\tilde{k}) = \frac{\alpha}{2 \tilde{k}} \frac{d}{d\tilde{k}} S_{n-1,out}(\tilde{k}).
\end{equation}

\subsubsection*{Inner solution}

The inner solution is not an analytical function of $Pe$. The boundary layer has a width $\delta$ that depends on $Pe$ \cite{Dhont_book}. We first have to introduce the inner transformation, allowing us to remove the singularity:
\begin{equation}
\begin{split}
    \tilde{k} = & \tilde{q} \delta(Pe)  \\
    S_{in}(\tilde{k}) & = S_{in}(\tilde{q}).
\end{split}
\end{equation}
Here, $\tilde{q}$ is a rescaled wavevector for describing the inner layer, where $\tilde{k}$ is small: $\tilde{k}<Pe^{1/2}$, corresponding to finite $\tilde{q}$ from $0$ to unity. The same notation $S_{in}$ is being used to indicate both the solution as a function of $\tilde{k}$ and as a function of $\tilde{q}$, even though they have distinct functional forms. Eq. (\ref{eq:smol_deltaS_adim}) now reads as
\begin{equation} \label{eq:in_transformed}
     \frac{Pe}{\delta(Pe)} \frac{d}{d\tilde{q}} S_{in}(\tilde{q}) = \delta(Pe) \frac{2\tilde{q}}{\alpha} \left[ S_{in}(\tilde{q}) - S_{eq}(\tilde{q}) \right].
\end{equation}

It is now possible to determine the width of the boundary layer, $\delta(Pe)$, through the principle of dominant balancing \cite{Bender_1978}. The procedure is particularly simple in this case: since there are only two terms in the last equation, they must be of the same order. The only value of $\delta(Pe)$ giving rise to a non trivial relation and removing the singularity is $\delta(Pe) = \mathcal{O} (Pe ^{1/2})$, which is the same result found by Dhont \cite{Dhont_1989, Dhont_book} (the same result can be obtained from dimensional analysis \cite{Zaccone_2009}). Thus, substitution of $\delta(Pe)=Pe^{1/2}$ into Eq. (\ref{eq:in_transformed}) gives
\begin{equation} \label{eq:smol_deltaS_in}
     \frac{d}{d\tilde{q}} S_{in}(\tilde{q}) = \frac{2\tilde{q}}{\alpha} \left[ S_{in}(\tilde{q}) - S_{eq}(\tilde{q}) \right] .
\end{equation}

The inner solution is valid for $\tilde{q} < 1$, which corresponds to $\tilde{k} < \delta(Pe)=Pe^{1/2}$. At the point $\tilde{k} = \delta(Pe)=Pe^{1/2}$ it must match the outer solution in a smooth (differentiable as many times as possible) way. In \cite{Banetta_2019, Banetta_2020} this was obtained by applying the ``patching'' procedure to determine suitable values for the integration constants in the inner solution. However, this procedure does not work in our case because of the requirements of second-order continuity being redundant, therefore we adopted a different approach. Eq. (\ref{eq:smol_deltaS_in}) does not depend on the perturbative parameter $Pe$, which implies that its solution gives the leading-order $Pe$-independent approximation for the structure factor.

There is a more appropriate condition than the continuity with the outer solution that can be imposed, which is that in the limit of zero shear rate the solution reduces to the equilibrium one, given by Eq. (\ref{eq:Se}). This is a fundamental requirement in a theory for low shear rates, which must return the correct limit when the shear rate vanishes. We calculated the inner solution with a procedure derived from the method of the variation of constants illustrated in \cite{Dhont_book}, in which such condition is intrinsically satisfied. It is possible to show that this solution automatically matches with the outer one in a smooth way even if it is not explicitly required. We refer to Appendix \ref{app:Sin} for the derivation. The resulting leading-order approximation in the extensional quadrants in terms of the original variable $\tilde{k} = Pe^{1/2} \tilde{q}$ is:
\begin{equation} \label{eq:S0_in}
   S_{in}^e(\tilde{k}) = \left[ \int_{\tilde{k}}^{\infty} dQ \frac{2Q}{Pe \, \alpha_e} S_{eq}(Q) e^{-\frac{Q^2}{Pe \, \alpha_e}} \right] e^{\frac{\tilde{k}^2}{Pe \, \alpha_e}}.
\end{equation}

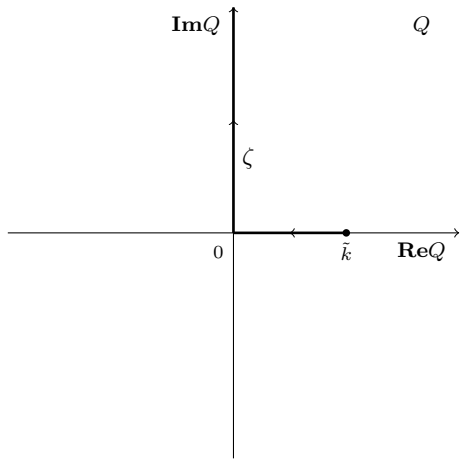
\begin{figure}
\centering

\begin{tikzpicture}

\draw [->] (0,3) -- (6,3);
\draw [->] (3,0) -- (3,6);

\node at (6-0.5,3-0.25) {\footnotesize {$\mathbf{Re}Q$}};
\node at (3-0.5,6-0.25) {\footnotesize {$\mathbf{Im}Q$}};
\node at (6-0.5,6-0.25) {\footnotesize {$Q$}};
\fill[black] (4.5,3) circle (0.5mm);
\node at (4.5-0,3-0.25) {\scriptsize {$\tilde{k}$}};
\node at (3-0.2,3-0.25) {\scriptsize {$0$}};

\draw [line width =1] (4.5,3) -- (3,3);
\draw [line width =1] (3,3) -- (3,6);
\draw [->] (4.5,3) -- (3.75,3);
\draw [->] (3,3) -- (3,4.5);

\node at (3+0.2,4.5-0.5) {\small {$\zeta$}};

\end{tikzpicture}

\caption{
The path in the complex $Q$-plane along which the integration in Eq.(\ref{eq:S0_in_complex}) is carried out. To calculate the integral it is useful to separate it into two straight lines, for details and explanation see the main text and Appendix \ref{app:compressing}.
} \label{fig:complex_path}
\end{figure}

We remind that $\alpha_e=\frac{1}{3\pi}$, while $\alpha_c=-\frac{1}{3\pi}$. The negative sign in the factor $\alpha$ in the compressing quadrants makes a new strategy necessary, as we can easily see from the integral in Eq. (\ref{eq:S0_in}), which diverges. In this case, the integration must be carried out on a broken line $\zeta$ in the complex $Q$-plane as illustrated in Fig. \ref{fig:complex_path}. Nevertheless, it is possible to show that the integral is real, by splitting the curve $\zeta$ into two straight lines and applying the trivial change of variable $Q=i t$ on the imaginary axis. For more details and justification of this procedure see also Appendix \ref{app:compressing}.

The resulting leading-order approximation in the compressing quadrants is: 
\begin{equation} \label{eq:S0_in_complex}
   S_{in}^c(\tilde{k}) = \left[ \int_{\zeta} dQ \frac{2Q}{Pe \, \alpha_c} S_{eq}(Q) e^{- \left( \frac{Q^2}{Pe \, \alpha_c} - \frac{\tilde{k}^2}{Pe \, \alpha_c} \right)} \right].
\end{equation}
The choice of the integration path $\zeta$ is made so that this integral is convergent for $Q = i t$, $t\rightarrow \infty$.

Eqs. (\ref{eq:Sout_approx}), (\ref{eq:S0_in}) and (\ref{eq:S0_in_complex}) constitute the final form of the solution. The first one is valid for $\tilde{k}>\delta(Pe)=Pe^{1/2}$ both in extension and compression, the other two are valid for $\tilde{k}<\delta(Pe)=Pe^{1/2}$. Their combination produces continuous and differentiable functions.

\section{Results} \label{results}

The resulting Eqs. (\ref{eq:Sout_approx}), (\ref{eq:S0_in}) and (\ref{eq:S0_in_complex}) have been applied to charge-stabilized colloids, for which we implemented the screened-Coulomb Yukawa (or Debye-H{\"u}ckel) two-body interaction potential from the DLVO theory. A hard-core repulsive wall is added to avoid particles interpenetration. The dimensionless potential $\tilde{V}=\beta V$ has the form,
\begin{equation} \label{eq:Yuk_potential}
    \tilde{V}(\tilde{r}) = \left \{\begin{array}{rl}
    \infty \quad \quad \quad \quad \quad \quad \quad \quad \quad \quad \quad \quad \quad \quad \quad & \mathrm{if} \, \tilde{r}<2 \\
    \frac{Z^2 L_B e^{2\tilde{\kappa}}}{a (1+\tilde{\kappa})^2} \frac{\exp(- \tilde{\kappa} \tilde{r})}{\tilde{r}} = \Gamma \frac{\exp(- \tilde{\kappa} \tilde{r})}{\tilde{r}} \quad \, & \mathrm{if} \, \tilde{r}>2
    
    \end{array}
    \right. ,
\end{equation}
where $r=\tilde{r}a$, $Z$ is the effective particle charge in units of the electron charge $e$, $L_{B}=\frac{\beta e^2}{4 \pi \epsilon_0 \epsilon_r}$ is the Bjerrum length with $\epsilon_0$ and $\epsilon_r$ the dielectric permittivity of vacuum and the medium relative dielectric permittivity, and $\kappa=\tilde{\kappa}/a$ is the Debye screening parameter, which is the inverse of the Debye screening length. In a water solution ($\epsilon_r = 78.3$) at room temperature ($T=298K$) $L_{B}\approx0.71 nm$.
$\Gamma$ and $\tilde{\kappa}$ are the potential parameters that can be freely adjusted in our model and they are both dimensionless. Increasing $\Gamma$ at fixed $\tilde{\kappa}$ has roughly the same effect as decreasing $\tilde{\kappa}$ at fixed $\Gamma$, making the electrostatic repulsion overall stronger.

The form of the potential influences the structure factor through the equilibrium solution, for which we have adopted the dilute approximation given by Eq. (\ref{eq:Se}). The integrals in Eqs. (\ref{eq:Se}), (\ref{eq:S0_in}) and (\ref{eq:S0_in_complex}) are evaluated numerically using adaptive quadrature methods.

Besides the structure factor for the Yukawa potential, whose results are presented in this Section, we also computed the structure factor for hard-sphere particles, for which we refer to Appendix \ref{app:HS}. Here we fix the volume fraction $\phi=0.04$, which is low enough for the dilute approximation to hold. The model is limited to Debye screening parameters between $1.5$ and $3$. Moreover the electrostatic potential at the particle surface $\Psi_0$ has to be lower than $25mV$ in order to apply the Debye-H\"uckel linear approximation. We choose $\tilde{\kappa}=2$ and $\Psi_0=10mV$ and calculate the coupling constant accordingly (see Appendix \ref{app:DLVO} for the relation between these three quantities and a discussion of the limitations above), finding $\Gamma=184.98$ for particles of radius $a=16nm$. We operate under the assumption of constant potential as opposed to constant surface charge. As stated in \cite{Verwey_1948}, this constitutes a more reasonable assumption since the potential needs to remain constant during the approach of two colloidal particles in order to maintain the thermodynamic equilibrium.

\subsection{Low Péclet numbers} \label{low_Pe}

At low Péclet numbers ($Pe<1$) the flow field causes a weak distortion of the structure factor with respect to equilibrium. For this reason we focus on this distortion, namely the difference between the sheared solution and the equilibrium solution: $\Delta S (\tilde{k}) = S (\tilde{k}) - S_{eq} (\tilde{k})$. We first study the distortion upon varying the Péclet number.

\begin{figure}[t]
    \centering
    \includegraphics[width=0.45 \textwidth, keepaspectratio]{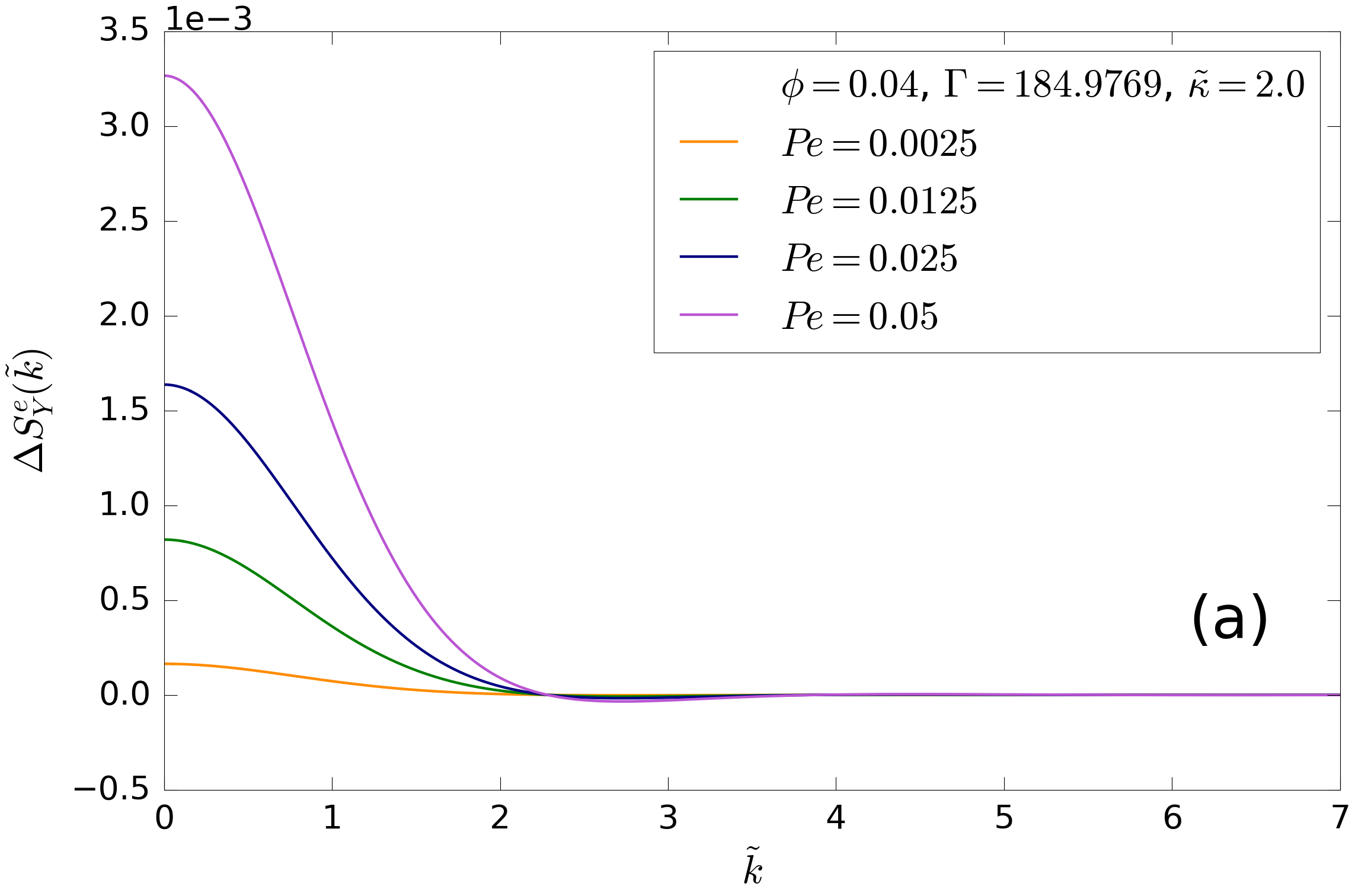}
    \\
    \vspace{0.4cm}
    \includegraphics[width=0.45 \textwidth, keepaspectratio]{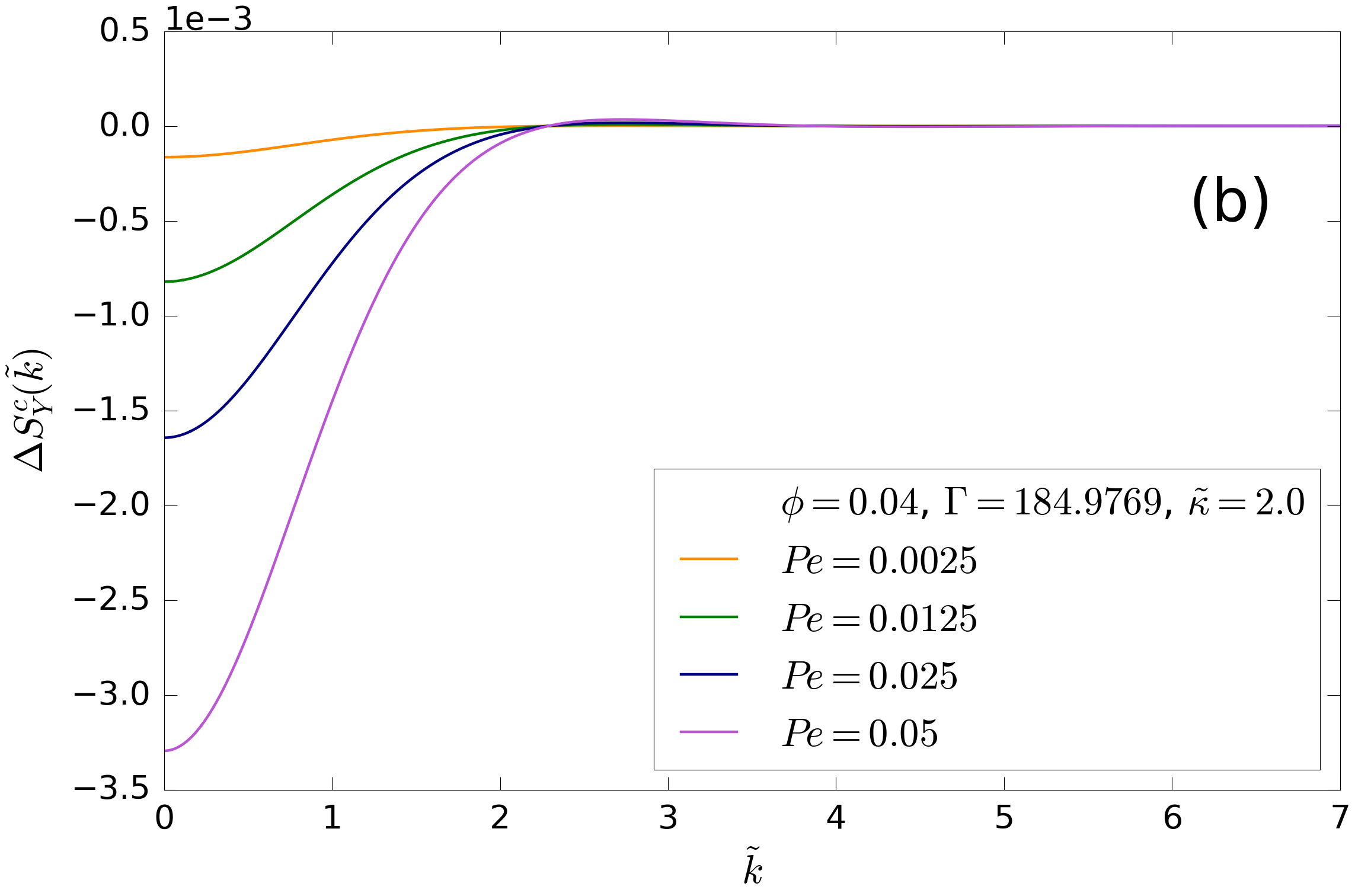}
    \caption{Angular average of the structure factor distortion for Yukawa interaction potential at fixed $\Gamma$, $\tilde{\kappa}$ and $\phi$ and varying $Pe$. \textbf{Panel (a)} shows the distortion averaged on the extensional quadrants. \textbf{Panel (b)} shows the distortion averaged on the compressing quadrants.}
    \label{fig:Yuk_varying_Pe_e_c}
\end{figure}

\begin{figure}[t]
    \centering
    \includegraphics[width=0.45 \textwidth, keepaspectratio]{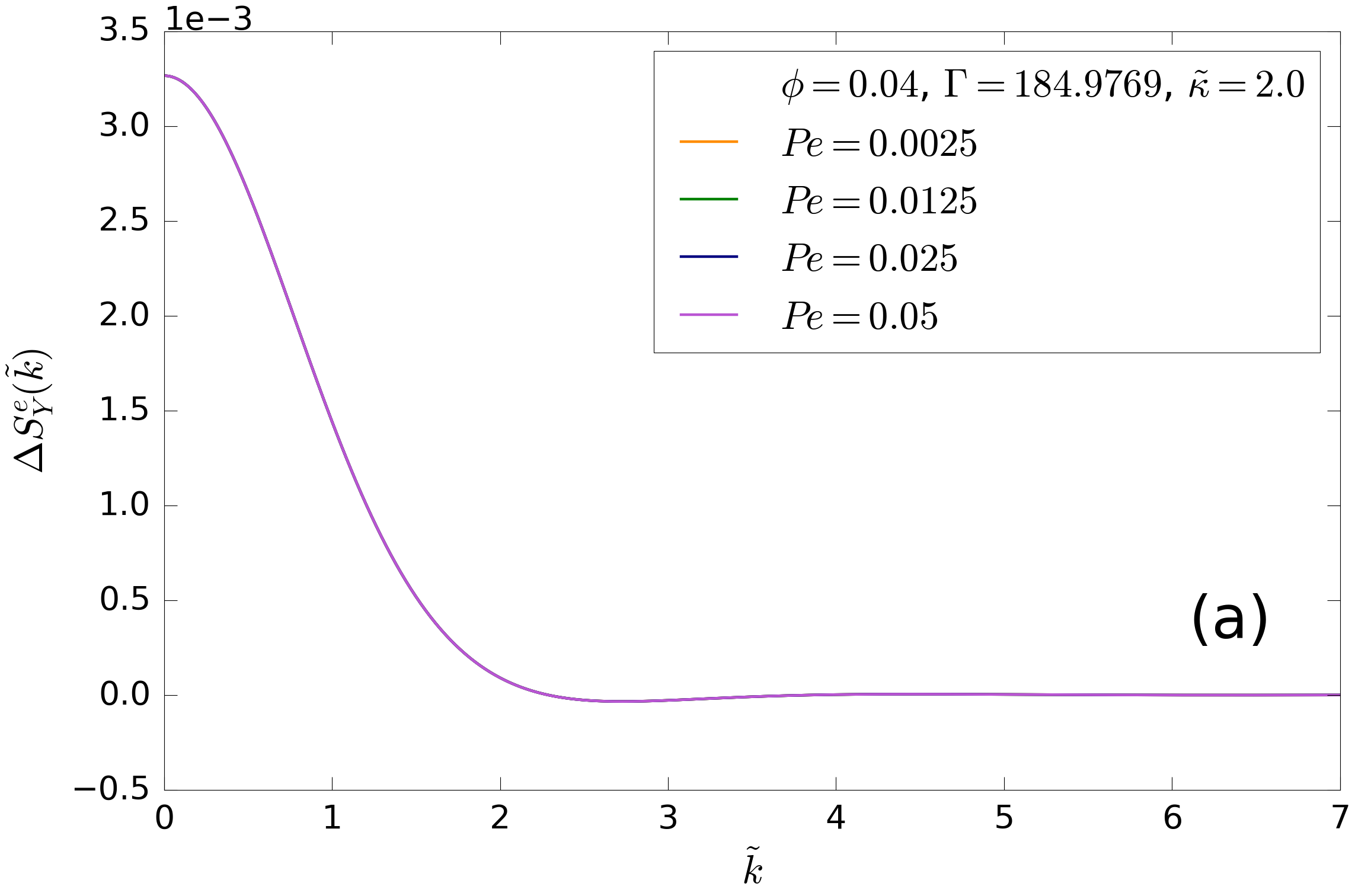}
     \\
    \vspace{0.4cm}
    \includegraphics[width=0.45 \textwidth, keepaspectratio]{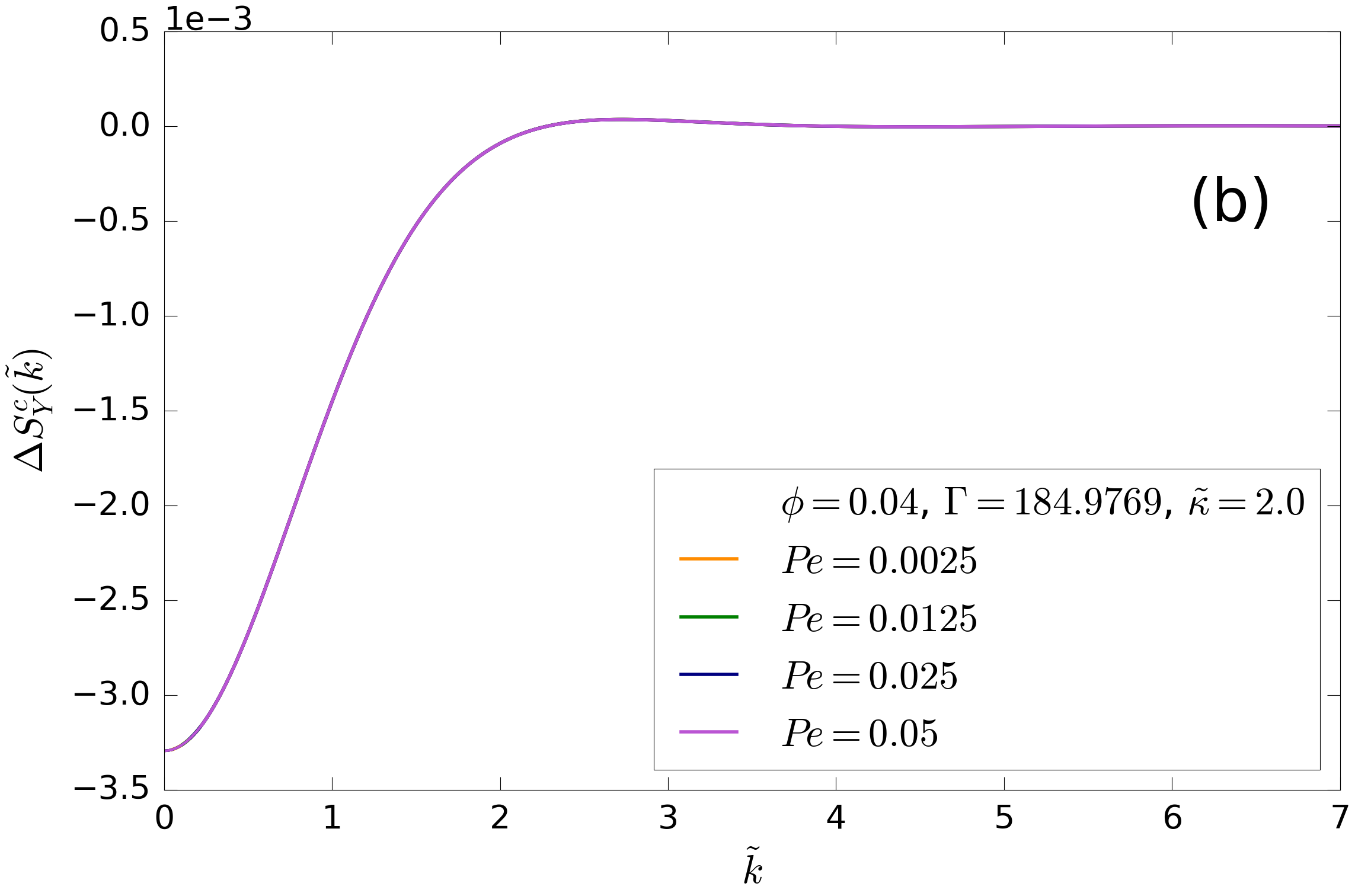}
    \caption{Angular average of the rescaled structure factor distortion for Yukawa interaction potential at fixed $\Gamma$, $\tilde{\kappa}$ and $\phi$ and varying $Pe$. \textbf{Panel (a)} shows the distortion averaged on the extensional quadrants. \textbf{Panel (b)} shows the distortion averaged on the compressing quadrants. All curves are rescaled by a factor proportional to $Pe$ to make them coincide at $\tilde{k}=0$ (see Tab. \ref{tab:scale_factors_e_c} for the scaling factors).}
    \label{fig:Yuk_varying_Pe_e_c_rescaled}
\end{figure}

In Fig. \ref{fig:Yuk_varying_Pe_e_c} we present the structure factor distortion for different $Pe$, while the potential parameters $\Gamma$ and $\tilde{\kappa}$ have the fixed values reported in the figure. The different curves are proportional to the corresponding $Pe$; indeed, if we rescale them to make their starting point ($\tilde{k}=0$) coincide with the curve for $Pe=0.05$ (purple), they all overlap as illustrated in Fig. \ref{fig:Yuk_varying_Pe_e_c_rescaled}. The scaling factors multiplied by $0.05$ are reported in Tab. \ref{tab:scale_factors_e_c} for the various curves, where they can be seen to be almost equal to the respective values of the Péclet number.

We conclude that the structure factor distortion averaged over the extensional and compressing quadrants is proportional to $Pe$. From what we have observed, all our results, including this, are valid for low $Pe$, the upper limiting value of the theory being $Pe \leq 0.75$.

This is a result we might have expected, since it means that the distortion, which is the response to the flow perturbation, is linear in the perturbation parameter ($Pe$), for weak shear flows. It can be explained for the outer part of the solution considering the expression we have obtained, Eq. (\ref{eq:Sout_approx}), where the linear term is the dominant correction at small $Pe$. A more general argument, which holds for the outer solution as well as for the inner, can be made by looking at the Smoluchowski equation in reduced units, Eq. (\ref{eq:smol_deltaS_adim}). At the leading order, $S(\tilde{k})$ in the left-hand side can be approximated by $S_{eq}(\tilde{k})$, since we have imposed this limit to be valid at $Pe=0$ for both solutions. This highlights that the correction $\Delta S(\tilde{k}) = S(\tilde{k}) - S_{eq}(\tilde{k})$, after the first iteration for solving the equation, is directly proportional to $Pe$, and so is the distortion in the extensional and compressing quadrants.

Another consideration from Fig. \ref{fig:Yuk_varying_Pe_e_c} is that the distortion appears to be symmetrical between extensional and compressing quadrants. This again is justified by Eq. (\ref{eq:Sout_approx}), where the dominant linear-order correction contains the factor $\alpha$, whose sign switches form positive to negative, while $S_{eq}$ is isotropic and thus does not change from one quadrant to the other. However, we expect the quadratic term to introduce a slight asymmetry, as it contains $\alpha^2$ and thus brings the same contribution in extension and compression.

\begin{table}[thb]
	\begin{center}
		\begin{tabular}{ccccc}
			\toprule
				$Pe$ & $0.0025$ & $0.0125$ & $0.025$ & $0.05$ \\
				extensional & $0.00251$ & $0.01254$ & $0.02506$ & $0.05$ \\
				compressing & $0.00249$ & $0.01246$ & $0.02494$ & $0.05$ \\
			\toprule
		\end{tabular}
			\caption{Proportionality factors between the curves at different $Pe$ in the extensional and compressing quadrants multiplied by $0.05$. They are basically the same as the corresponding Péclet numbers. Values in the last column are exact as the curve is rescaled on itself.}
\label{tab:scale_factors_e_c}
	\end{center}
\end{table}

\begin{figure}[t]
    \centering
    \includegraphics[width=0.45 \textwidth, keepaspectratio]{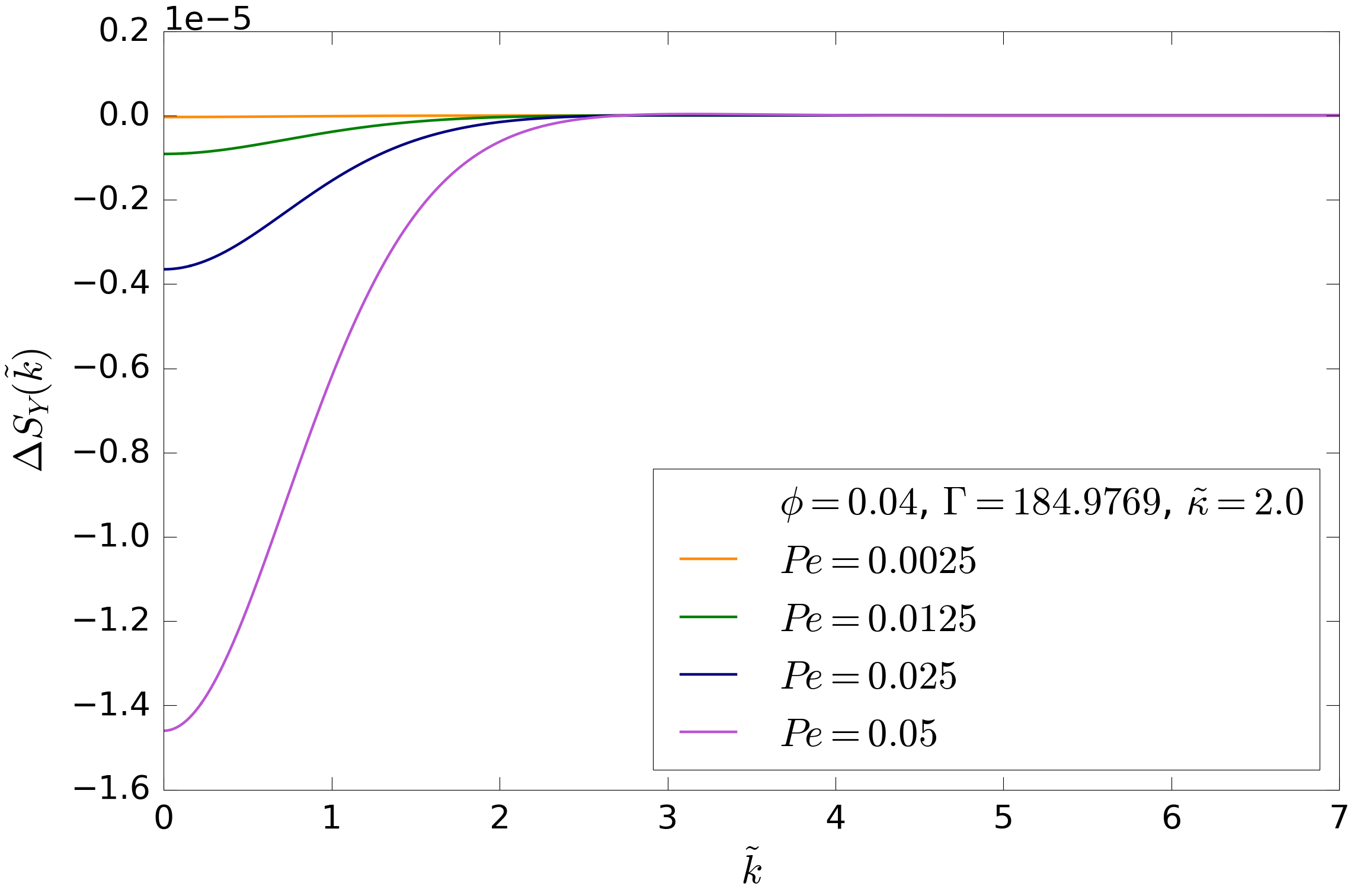}
    \caption{Angular average, taken over the whole solid angle, of the structure factor distortion for Yukawa interaction potential at fixed $\Gamma$, $\tilde{\kappa}$ and $\phi$ and upon varying $Pe$.}
    \label{fig:Yuk_varying_Pe_ave}
\end{figure}

To confirm this, we have calculated the algebraic mean of the two panels of Fig. \ref{fig:Yuk_varying_Pe_e_c}, which represents the distortion averaged over all directions in the $4\pi$ solid angle. The result is shown in Fig. \ref{fig:Yuk_varying_Pe_ave}. We can notice that, due to this asymmetry, the average is non-zero and the distortion is proportional to $Pe^2$.

\begin{figure}[t]
    \centering
    \includegraphics[width=0.45 \textwidth, keepaspectratio]{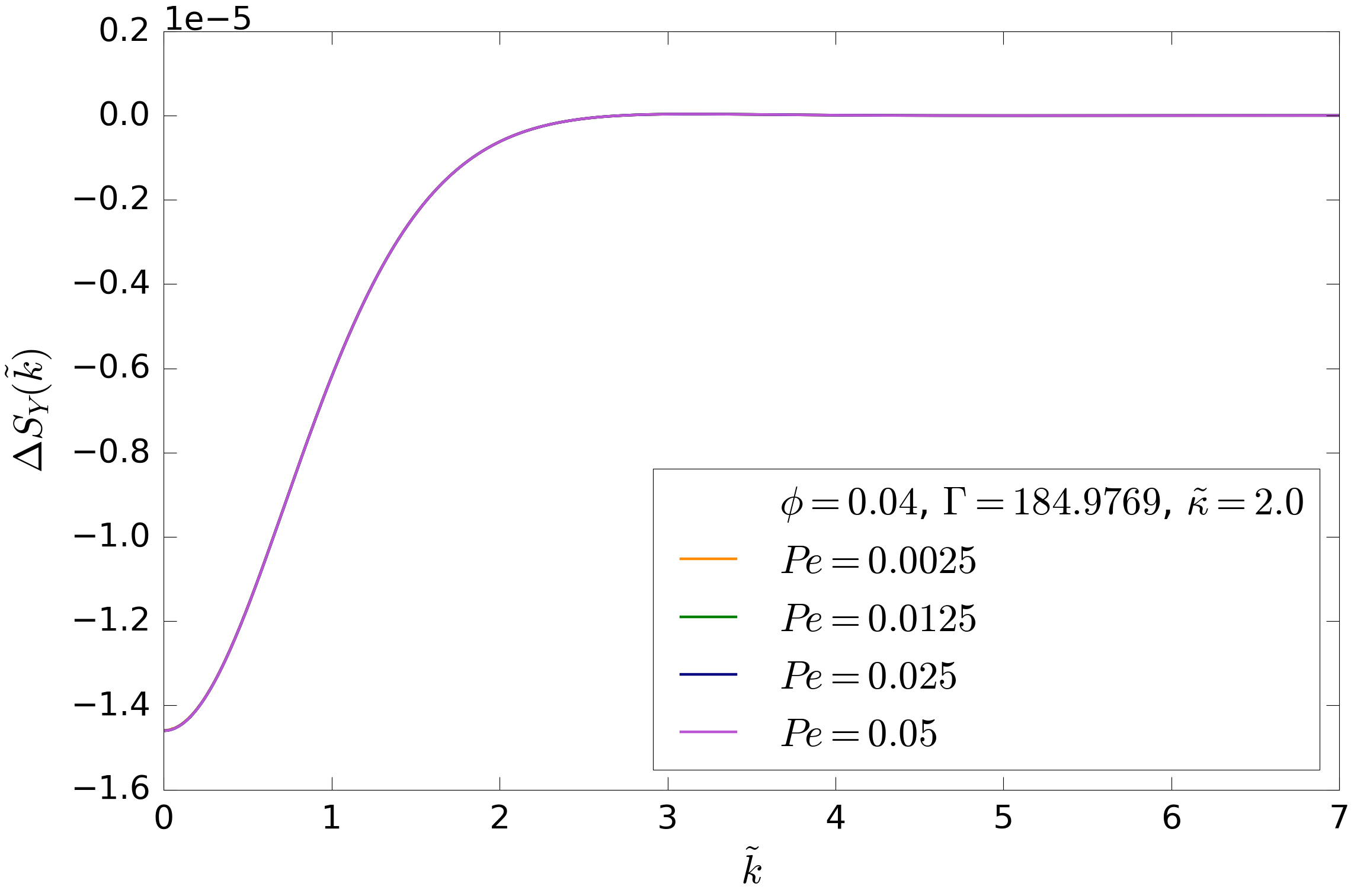}
    \caption{Angular average, taken over the whole solid angle, of the rescaled structure factor distortion for Yukawa interaction potential at fixed $\Gamma$, $\tilde{\kappa}$ and $\phi$ and varying $Pe$. All curves are rescaled by a factor proportional to $Pe^2$ to make them coincide at $\tilde{k}=0$ (see Tab. \ref{tab:scale_factors_ave} for the square roots of the scaling factors).}
    \label{fig:Yuk_varying_Pe_ave_rescaled}
\end{figure}

In Fig. \ref{fig:Yuk_varying_Pe_ave_rescaled} we apply the same rescaling procedure used in the extensional and compressing case to make the curves collapse onto each other. The square roots of the scaling factors multiplied by $0.05$ are reported in Tab. \ref{tab:scale_factors_ave} and are even more accurately close to the values of the Péclet number.

\begin{table}[thb]
	\begin{center}
		\begin{tabular}{ccccc}
			\toprule
				$Pe$ & $0.0025$ & $0.0125$ & $0.025$ & $0.05$ \\
				average & $0.00250$ & $0.01250$ & $0.02500$ & $0.05$ \\
			\toprule
		\end{tabular}
			\caption{Square roots of the proportionality factors between the averaged curves at different $Pe$ multiplied by $0.05$. The last value is exact.}
\label{tab:scale_factors_ave}
	\end{center}
\end{table}

With the linear response method used by Dhont \cite{Dhont_1989} and Ronis \cite{Ronis}, the average of the distortion on the whole solid angle would be intrinsically null, because mirror-image directions with respect to the $k_2$-$k_3$ plane bring opposite contributions (in other words, there is odd parity with respect to the $k_2$-$k_3$ plane). In our theory, the linear term in $Pe$ vanishes as well, again due to symmetry, while the averaged distortion is controlled by the quadratic term $\sim Pe^{2}$.
This fact bears an important consequence for the overall shape of the structure factor peak (see the next Section) because the quadratic term $\sim Pe^{2}$ is also proportional to $\alpha^{2}$, but with a negative numerical prefactor. Hence, this  gives the same negative sign to the distortion as in the compressing quadrants (where $\alpha_{c}<0$), which is thus responsible for the overall narrowing of the peak with increasing $Pe$ observed for the fully-averaged structure factor.

This is in agreement with Ronis \cite{Ronis}, who also calculated this quadratic term from a fluctuating-diffusive equation, however, here we adopted a perturbative approach, which can, in theory, be iterated to any expansion order.

\subsection{Structure factor peak}

Having discussed the influence of the shear flow on the distortion $\Delta S (\tilde{k})$, we now study the bare structure factor $S (\tilde{k})$. We use higher values of $Pe$ in order to make the deviations from the equilibrium solution, which are a small fraction when $Pe < 1$, more evident.

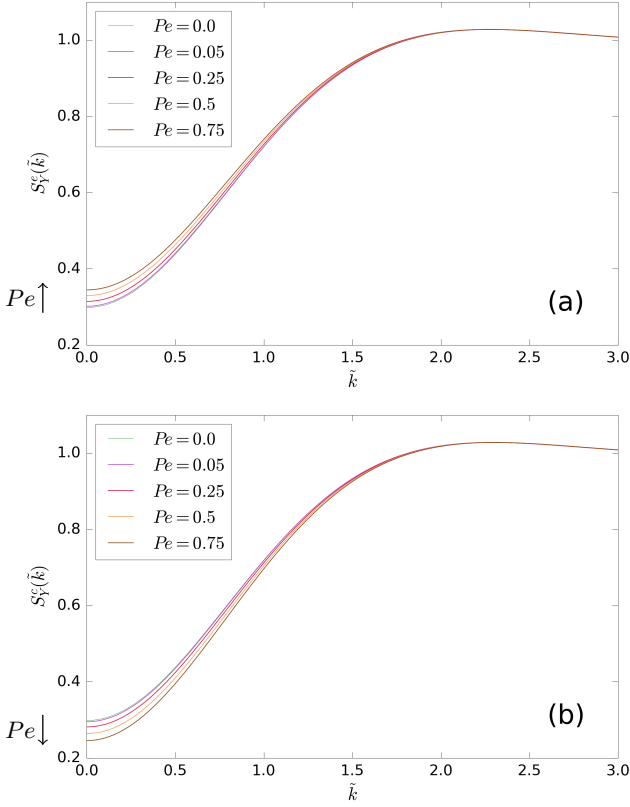
\begin{figure}[ht]
    \centering
    \begin{tikzpicture}
    \begin{scope}
    \node[inner sep=0] at (0,0) {\includegraphics[width=0.45 \textwidth, keepaspectratio]{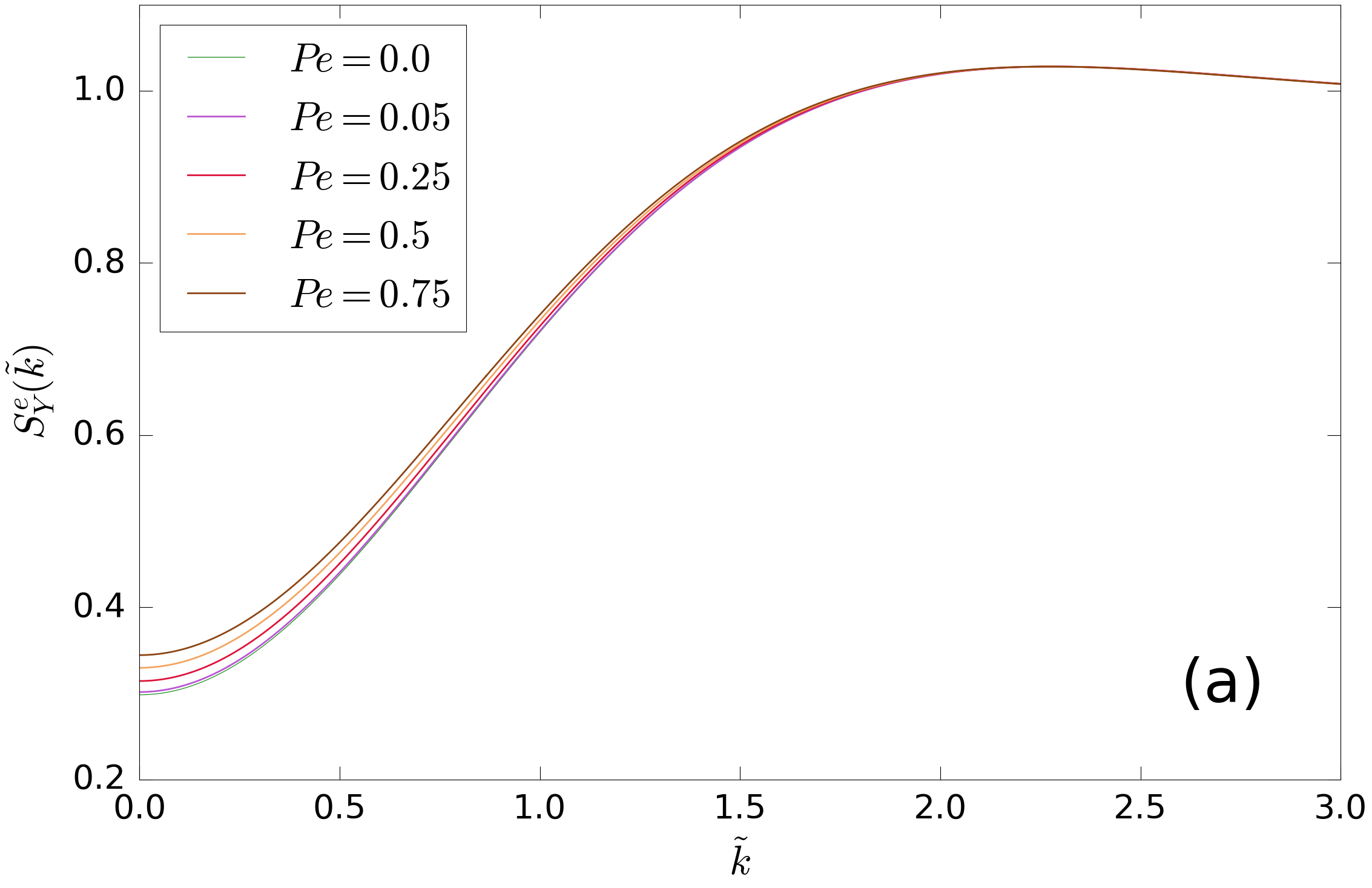}};
    \node[inner sep=0] at (0,-5.5) {\includegraphics[width=0.45 \textwidth, keepaspectratio]{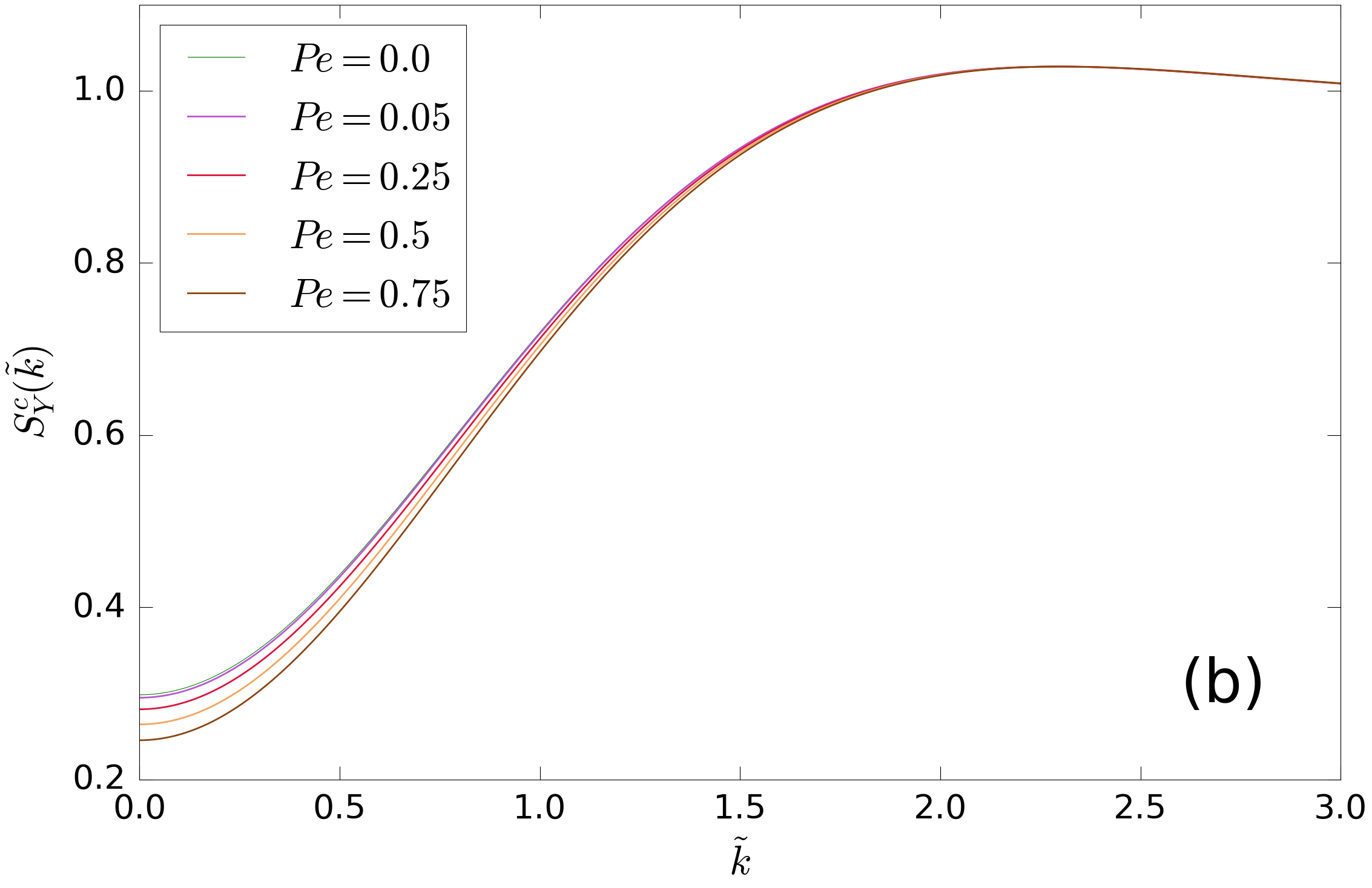}};
    \draw [->] [line width =0.5] (-3.8,-1.55) -- (-3.8,-1.15);
    \node at (-4.1,-1.35) { {$Pe$}};
    \draw [->] [line width =0.5] (-3.8,-6.9) -- (-3.8,-7.3);
    \node at (-4.1,-7.1) { {$Pe$}};
    \end{scope}
    \end{tikzpicture}
    \caption{Structure factor around the peak and at small $\tilde{k}$ at varying $Pe$. \textbf{Panel (a)} shows the structure factor averaged on the extensional quadrants, where a broadening of the peak is visible. \textbf{Panel (b)} shows the structure factor averaged on the compressing quadrants, where a narrowing of the peak is visible. $\Gamma$, $\tilde{\kappa}$ and $\phi$ are fixed as in Fig. \ref{fig:Yuk_varying_Pe_e_c}}
\label{fig:broadening_e_c}
\end{figure}

Fig. \ref{fig:broadening_e_c} shows $S(\tilde{k})$ at small wavevector $\tilde{k}$ and around the first peak. At small $\tilde{k}$, upon increasing $Pe$, the structure factor increases in the extensional quadrants and decreases in the compressing quadrants with respect to the equilibrium solution (green line). This is a consequence of the difference $\Delta S$ being positive and negative, for extension and compression, respectively  (see Fig. \ref{fig:Yuk_varying_Pe_e_c}), at $\tilde{k}=0$, and thus causes a slight broadening and narrowing of the peak, respectively.

\begin{figure}[ht]
    \centering
    \includegraphics[width=0.5 \textwidth, keepaspectratio]{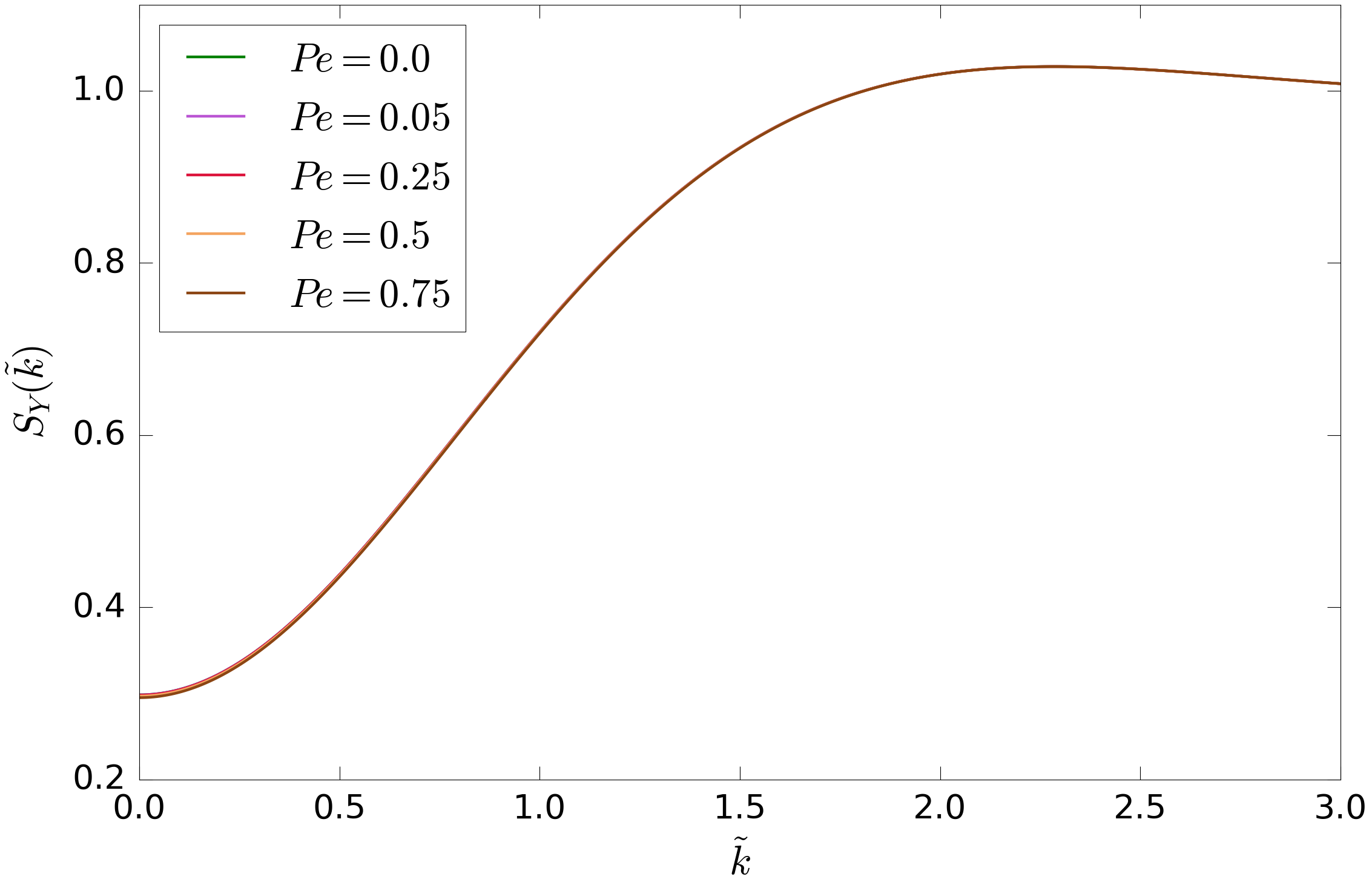}
    \begin{picture}(0,0)
    \put(-5,33){\includegraphics[height=3.7cm]{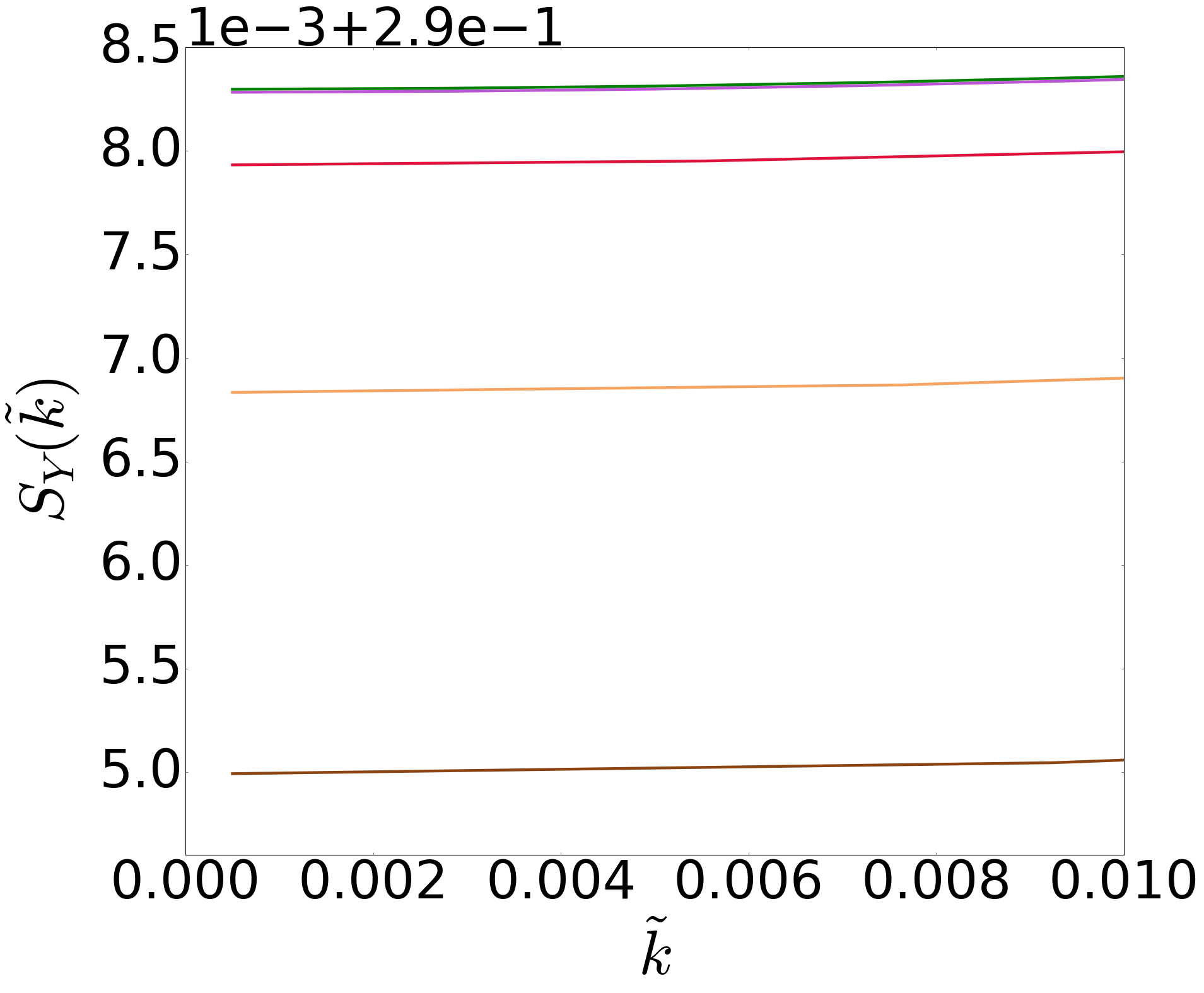}}
    \end{picture}
    \caption{Average of the structure factor, around the peak, for varying $Pe$, taken on the whole solid angle. $\Gamma$, $\tilde{\kappa}$ and $\phi$ are fixed as in Fig. \ref{fig:Yuk_varying_Pe_e_c}. The inset shows the structure factor at small $\tilde{k}$, highlighting a narrowing of the peak with increasing $Pe$.}
    \label{fig:broadening_ave}
\end{figure}

From Fig. \ref{fig:broadening_ave} the structure factor averaged over the whole solid angle is seen to have a similar behaviour to the compressing one, even though differences upon varying $Pe$ are less evident. The inset makes the different curves distinguishable near $\tilde{k}=0$. We find again a narrowing of the peak upon increasing $Pe$, similar to what was found in the compressing quadrants.
The reason was mentioned in the previous Section: due to symmetry, the wholly averaged structure factor is dominated by the term $\sim Pe^{2}$, which contains the factor $\alpha^{2}$ with a negative prefactor, thus having the same sign as the distortion in the compression quadrants ($\alpha_{c}<0$). 

\begin{figure}[ht]
    \centering
    \includegraphics[width=0.45 \textwidth, keepaspectratio]{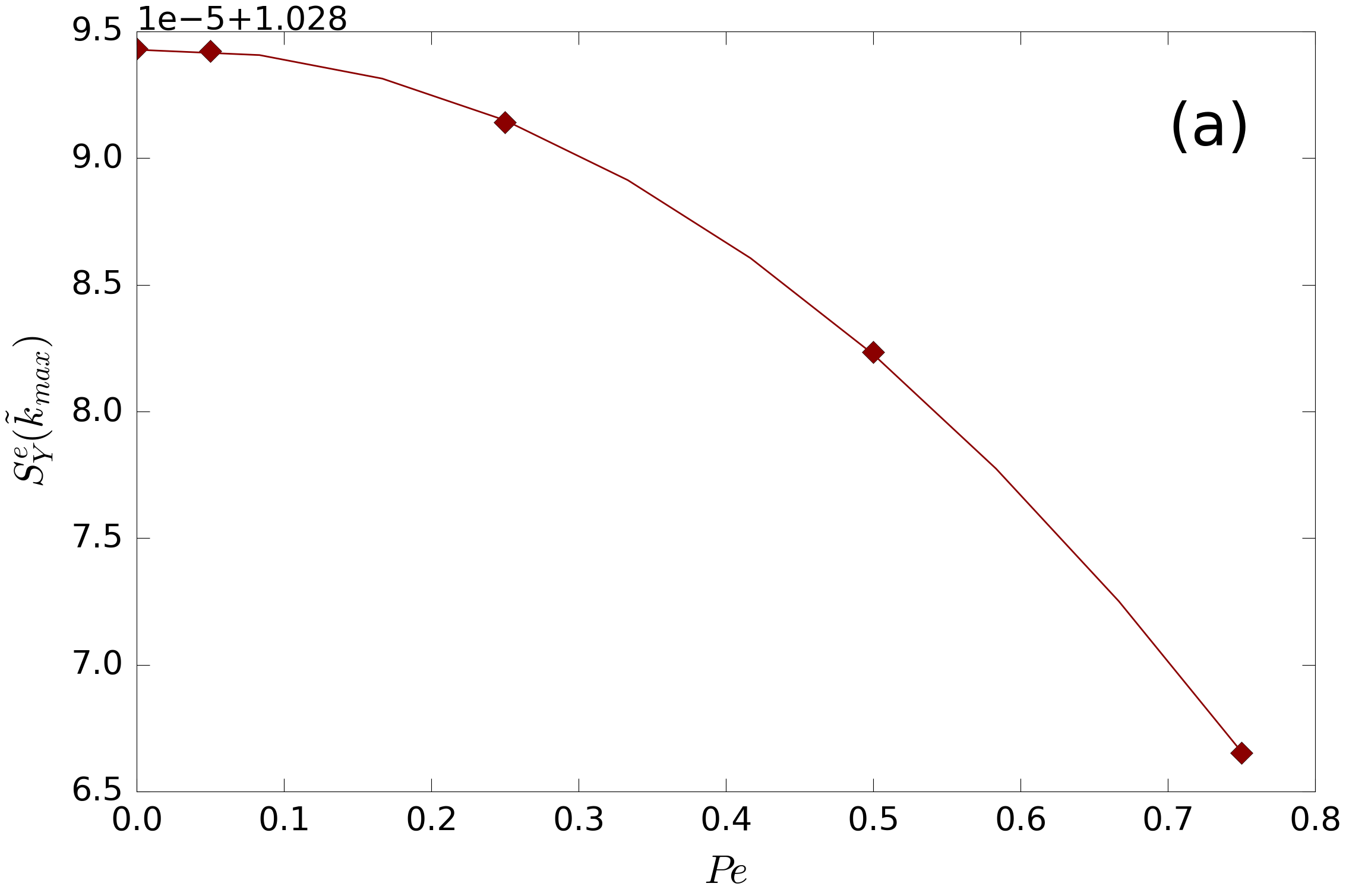}
    \\
    \vspace{0.4cm}
    \includegraphics[width=0.45 \textwidth, keepaspectratio]{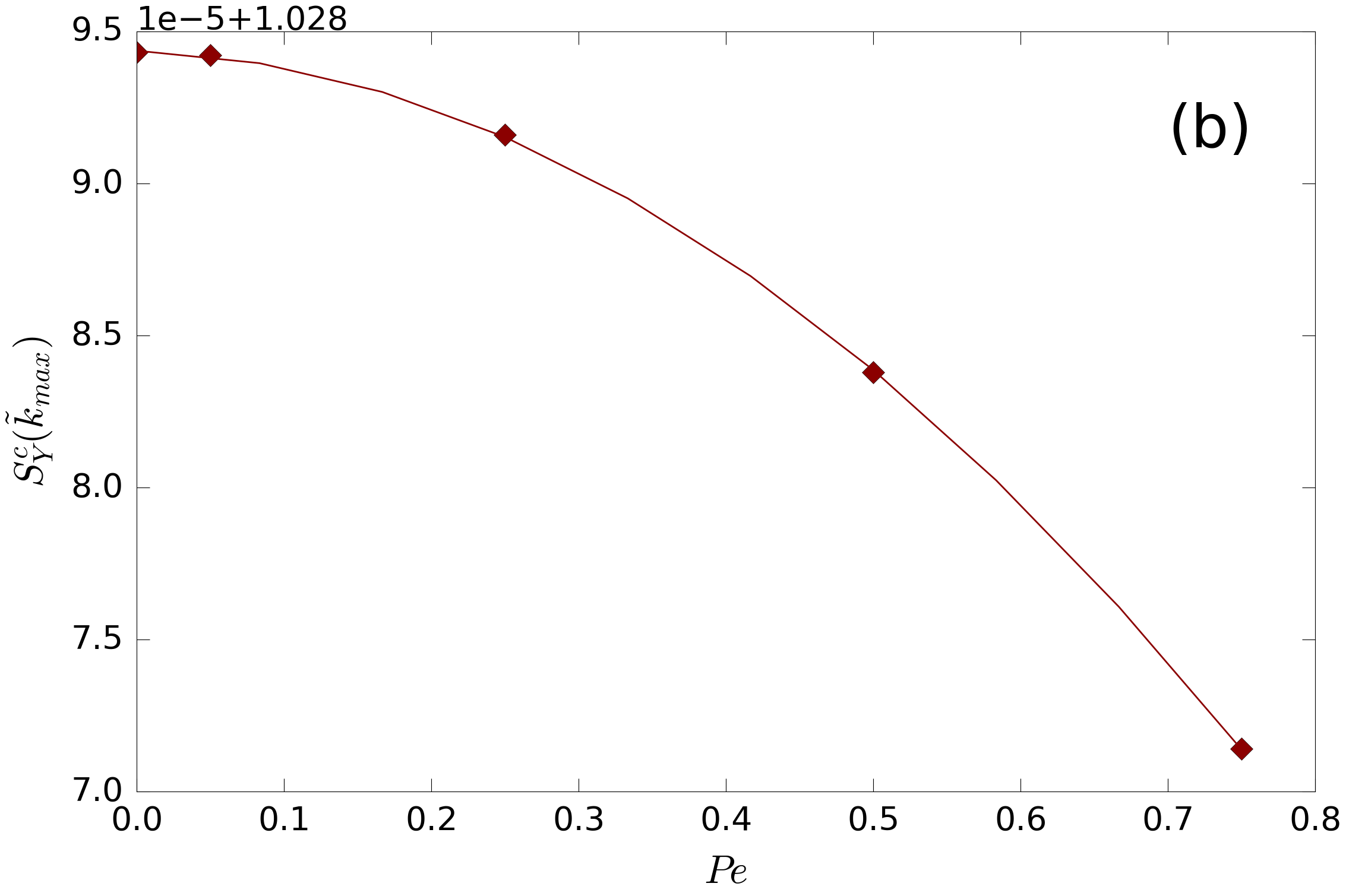}
    \\
    \vspace{0.4cm}
    \includegraphics[width=0.45 \textwidth, keepaspectratio]{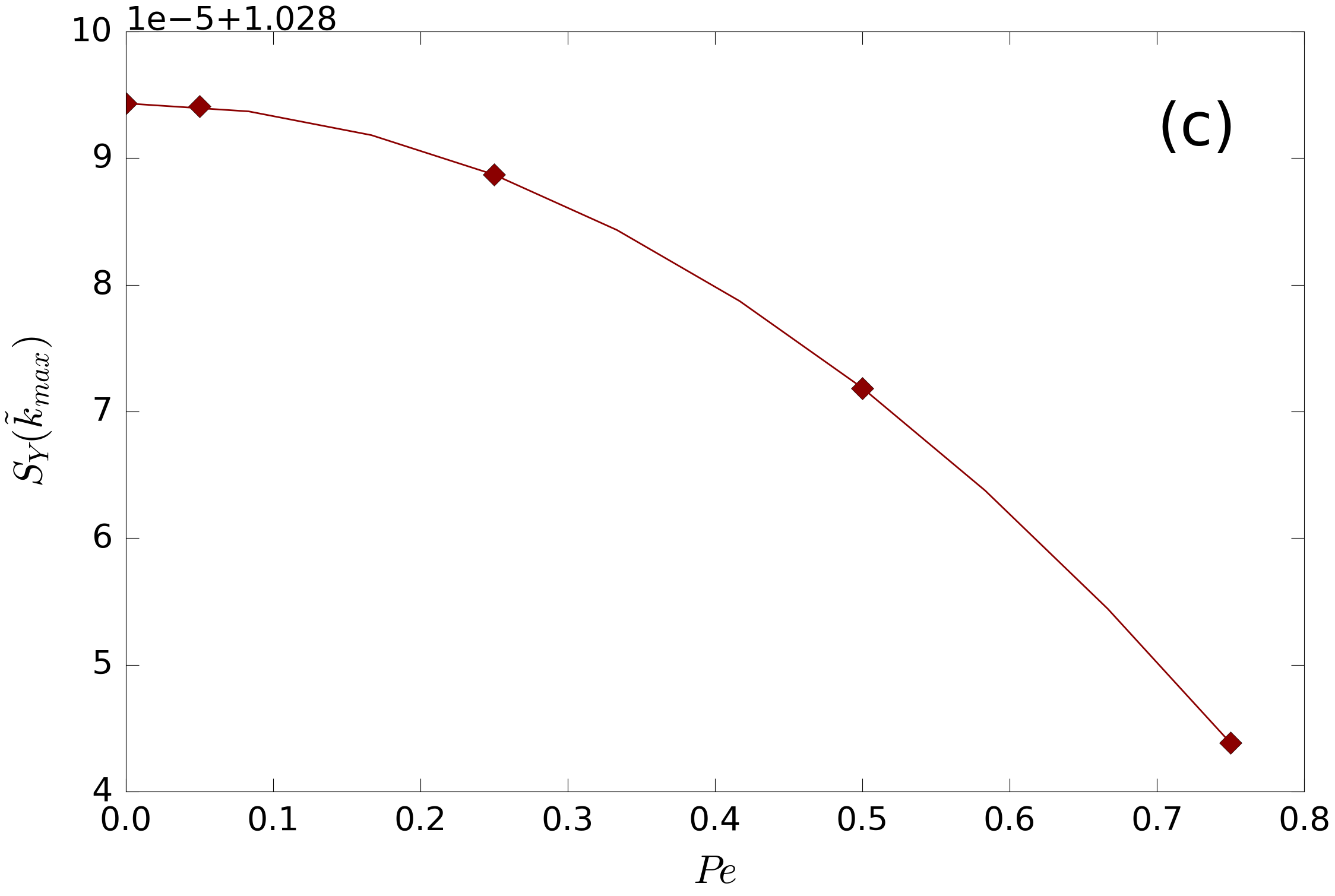}
    \caption{Maximum value of the structure factor first peak as a function of $Pe$. \textbf{Panel (a)} shows the maximum in the extensional quadrants, which decreases quadratically with prefactor $-5.16 \times 10^{-5}$. \textbf{Panel (b)} shows the maximum in the compressing quadrants, which decreases quadratically with a similar prefactor, $-3.86 \times 10^{-5}$. \textbf{Panel (c)} shows the total average on the solid angle, which decreases quadratically, but with a rather different prefactor, $-8.96 \times 10^{-5}$.}
\label{fig:broadening_S_max_vs_Pe_small_Pe}
\end{figure}

It is not obvious to determine whether the shear flow increases or decreases the height of the first peak, as the distortion passes through zero approximately in correspondence of the equilibrium structure factor peak, which is around $\tilde{k}=2.3$. From a zoom-in of the peak position we observe that its height is reduced by the flow in both the extensional and compressing quadrants. In Fig. \ref{fig:broadening_S_max_vs_Pe_small_Pe} the peak maximum is plotted as a function of $Pe$. The fit shows that the decrease is quadratic up to $Pe=0.75$ and the quadratic law is similar in extension (a) and compression (b), the quadratic coefficients being $-5.16 \times 10^{-5}$ and $-3.86 \times 10^{-5}$ respectively. In the fully averaged case (c), the peak still decreases with a quadratic law but with a larger multiplicative prefactor: $-8.96 \times 10^{-5}$.

\begin{figure}[h!]
    \centering
    \includegraphics[width=0.45 \textwidth, keepaspectratio]{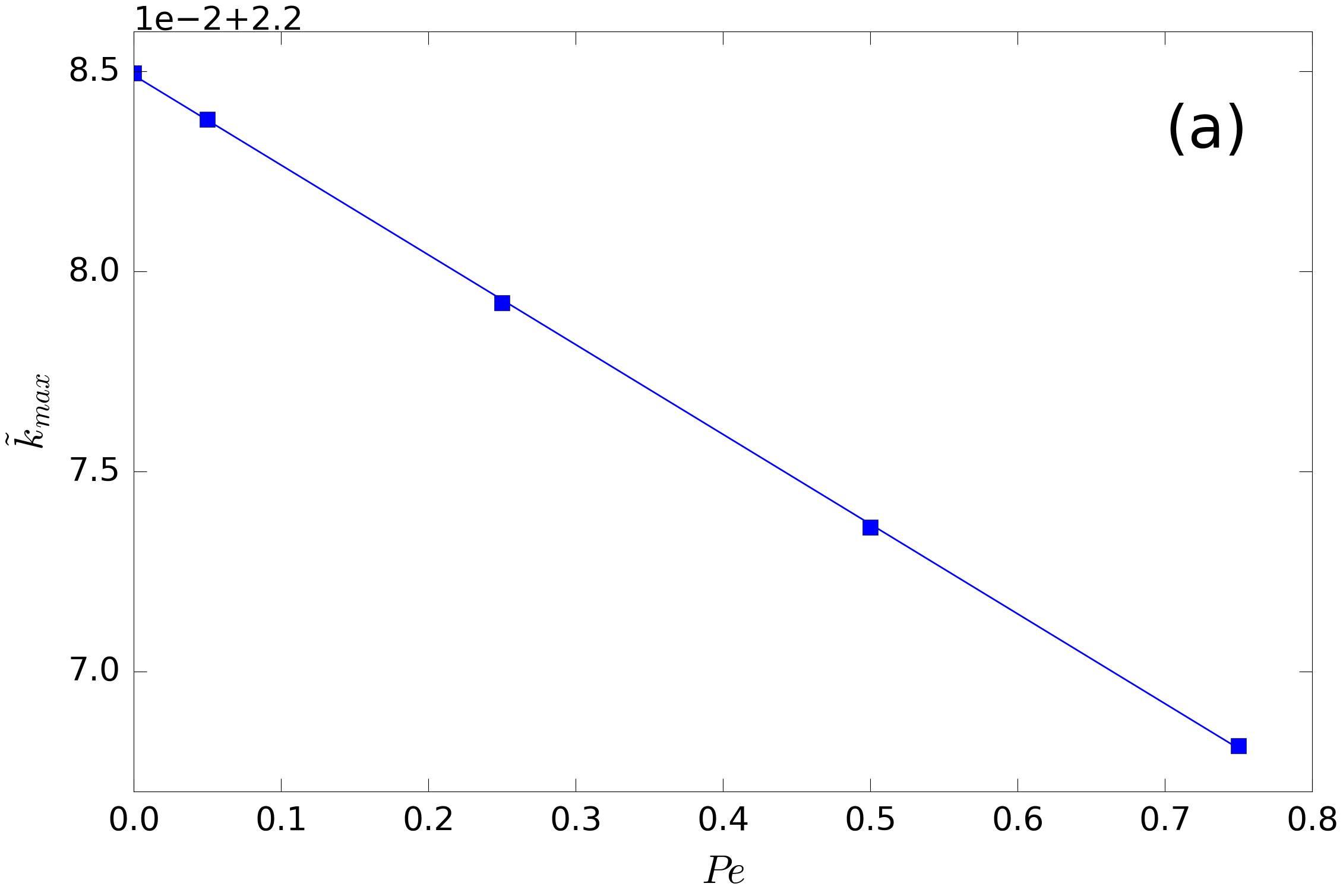}
    \\
    \vspace{0.4cm}
    \includegraphics[width=0.45 \textwidth, keepaspectratio]{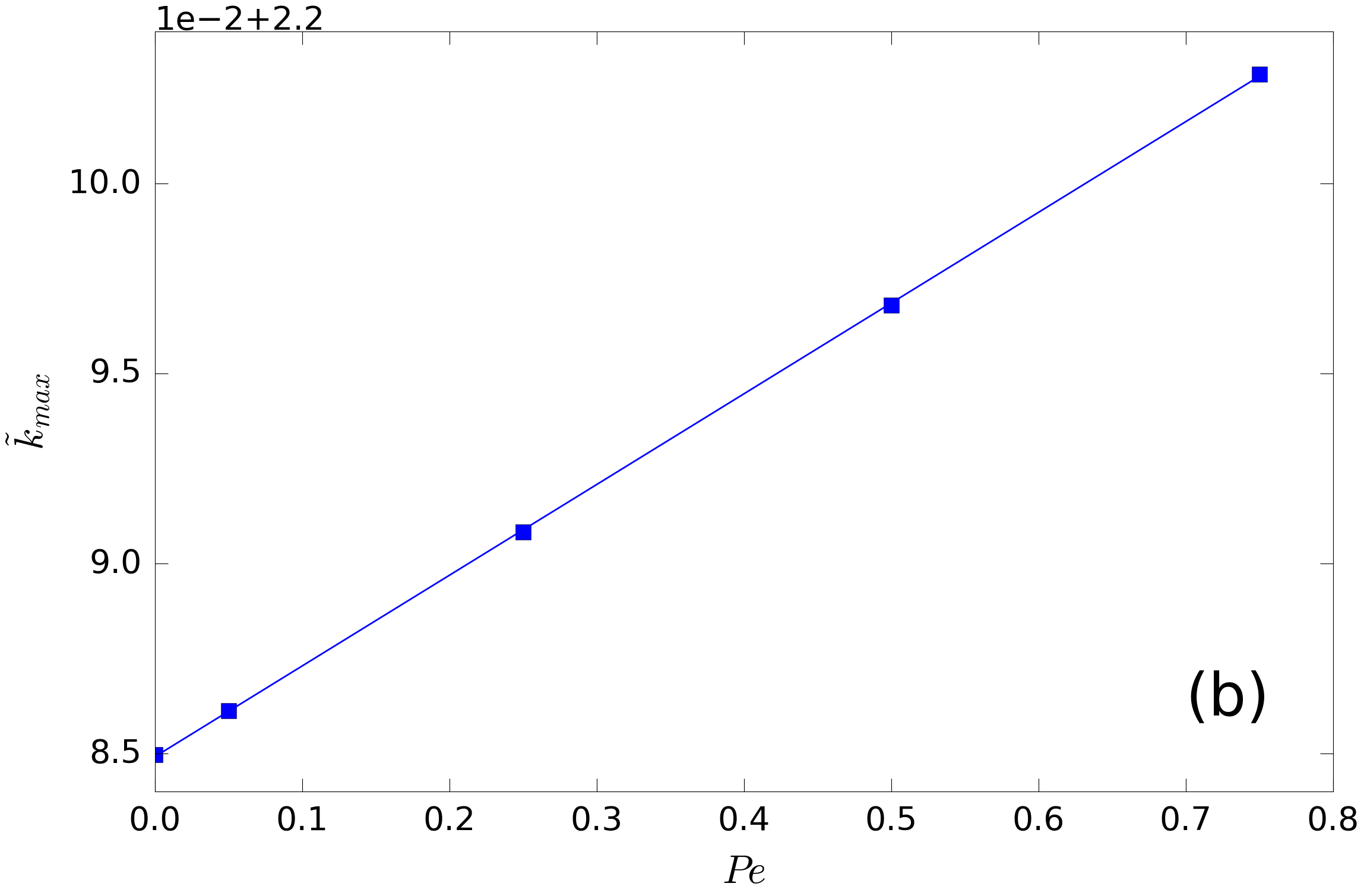}
    \\
    \vspace{0.4cm}
    \includegraphics[width=0.45 \textwidth, keepaspectratio]{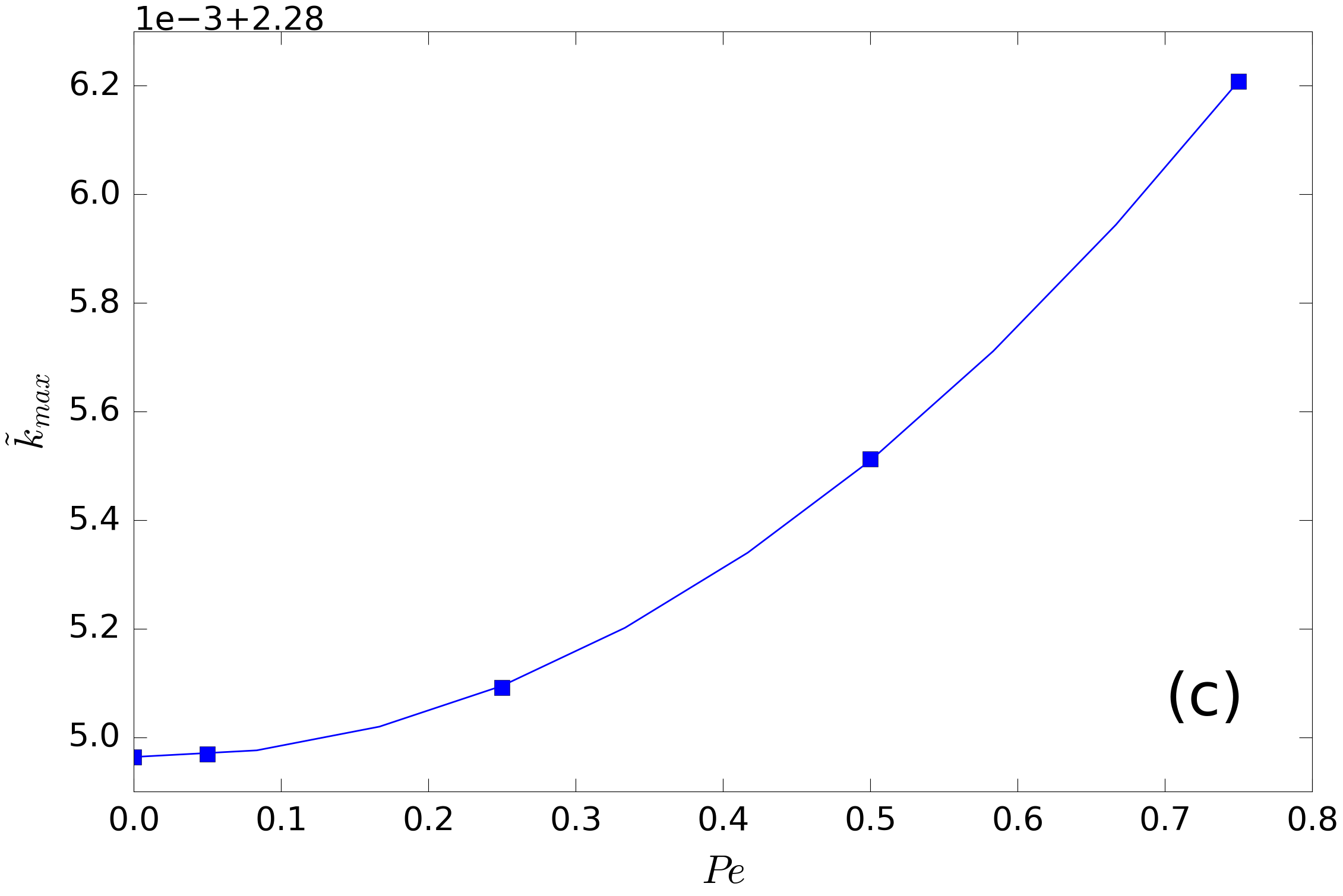}
    \caption{Position of the structure factor first peak as a function of $Pe$. \textbf{Panel (a)} shows the peak position in the extensional quadrants, which is shifted linearly towards lower $\tilde{k}$ with linear coefficient $-0.02244$. \textbf{Panel (b)} shows the peak position in the compressing quadrants, which is shifted linearly towards larger $\tilde{k}$ with linear coefficient $0.02388$. \textbf{Panel (c)} shows the total average on the solid angle, where the shift is quadratic with quadratic coefficient $0.00227$.}
\label{fig:broadening_k_max_vs_Pe_small_Pe}
\end{figure}

The flow also causes a shift of the peak, as shown in Fig. \ref{fig:broadening_k_max_vs_Pe_small_Pe}, where we report the position of the maximum, $\tilde{k}_{max}$ as a function of $Pe$. We notice that the peak is shifted towards lower values of $\tilde{k}$ in extension (a) and towards higher values in compression (b). In both cases, the shift is linear with the Péclet number and the linear coefficients are similar, apart from the sign: $-0.02244$ and $0.02388$. However, on average (c) the peak shifts to higher $\tilde{k}$ with a quadratic law, with quadratic coefficient $0.00227$.

\begin{figure}[ht]
    \centering
    \includegraphics[width=0.45\textwidth, keepaspectratio]{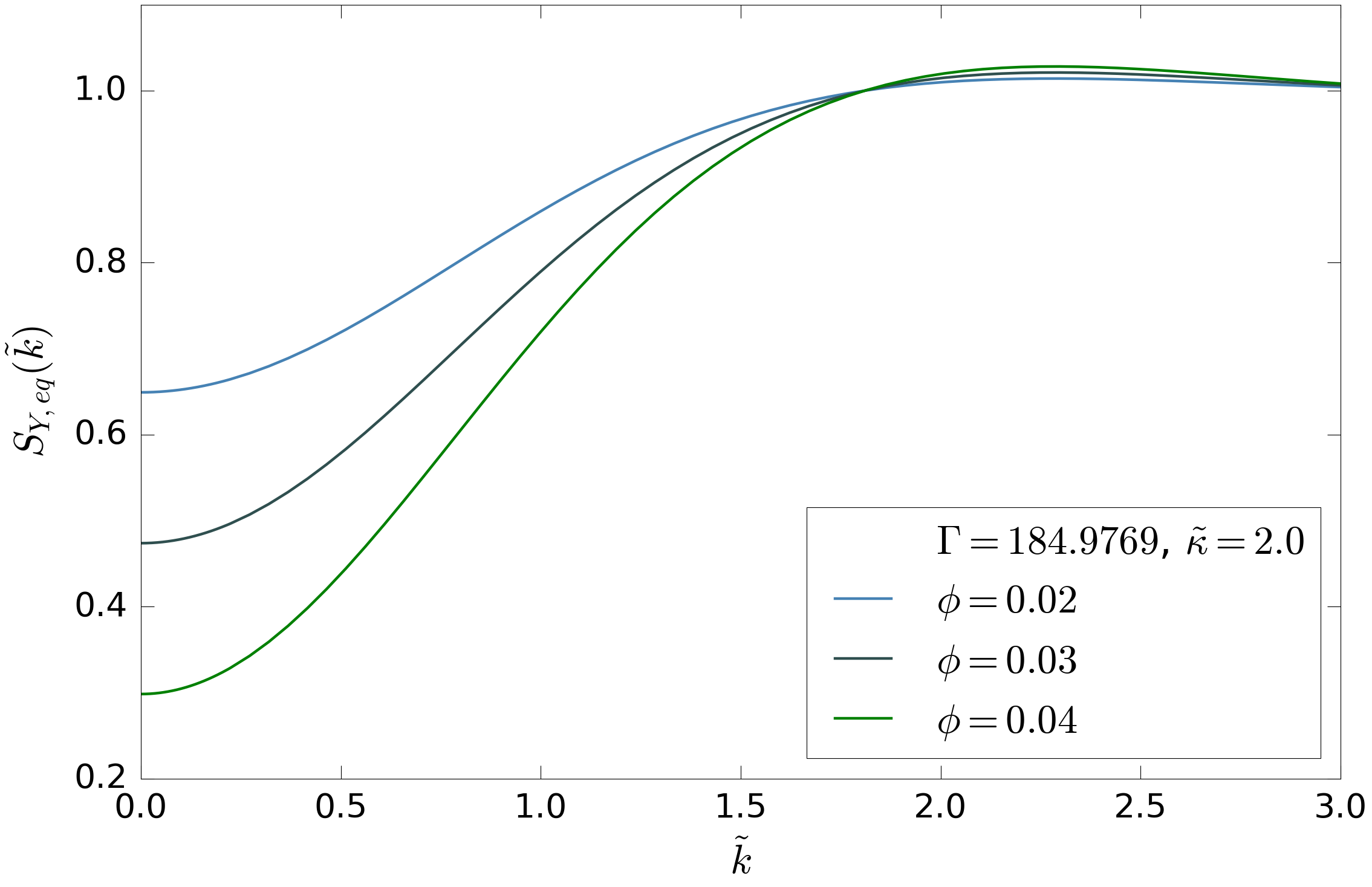}
    \caption{Equilibrium structure factor for Yukawa interaction potential at different volume fractions $\phi$ with fixed $\Gamma$ and $\tilde{\kappa}$. $Pe$ is kept equal to $0$. A growth at small $\tilde{k}$ and a decrease at the peak maximum can be observed when $\phi$ is reduced.}
    \label{fig:S_Y_phi}
\end{figure}

It is possible to observe a similarity between some features appearing when introducing the flow in the extensional and compressing quadrants and the behaviour of the equilibrium structure factor upon varying the particles density $\rho$ or the volume fraction $\phi=\frac{4}{3}\pi a^3 \rho$ of the solution. Fig. \ref{fig:S_Y_phi} shows the equilibrium solution obtained from Eq. (\ref{eq:Se}) with different volume fractions; we have chosen a wide range of $\phi$ to enhance the differences. An increasing volume fraction has the effect of reducing the structure factor at small $\tilde{k}$, thus leading to a narrowing of the peak, and at the same time increasing the maximum of the peak. On the contrary, decreasing $\phi$ makes $S(\tilde{k})$ larger at small $\tilde{k}$ and lowers the peak.

We notice that some of the effects of decreasing the concentration of particles, namely the increase of $S(\tilde{k})$ at small $\tilde{k}$ and the decrease of the peak, have also been predicted in the extensional quadrants upon increasing the shear rate. On the other hand, the decrease at small $\tilde{k}$ upon increasing $\phi$ is observed in the compressing quadrants, but not the rise of the peak, which, instead, gets reduced.

This suggests an explanation for the origin of the features observed in Fig. \ref{fig:broadening_e_c}. In the extensional quadrants, the flow drags particles away from each other, thus locally decreasing the concentration of particles, while in the compressing quadrants it pushes them against each other, thus locally increasing the concentration. As already pointed out by Johnson et al. \cite{Johnson}, the lowering of the peak in compression, where we would expect a rise, might be explained by the formation of small transient accumulations of particles near contact. This is because the peak in $S(\tilde{k})$ reflects the particle concentration/ordering in the medium-range, a few (2-3) particles away from contact. In the compressing quadrants, accumulation of particles near contact due to the local compressing effect of the shear, acting as an external forcing,  would deplete the medium-range region (thus lowering the peak) and push the peak towards contact. The concomitant narrowing predicted here is a manifestation of the shear-induced ``ordering'' in the compression quadrants.

Indeed, the Baxter theory \cite{Baxter} predicts a similar lowering, as well as a shift of the peak towards large $k$, for adhesive particles compared to hard spheres at the same concentration, due to the formation of small clusters of particles in contact. Although rigorously speaking our model, which is purely repulsive, cannot exhibit clusterization, the fact that we observe qualitatively similar features in such systems, suggests that an interplay of two effects, i. e. increasing local concentration near contact in compressing sectors due to the flow field and near-contact accumulation, might occur. These transient stacks of particles are different from ``hydro-clusters'' observed in systems where HIs are present \cite{Ball}, since those originate from the effect of lubrication forces, which make particles stick together. On the contrary, in our model the shear acts similarly to an external force (rather than an inter-particle attraction) in the compressing regions, causing the particles to get closer to each other, on average. As a consequence, such particle accumulations are expected to have a very limited lifespan, as opposed to ``hydro-clusters'', which are made long-lived by HIs.

In \cite{Johnson}, it was also shown, based on the Percus-Yevick theory (which makes more sensible predictions for the equilibrium pcf at higher concentrations), that, at least in the case of hard spheres, the peak is shifted towards lower $k$, besides being lowered, upon decreasing the particle concentration. This is what we also observed in the extensional quadrants and confirms our hypothesis of the system becoming locally less concentrated in these regions.

\subsection{Scaling with $Pe$}

In \cite{Johnson}, scattering measurements were conducted in the $k_1-k_3$ plane, thus for extremely small $k_2$. It is pointed out that the linear-response term in the structure factor expansion of previous theories \cite{Dhont_1989, Ronis} can be neglected in this case as it is proportional to $k_2$, thus the larger contribution to the distortion is given by the second-order term and $\Delta S$ should be proportional to $Pe^2$. This is contradicted by experimental results, which show a scaling exponent $n(\phi)$ depending on the volume fraction and much smaller than $2$. Other studies \cite{Wagner} highlighted that even the linearity might not hold for the minimum values of $Pe$ at which the distortion can be detected, thus questioning the validity of the perturbative expansion at low but non-zero $Pe$.

This aspect could not be investigated in our study, since it is not possible, within our framework, to calculate the distortion along a specific direction. Furthermore, the direction $k_2=0$, along which the measures are performed, is precisely what separates extensional and compressing quadrants in our model, and therefore its nature is not defined. The predictions for the scaling law of the structure factor distortion averaged on the different quadrants are those obtained in Section \ref{low_Pe}: the exponent is $1$ in extension and compression, and $2$ upon taking the full angular average.

\section{Discussion and conclusion} \label{conclusions}

In the present work, we have analytically solved the two-body Smoluchowski equation with shear to determine the microstructure of dilute colloidal dispersions in the limit of low-to-moderate Péclet numbers. The main issue that is encountered when addressing this problem is the singular perturbation of the equation, which makes the solution a non-analytic function of $Pe$ at sufficiently large separations, $r \approx \mathcal{O}(Pe^{-1/2})$. Indeed, in this region, the shear convective term, which grows linear with radial separation $r$, due to the great distance form the origin gives rise to a non-vanishing contribution even when $Pe$, and the shear rate, are low. The mathematical steps adopted for rearranging the equation into an analytically solvable form are: (i) the Fourier transform to reciprocal $k$ space, where the boundary-layer problem has a standard form, and (ii) the angular averaging, which simplifies the form of the equation, making it solvable to higher perturbative orders than in it was possible in the past, thereby extending the range of $Pe$ in which the theory can be applied. We have indeed observed that our second-order in $Pe$ outer expansion matches continuously with the inner solution up to $Pe=0.75$, making the solution consistent, while for a linear expansion, like the one calculated by Dhont \cite{Dhont_1989}, this happens only for $Pe \leq 0.1$. This benefit comes at the cost of being able to determine the structure factor only as averaged over the quadrants where the flow has the same extensional or compressing behaviour.

As a result, we derived analytical, or semi-analytical, expressions for the static structure factor at small $Pe$, which we implemented for a Yukawa repulsive inter-particle interaction. We calculated the structure factor distortion $\Delta S$ due to the flow disturbance, which represents a small correction to $S_{eq}$ and is proportional to $Pe$ in the extensional and compressing quadrants and to $Pe^2$ upon taking the full angular average. Furthermore, the distortion is almost symmetrical, with opposite sign, in extension and compression, the leading correction differing only by its sign.

A broadening of the structure factor peak is predicted in the extensional quadrants and a narrowing is predicted in the compressing quadrants as well as upon averaging over the whole solid angle. The height of the peak diminishes in all cases, and its position is shifted to lower $k$ in extension and to higher $k$ in compression and in the fully-averaged case. Since the main peak is associated with medium-range order in the microstructure, these effects indicate that the flow perturbs the equilibrium microstructure causing redistribution of particles between the near-contact and the second/third shells of the pair correlation function. In particular, these features can be attributed to a spreading-out of the particles concentration in the medium-range of extensional quadrants and a corresponding increase of local order/concentration, (possibly associated with near-contact accumulation) in the short-range region of compressing sectors, as also observed in scattering experiments on hard spheres \cite{deKruif}. Analogous results for hard spheres are presented more in depth in Appendix \ref{app:HS}.

The predicted broadening/narrowing and lowering of the structure factor peak due to an applied shear, even though observed in experiments on colloids or dusty plasmas \cite{Nosenko}, was not present, to our knowledge, in previous theoretical works on particles interacting via the DLVO potential \cite{Dhont_1989,Dhont_1987} or was obtained from a more phenomenological description \cite{Ronis}.
These effects have been predicted here for the first time using an analytical solution to the Smoluchowski equation with shear.
In previous work, a similar result was derived with a very different approach by Schwarzl and Hess \cite{Schwarzl}, who solved a kinetic equation to calculate the structure of a soft-sphere fluid with a $r^{-12}$ potential in concentrated conditions ($\phi \approx 0.4$). They obtained a decreasing peak in both the extensional and compressing directions shifted to left (low $k$) and right (large $k$), respectively, exactly as shown in our work. They also pointed out that the shift was linear in $Pe$ up to $Pe=0.12$ and was given by $\pm 0.4 \tilde{k}_{max} Pe$, with $\tilde{k}_{max}$ being the position of the peak at equilibrium.

The procedure we have developed is completely general and can be used with any kind of inter-particle interactions. Moreover, the simplified mathematical treatment due to the reduction of the PDE to an ODE should make it more feasible in the future to include further higher-order terms in the expansion, such as the calculation of one more term of order $Pe^{3/2}$ in the inner solution, and the introduction of hydrodynamic interactions. These two improvements will make it possible to (i) extend the theory to higher values of $Pe$, to fill the gap with the theory for high $Pe$ ($Pe>10$) \cite{Banetta_2020}, and (ii) pave the way for extensions of the theory to larger concentrations, respectively.

Finally, as proved by Grmela et al. \cite{Grmela} based on mechanical and thermodynamic arguments, the microstructural state of suspensions influences their rheology through the stress tensor. It would be interesting in a future development to derive the extra stress tensor from the structure distortion calculated in this paper in order to make predictions on the influence on the suspension's rheology and possibly confirm them through comparison with experimental results and Stokesian Dynamics simulations. Moreover, Brady derived a useful representation of the particle contribution to the bulk stress, which can be calculated once the pair distribution function is known \cite{Brady_1993,Brady_1995} and obtained the scaling law with which the stress tensor, whose main contribution comes form the Brownian stress, diverges at concentration approaching the maximum random packing fraction. A similar analysis would also be feasible starting from the results of the present work to obtain a scaling law with respect to the Péclet number and the concentration in dilute systems.

\subsection*{Acknowledgments} 
A.Z. acknowledges financial support from US Army Research Laboratory and US Army Research Office through contract nr. W911NF-19-2-0055. 

\appendix

\section{Derivation of the inner solution} \label{app:Sin}

We now derive the leading order in $Pe$ for $S_{in}(\tilde{k})$ through the method of the variation of constants, illustrated by Dhont in \cite{Dhont_book}, requiring that it reduces to the equilibrium solution in the zero-shear limit. We do this separately for the extensional and compressing quadrants, for which slightly different approaches (resulting in different formulations of the solution) are needed, due to the opposite sign of the constant $\alpha$. This procedure leads to the expressions of Eqs. (\ref{eq:S0_in}) and (\ref{eq:S0_in_complex}).

\subsection{Extensional quadrants}

Let us first consider the extensional quadrants, for which $\alpha_e=\frac{1}{3\pi} > 0$. Starting from the Smoluchowski equation in the inner variable, Eq. (\ref{eq:smol_deltaS_in}), we can write the corresponding homogeneous equation,
\begin{equation}
    \frac{d}{d\tilde{q}} S_{in} (\tilde{q}) = \frac{2\tilde{q}}{\alpha} S_{in}(\tilde{q}),
\end{equation}
whose solution is trivially $S_{in}(\tilde{q})=C e^{\frac{\tilde{q}^2}{\alpha}}$. Let us now assume that the constant $C$ is a function of $\tilde{q}$: $S_{in}(\tilde{q})=C(\tilde{q}) e^{\frac{\tilde{q}^2}{\alpha}}$. By substituting this into Eq. (\ref{eq:smol_deltaS_in}), we obtain an ODE for determining $C(\tilde{q})$:
\begin{equation} \nonumber
    \frac{d C(\tilde{q})}{d\tilde{q}} e^{\frac{\tilde{q}^2}{\alpha}} + C(\tilde{q}) \frac{2\tilde{q}}{\alpha} e^{\frac{\tilde{q}^2}{\alpha}} =   C(\tilde{q}) \frac{2\tilde{q}}{\alpha} e^{\frac{\tilde{q}^2}{\alpha}} - \frac{2\tilde{q}}{\alpha} S_{eq}(\tilde{q})
\end{equation}
\begin{equation} \label{eq:C_0_app}
    \frac{d C(\tilde{q})}{d\tilde{q}} = - \frac{2\tilde{q}}{\alpha} S_{eq}(\tilde{q}) e^{-\frac{\tilde{q}^2}{\alpha}}.
\end{equation}

By integrating from an undefined value to $\tilde{q}$, we find
\begin{equation} \label{eq:C_0_1_app}
    C(\tilde{q}) = \int_{\tilde{q}} dQ \frac{2Q}{\alpha} S_{eq}(Q) e^{-\frac{Q^2}{\alpha}} + C',
\end{equation}
where the constant $C'$ has to be determined according to the extreme of integration. The idea of this procedure is that the most suitable extreme should be found so that the integration constant takes a value that can be determined.

Let us assume that the integral is from $\tilde{q}$ to $\infty$ and write $C(\tilde{q})$ in terms of the original variable $\tilde{k}=Pe ^{1/2} \tilde{q}$:
\begin{equation} \label{eq:Soin_integ_app}
\begin{split}
   S_{in} & (\tilde{k}) = \left[ C' + \int_{\tilde{k}}^{\infty} dQ \frac{2Q}{Pe \, \alpha} S_{eq}(Q) e^{-\frac{Q^2}{Pe \, \alpha}} \right] e^{\frac{\tilde{k}^2}{Pe \, \alpha}} = \\
   & =  \left[ C' + \int_{-\infty}^{\infty} dQ \, \Theta(Q-\tilde{k}) \frac{2Q}{Pe \, \alpha} S_{eq}(Q) e^{-\frac{Q^2}{Pe \, \alpha}} \right] e^{\frac{\tilde{k}^2}{Pe \, \alpha}}.
\end{split}
\end{equation}

We now make use of the following lemma:
\\

\emph{Being $f(x)$ a real function such that $f'(x) > 0$ and $\lim_{x \rightarrow \infty}f(x)=\infty$, then}
\begin{equation}
    \delta (x-x_0) = \Theta(x-x_0) \lim_{\epsilon \rightarrow 0^+} \frac{f'(x)}{\epsilon} e^{-\frac{f(x)-f(x_0)}{\epsilon}}.
\end{equation}
\\
In our case with $\epsilon=Pe$, $x=Q$, $x_0=\tilde{k}$ and $f(Q)=\frac{Q^2}{\alpha}$ (so that $f'(Q)=\frac{2Q}{\alpha}$), the integral in Eq. (\ref{eq:Soin_integ_app}) for vanishing $Pe$ becomes
\begin{equation}
\begin{split}
    \int_{-\infty}^{\infty} dQ \, \Theta(Q-\tilde{k}) \frac{2Q}{Pe \, \alpha} S_{eq}(Q) e^{- \left( \frac{Q^2}{Pe \, \alpha} - \frac{\tilde{k}^2}{Pe \, \alpha} \right)} = \\
    = \int_{-\infty}^{\infty} dQ \, \delta (Q - \tilde{k}) S_{eq}(Q) = S_{eq}(\tilde{k}).
\end{split}
\end{equation}
This is exactly the limit we want to have for $S_{0,in}$: when $Pe \rightarrow 0$, it must reduce to $S_{eq}$, thus the above expression given by Eq. (\ref{eq:Soin_integ_app}) gives the correct limit if $C'=0$, which is Eq. (\ref{eq:S0_in}).

Here we have followed the same steps as in \cite{Dhont_1989} and \cite{Dhont_book} and obtained a very similar integral expression for $S_{0,in}(\tilde{k})$, the main difference being the argument of the exponential, which is an even power of $Q$ instead of an odd power. This is due to the angular average we have performed earlier which makes our solution depend only on the magnitude of the wavevector and has a significant consequence, as we shall see below.

It is important to highlight that, in this procedure, while the lower extreme of the integral is fixed ($\tilde{k}$), the upper extreme can be chosen arbitrarily and the integration constant must be adjusted accordingly. The choice of integrating from $\tilde{k}$ to $\infty$ was made so that the exponential lets the integral converge at $\infty$ when operating in the extensional quadrants, for which $\alpha_e > 0$. Luckily with this choice the integration constant turns out to be null if we want to fulfil the condition of retrieving the equilibrium solution for $Pe \rightarrow 0$.

\subsection{Compressing quadrants} \label{app:compressing}

In the compressing quadrants, one has to keep in mind that $\alpha_c=-\frac{1}{3\pi} < 0$, therefore the integral in Eq. (\ref{eq:S0_in}) diverges and we have to find a slightly different method to compute $S_{0,in}$. This issue does not have a parallel in \cite{Dhont_1989,Dhont_book}, where a change of sign in the exponential can be dealt with by choosing $- \infty$ as the integration extreme, which makes the integral converge thanks to the argument of the exponential being odd in $Q$. This obviously can not be applied to our case, because the exponential contains $Q^2$, so it diverges at both $\infty$ and $- \infty$ if $\alpha <0$. Our procedure is, however, inspired by the demonstration of the previous lemma, which is presented in \cite{Dhont_book}.

The fact that we have a quadratic function and want it to tend to $- \infty$ suggests that $Q$ might be treated as a complex number. For this reason we choose as the integration path the curve $\zeta$ in the complex $Q$-plane previously illustrated in Fig. \ref{fig:complex_path} in the main article.

As seen in the extensional case (in the passage from Eq. (\ref{eq:C_0_app}) to Eq. (\ref{eq:C_0_1_app})), the fundamental aspect in choosing the integration path is that, while the starting point is $\tilde{k}$, the other extreme is arbitrary, provided that a suitable integration constant is added to the result. With this choice of the path we are now going to show that this constant is zero. We suppose that the structure factor has the following form:
\begin{equation} \label{eq:S0_in_complex_app}
   S_{in}(\tilde{k}) = \left[ \int_{\zeta} dQ \frac{2Q}{Pe \, \alpha} S_{eq}(Q) e^{- \left( \frac{Q^2}{Pe \, \alpha} - \frac{\tilde{k}^2}{Pe \, \alpha} \right)} \right].
\end{equation}

Through the change of variable $y=f(Q)=\frac{Q^2}{\alpha}$, this becomes an integral on the real axis in the variable $y$ (not to be confused with the Cartesian dimension), which now converges:
\begin{equation} \label{eq:S0_in_real}
   S_{in}(\tilde{k}) = \frac{1}{Pe} \int_{\frac{\tilde{k} ^2}{\alpha}}^{\infty} dy \, S_{eq}(f^{-1}(y)) e^{- \frac{y-\frac{\tilde{k}^2}{\alpha}}{Pe}}.
\end{equation}

In the limit $Pe \rightarrow 0$, the exponential is non zero only when $y$ is close to $\frac{\tilde{k}^2}{\alpha}$, thus we can evaluate $S_{eq}$ in $y=\frac{\tilde{k}^2}{\alpha}$ and bring it out of the integral:
\begin{equation}
   S_{in}(\tilde{k}) = \frac{1}{Pe} S_{eq} \left( f^{-1} \left(\frac{\tilde{k}^2}{\alpha} \right) \right) \int_{\frac{\tilde{k} ^2}{\alpha}}^{\infty} dy \, e^{- \frac{y-\frac{\tilde{k}^2}{\alpha}}{Pe}} = S_{eq}(\tilde{k}).
\end{equation}
With this result we have proved that our starting ansatz, Eq. (\ref{eq:S0_in_complex_app}), reproduces the equilibrium structure factor in the $Pe \rightarrow 0$ limit without adding any integration constant.

It should be noticed that the integrand in Eq. (\ref{eq:S0_in_real}) is actually a real function even though $f^{-1}(y)$ is purely imaginary when $y > 0$, because $S_{eq}(Q)$ is real even if its argument is imaginary, as one can easily see from Eq. (\ref{eq:Se}). Therefore, all these integrals must be real functions of $\tilde{k}$.

For practical evaluation, the integral in Eq. (\ref{eq:S0_in_complex_app}), which is in the complex plane, can be divided into two integrals following the two straight lines of which the described path $\zeta$ is composed (see Fig. \ref{fig:complex_path}). The first one is an integral on a portion of the real axis; the second one is on the positive imaginary axis, thus it can be computed as an integral from $0$ to $\infty$ just by replacing $Q$ with $i Q$ in the integrand. As we expected, after applying this transformation, the integrand is still a real function, resulting in the integral being real as well.

\section{Results for hard spheres} \label{app:HS}

\subsection{Theory and results}
Hard spheres are a landmark for many statistical physics studies and applications and have been investigated in most previous theoretical \cite{Dhont_1989, Batchelor, Brady_1995, Brady_1997, Nazockdast_2012} and experimental studies \cite{Johnson, deKruif}. In this Appendix we show the results obtained implementing the following potential:
\begin{equation} \label{eq:HS_potential}
        \tilde{V}(\tilde{r}) = \left \{ \begin{array}{rl}
        \infty \quad \quad & \mathrm{if} \quad \tilde{r}<2 \\
        0 \quad \qquad & \mathrm{if} \quad \tilde{r}>2
        \end{array}
        \right. 
\end{equation}
This potential can be obtained by choosing either $\Gamma=0$ or $\tilde{\kappa}= \infty$ in the previous parametrization for the Yukawa potential, Eq. (\ref{eq:Yuk_potential}). The qualitative features are mainly the same predicted for charge-stabilized colloids.

Notice that the integral in Eq. (\ref{eq:Se}) can be calculated exactly with this interaction potential, giving the known result \cite{Dhont_1989}:
\begin{equation} \label{eq:Seq_HS}
    S_{HS,eq}(\tilde{k}) = 1 - 3 \phi \frac{\sin(2\tilde{k}) - 2\tilde{k} \cos(2\tilde{k})}{\tilde{k}^3}.
\end{equation}
This analytical expression for the equilibrium solution allows us to explicitly derive the form of the outer solution, Eq. (\ref{eq:Sout_approx}), which involves its first and second derivatives, leading to:
\begin{equation} \label{eq:Sout_HS}
\begin{split}
    S & _{HS,out}(\tilde{k}) = 1 - 3 \phi \frac{\sin(2\tilde{k}) - 2\tilde{k} \cos(2\tilde{k})}{\tilde{k}^3} \\
    & - 3 \phi \frac{\alpha}{2} \frac{(8\tilde{k}^3-3) \sin(2\tilde{k}) +6 \tilde{k} \cos(2\tilde{k})}{4\tilde{k}^5} Pe \\
    & - 3 \phi \frac{\alpha^2}{4} \frac{(15-24\tilde{k}^2) \sin(2\tilde{k}) + 2\tilde{k} (4\tilde{k}^2-15) \cos(2\tilde{k})}{16\tilde{k}^7} Pe^2.
\end{split}
\end{equation}

It is therefore possible to write explicitly the expression of the outer solution for hard spheres. For the inner solution Eq. (\ref{eq:Seq_HS}) has to be substituted into Eqs. (\ref{eq:S0_in}) and (\ref{eq:S0_in_complex}) and the integrals are evaluated numerically.

\begin{figure}[t]
    \centering
    \includegraphics[width=0.45 \textwidth, keepaspectratio]{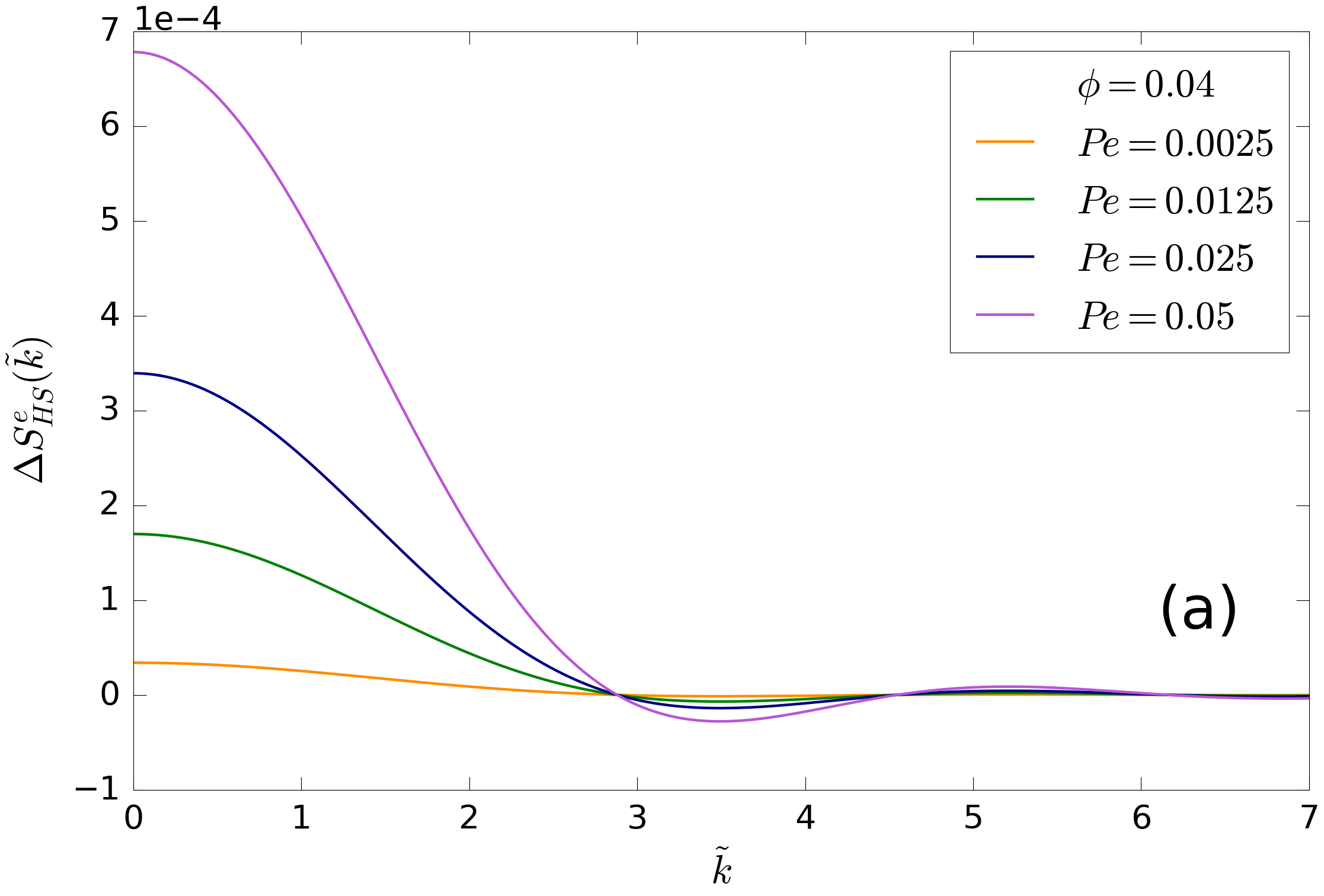}
     \\
    \vspace{0.4cm}
    \includegraphics[width=0.45 \textwidth, keepaspectratio]{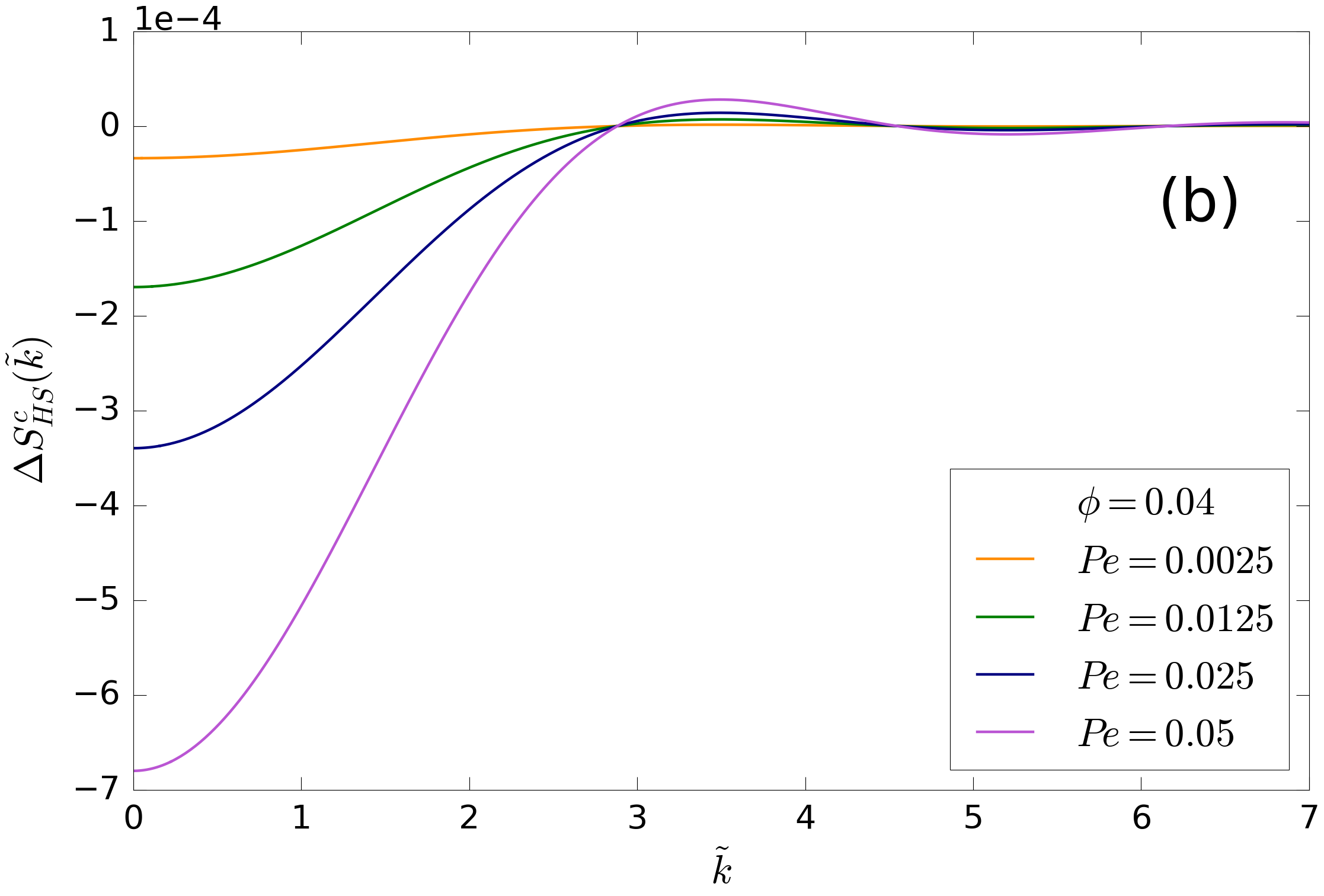}
    \\
    \vspace{0.4cm}
    \includegraphics[width=0.45 \textwidth, keepaspectratio]{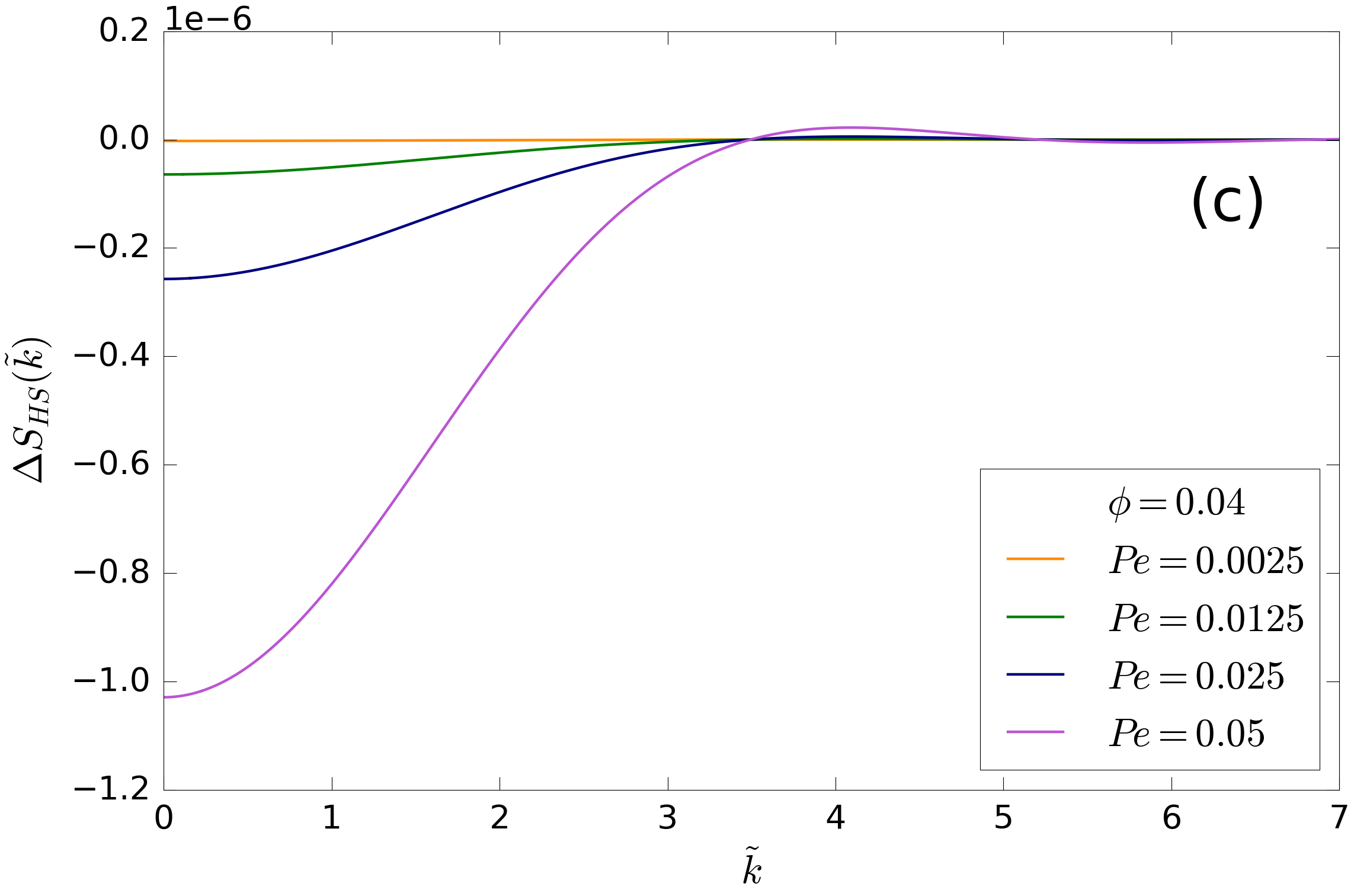}
    \caption{Angular average of the structure factor distortion for hard-sphere interaction potential at fixed $\phi$ and varying $Pe$. \textbf{Panel (a)} shows the distortion averaged on the extensional quadrants. \textbf{Panel (b)} shows the distortion averaged on the compressing quadrants. \textbf{Panel (c)} shows the total average on the solid angle.}
    \label{fig:HS_varying_Pe}
\end{figure}

In Fig. \ref{fig:HS_varying_Pe} we report the structure factor distortion for different $Pe$ values in the extensional and compressing quadrants, as well as their average over the whole solid angle. As in the case of Yukawa interacting particles, the different curves are proportional to the corresponding $Pe$ in Panels (a) and (b) and to $Pe^2$ in Panel (c). The distortion is much smaller than in charged systems at small $\tilde{k}$, as we can see from the scale on the ordinate axis, but the shape of $\Delta S$ is qualitatively the same, with the plot being almost symmetrical in extension and compression. The averaged distortion is orders of magnitude smaller.

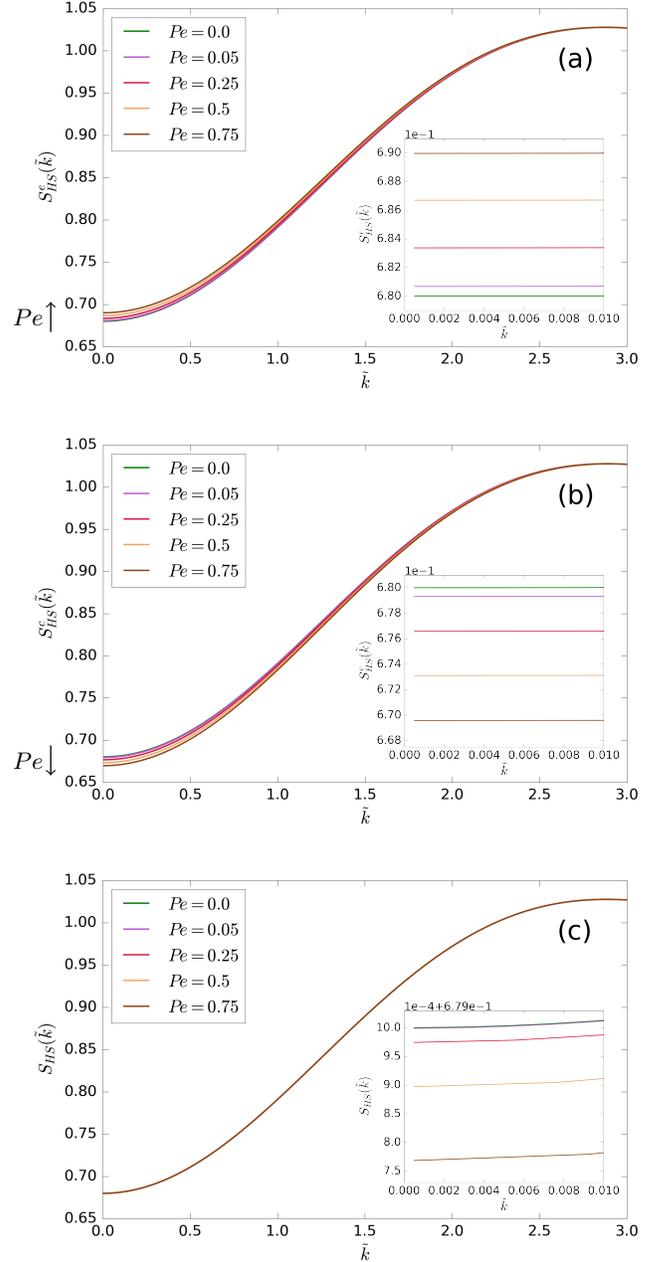
\begin{figure}[ht]
    \centering
    \begin{tikzpicture}
    \begin{scope}
    
    \node[inner sep=0] at (0,0) {\includegraphics[width=0.45 \textwidth, keepaspectratio]{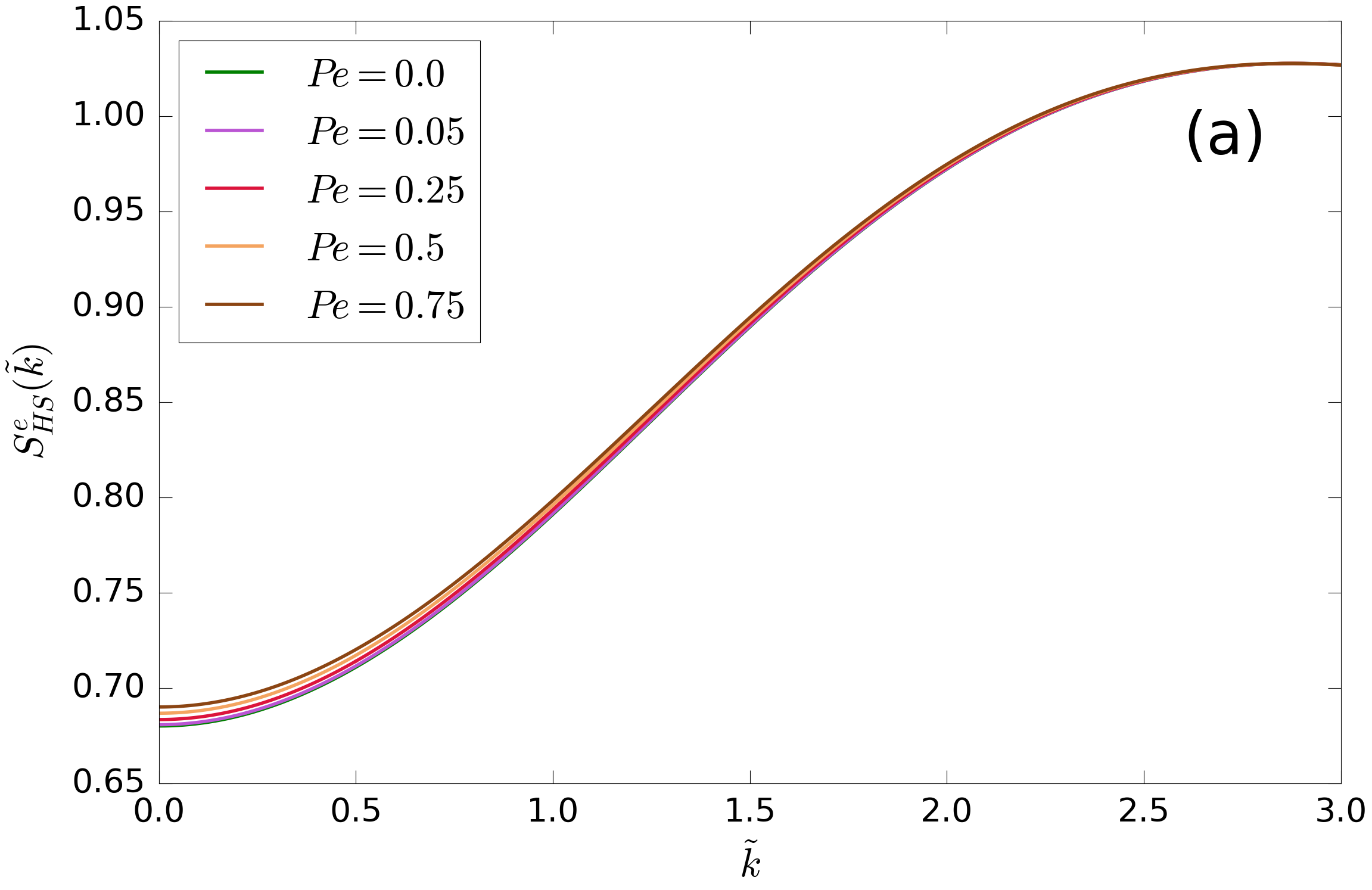}};
    
    \node[inner sep=0] at (0,-5.8) {\includegraphics[width=0.45 \textwidth, keepaspectratio]{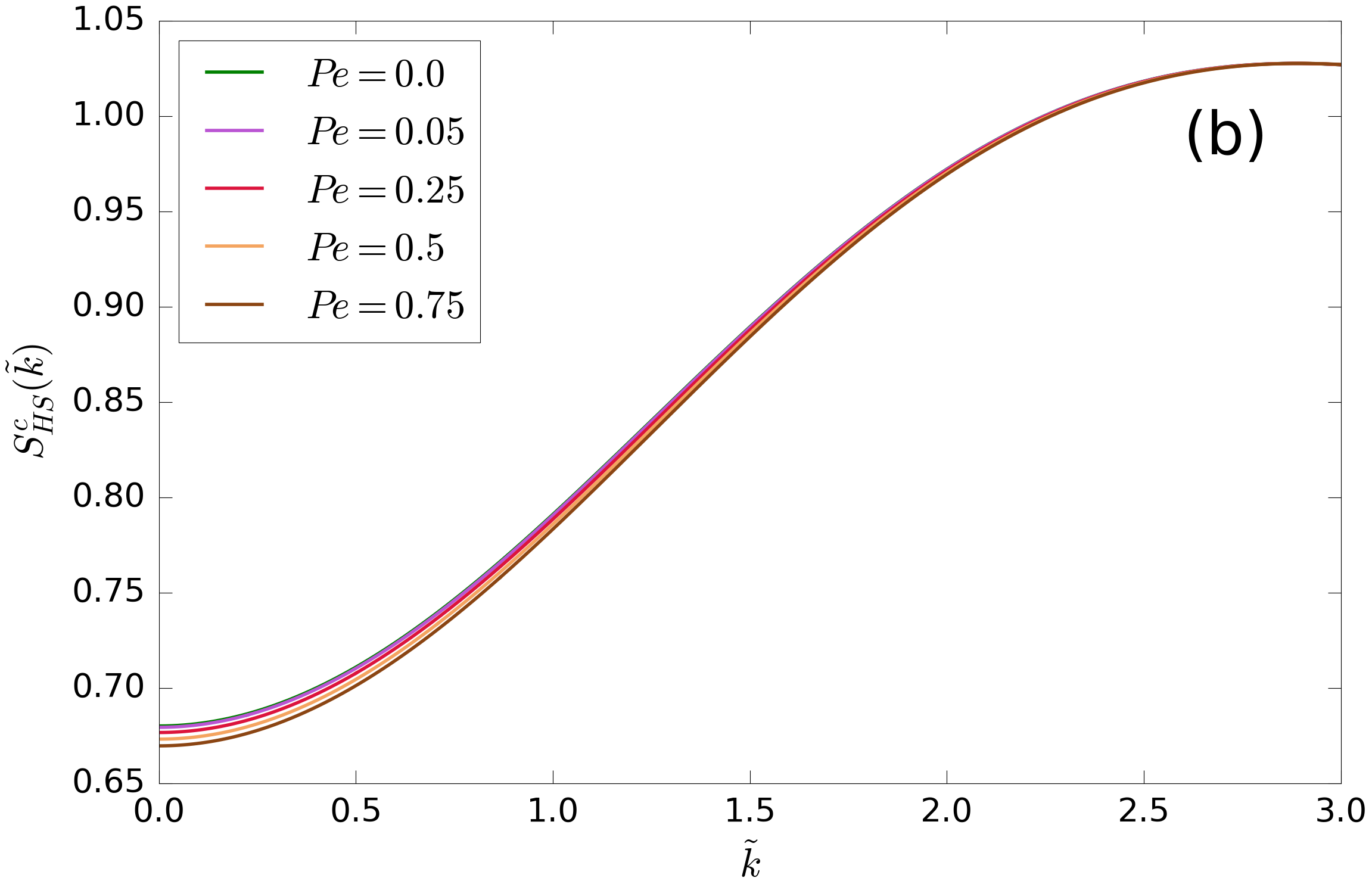}};
    
    \draw [->] [line width =0.5] (-3.8,-1.8) -- (-3.8,-1.4);
    \node at (-4.1,-1.6) { {$Pe$}};
    \draw [->] [line width =0.5] (-3.8,-7.3) -- (-3.8,-7.7);
    \node at (-4.1,-7.5) { {$Pe$}};
    
    \node[inner sep=0] at (0,-11.6) {\includegraphics[width=0.45 \textwidth, keepaspectratio]{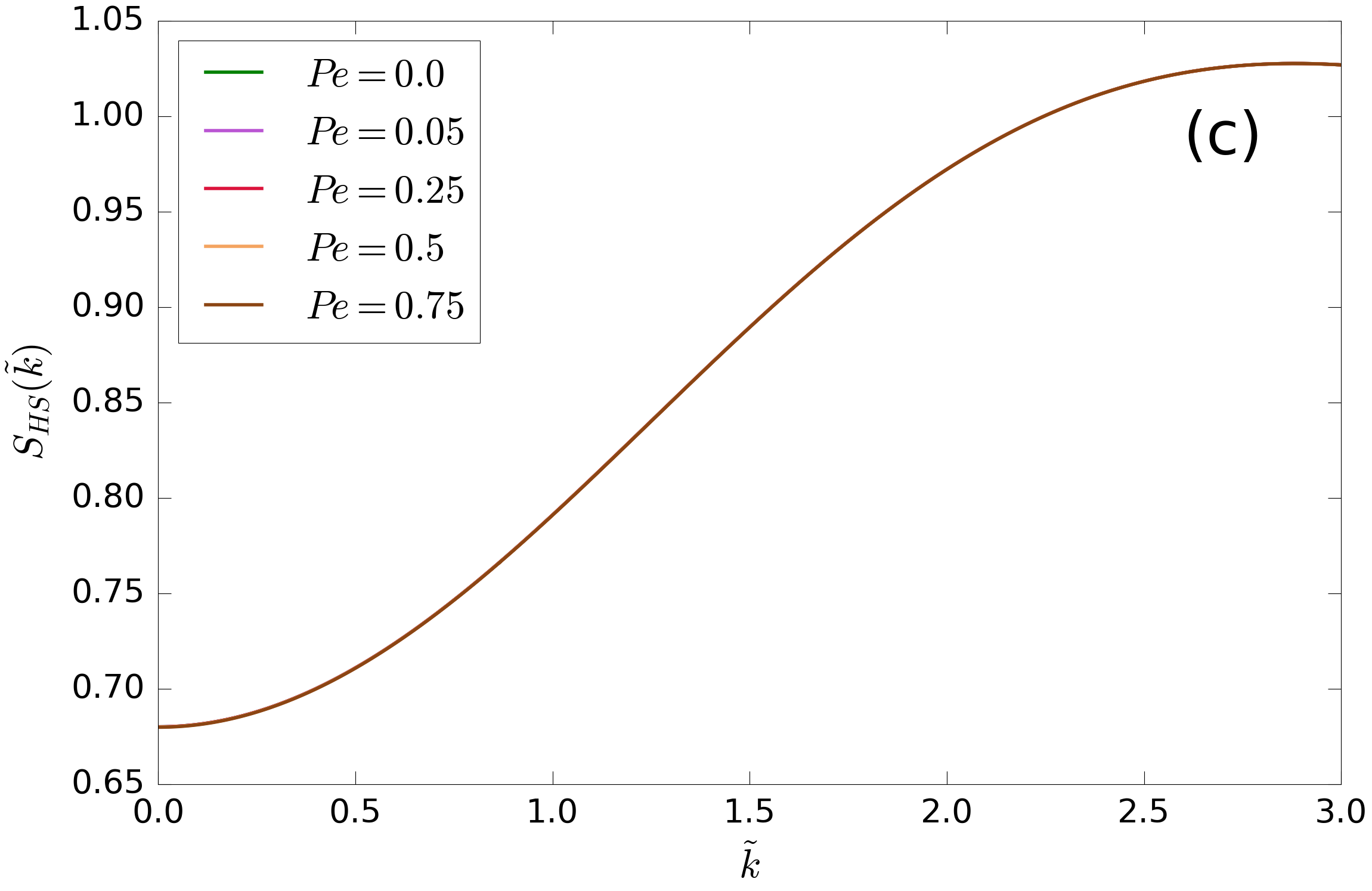}};
    
    \begin{picture}(0,0)
    \put(7,-55){\includegraphics[height=2.83cm]{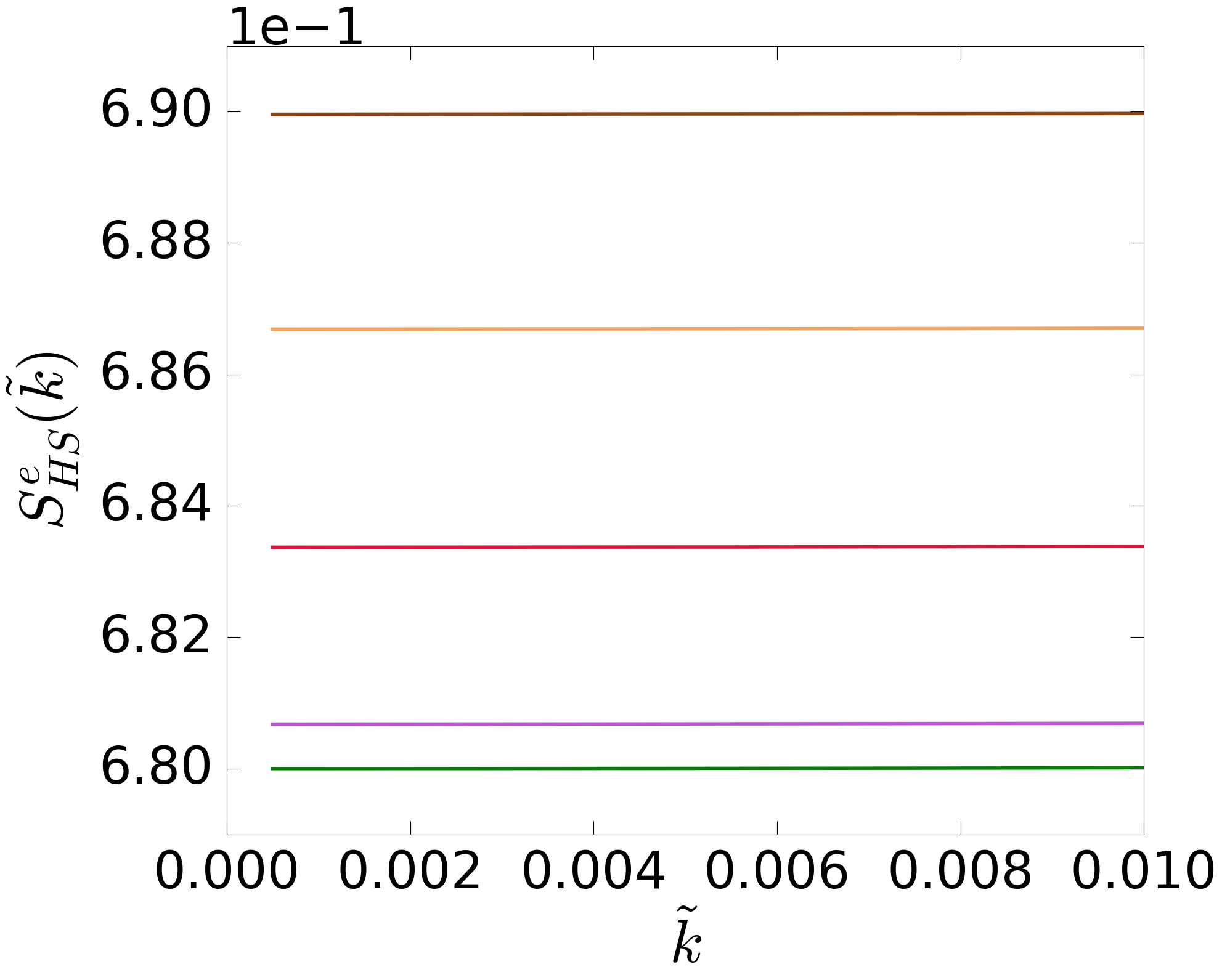}};
    \end{picture}
    
    \begin{picture}(0,0)
    \put(7,-220){\includegraphics[height=2.83cm]{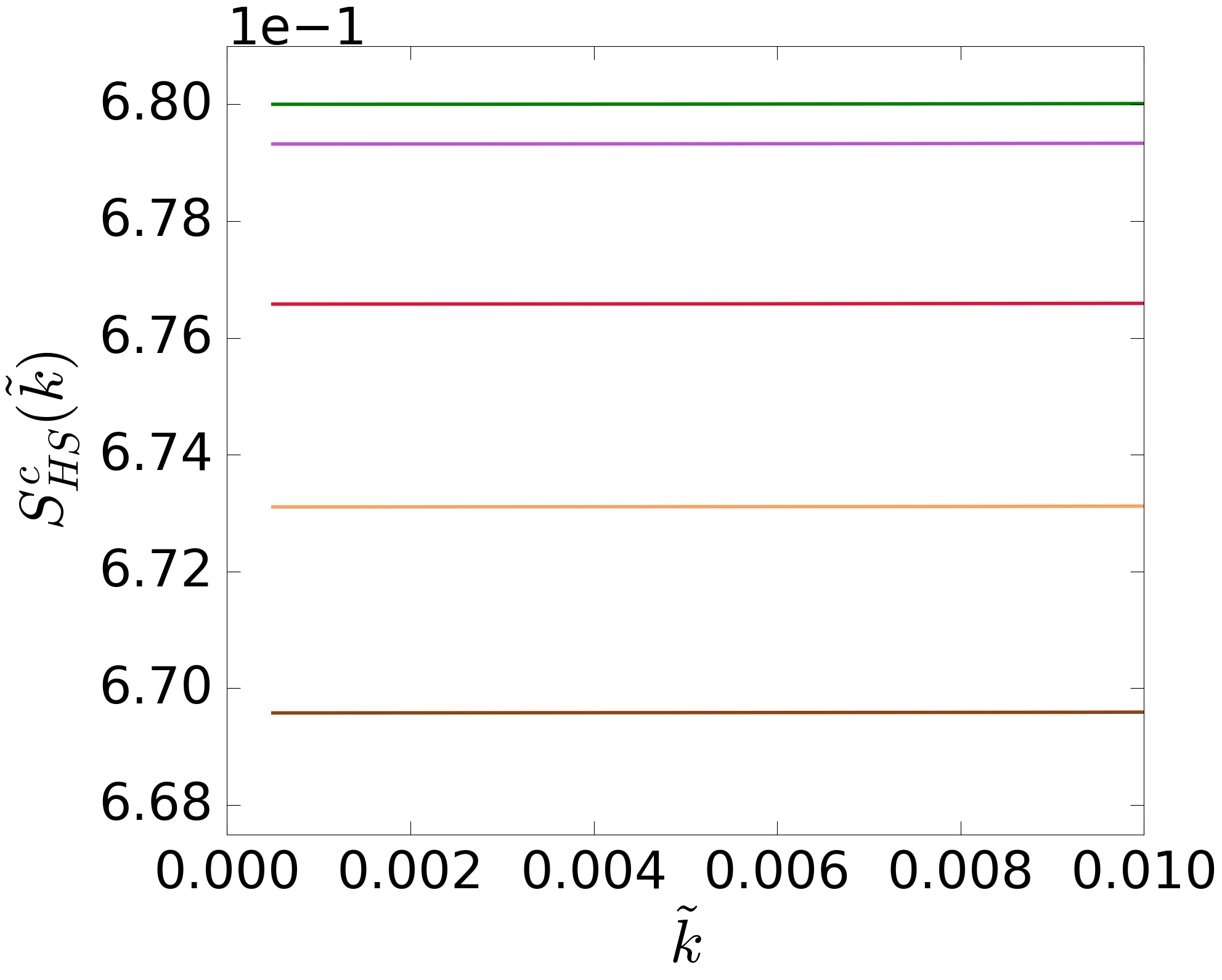}};
    \end{picture}
    
    \begin{picture}(0,-11.6)
    \put(7,-385){\includegraphics[height=2.83cm]{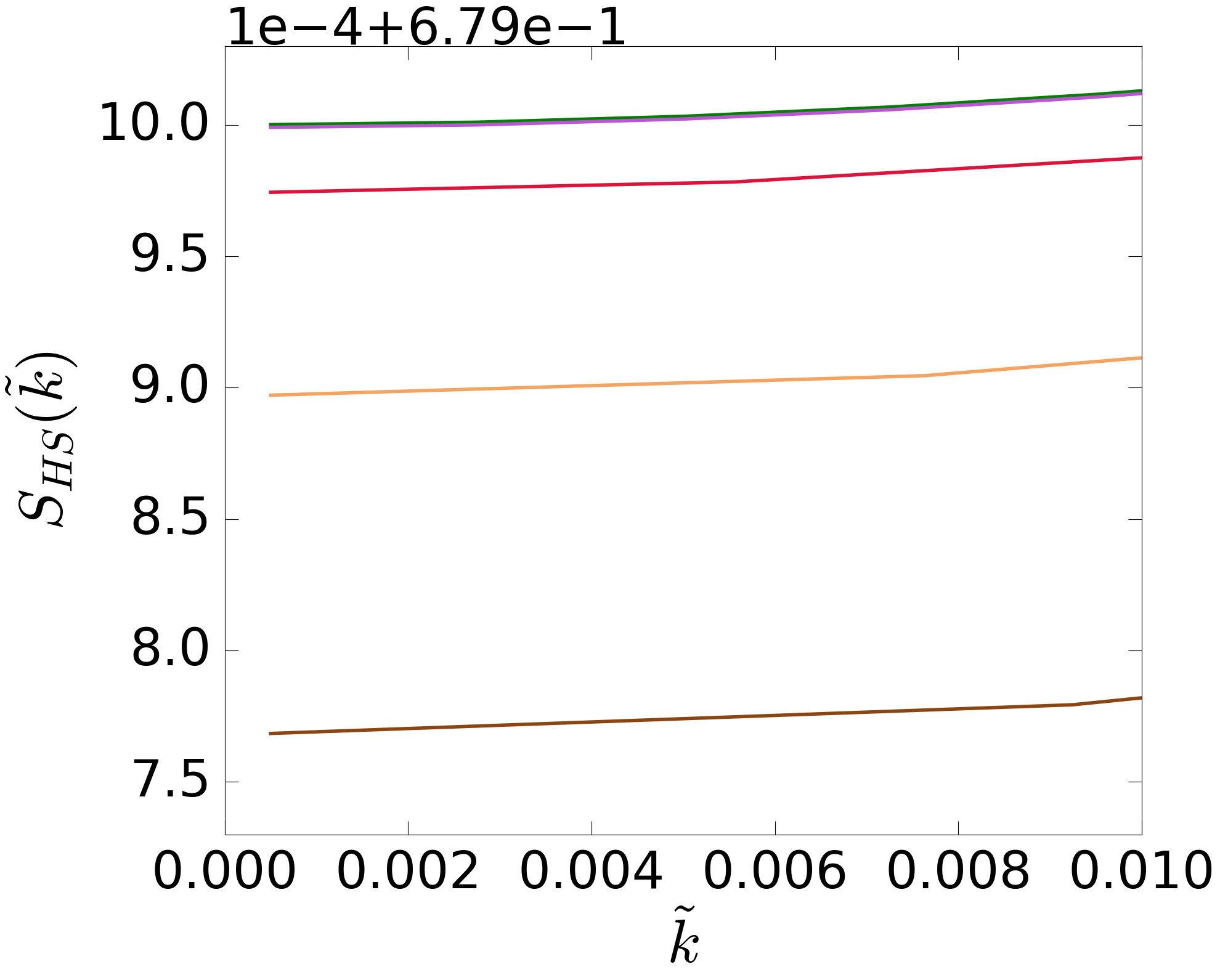}};
    \end{picture}
    
    \end{scope}
    \end{tikzpicture}
    \caption{Structure factor around the peak for varying $Pe$. $\phi$ is fixed as in Fig. \ref{fig:HS_varying_Pe}. The insets show the structure factor at small $\tilde{k}$. \textbf{Panel (a)} shows the structure factor averaged on the extensional quadrants, where a broadening of the peak is visible. \textbf{Panel (b)} shows the structure factor averaged on the compressing quadrants, where a narrowing of the peak is visible. \textbf{Panel (c)} shows the structure factor averaged on the whole solid angle, where a narrowing of the peak is visible.}
    \label{fig:broadening_HS}
\end{figure}

Fig. \ref{fig:broadening_HS} shows the structure factor at small wavevector and around the peak. An increasing shear rate causes the structure factor to grow at small $\tilde{k}$ and the peak to broaden consequently in the extensional quadrants (a). The opposite occurs in the compressing quadrants and, to a lesser extent, in the fully averaged structure factor, as we observed with Yukawa interacting particles as well.

\begin{figure}[h!]
    \centering
    \includegraphics[width=0.45 \textwidth, keepaspectratio]{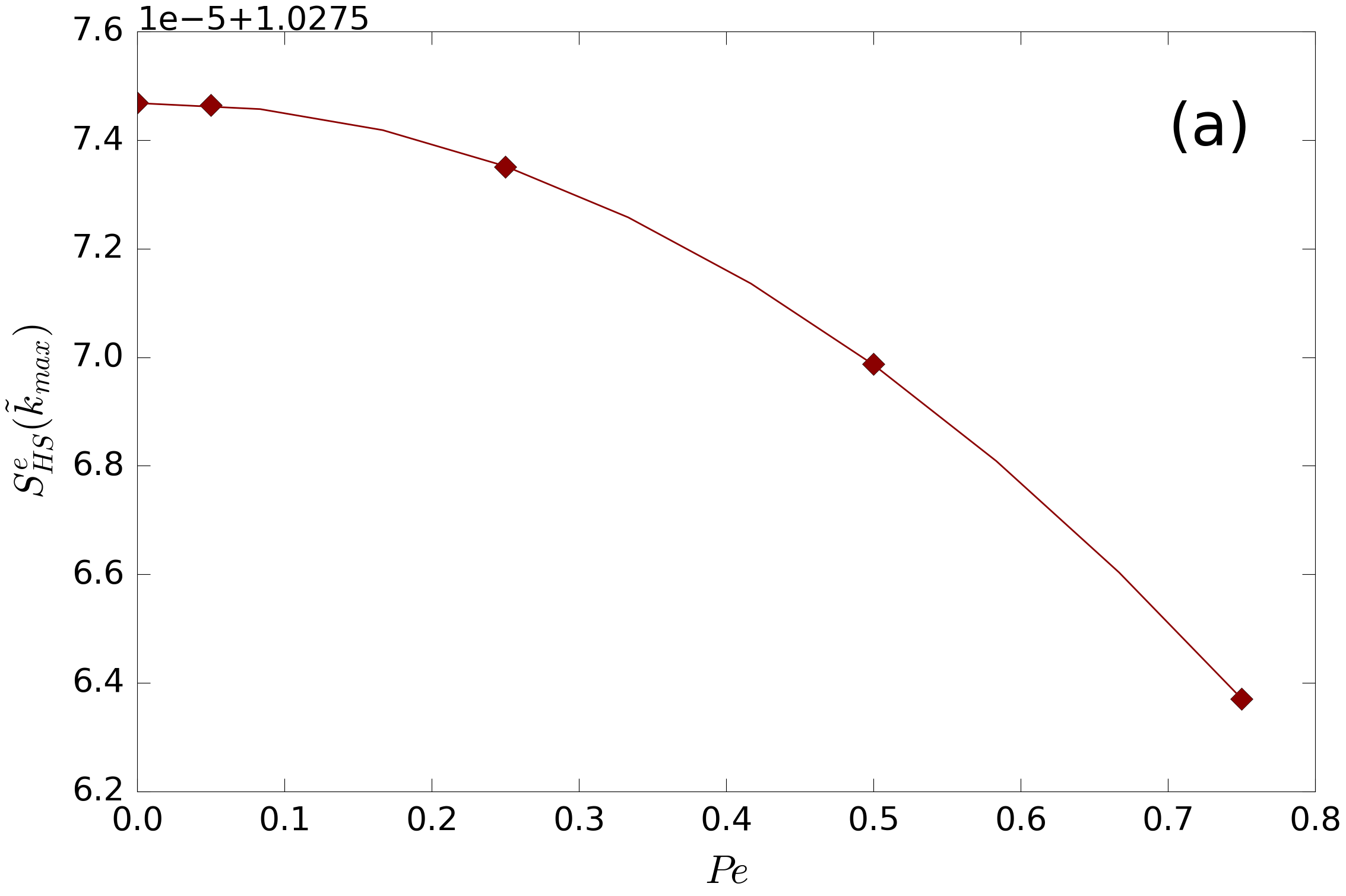}
    \\
    \vspace{0.4cm}
    \includegraphics[width=0.45 \textwidth, keepaspectratio]{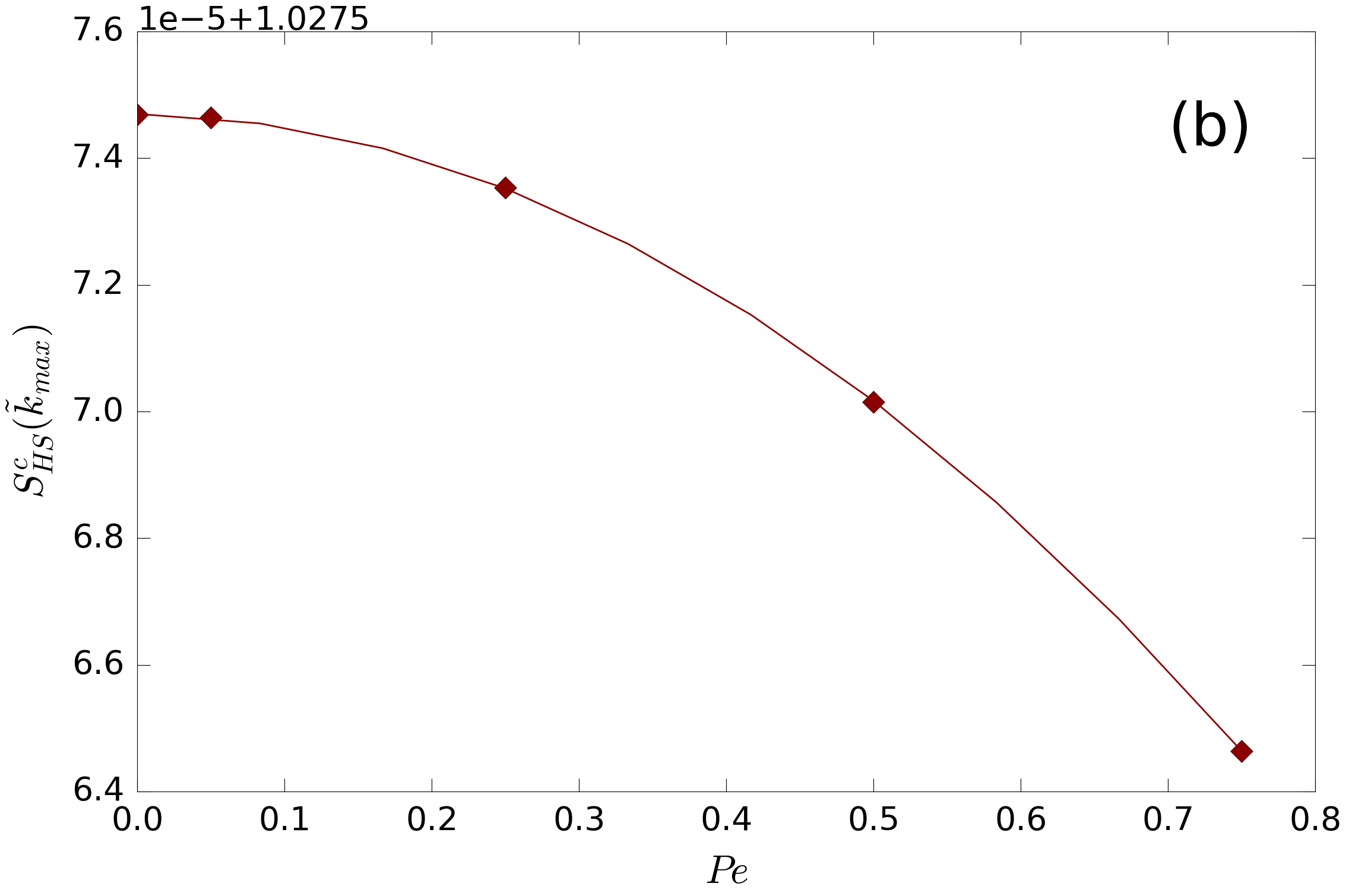}
    \\
    \vspace{0.4cm}
    \includegraphics[width=0.45 \textwidth, keepaspectratio]{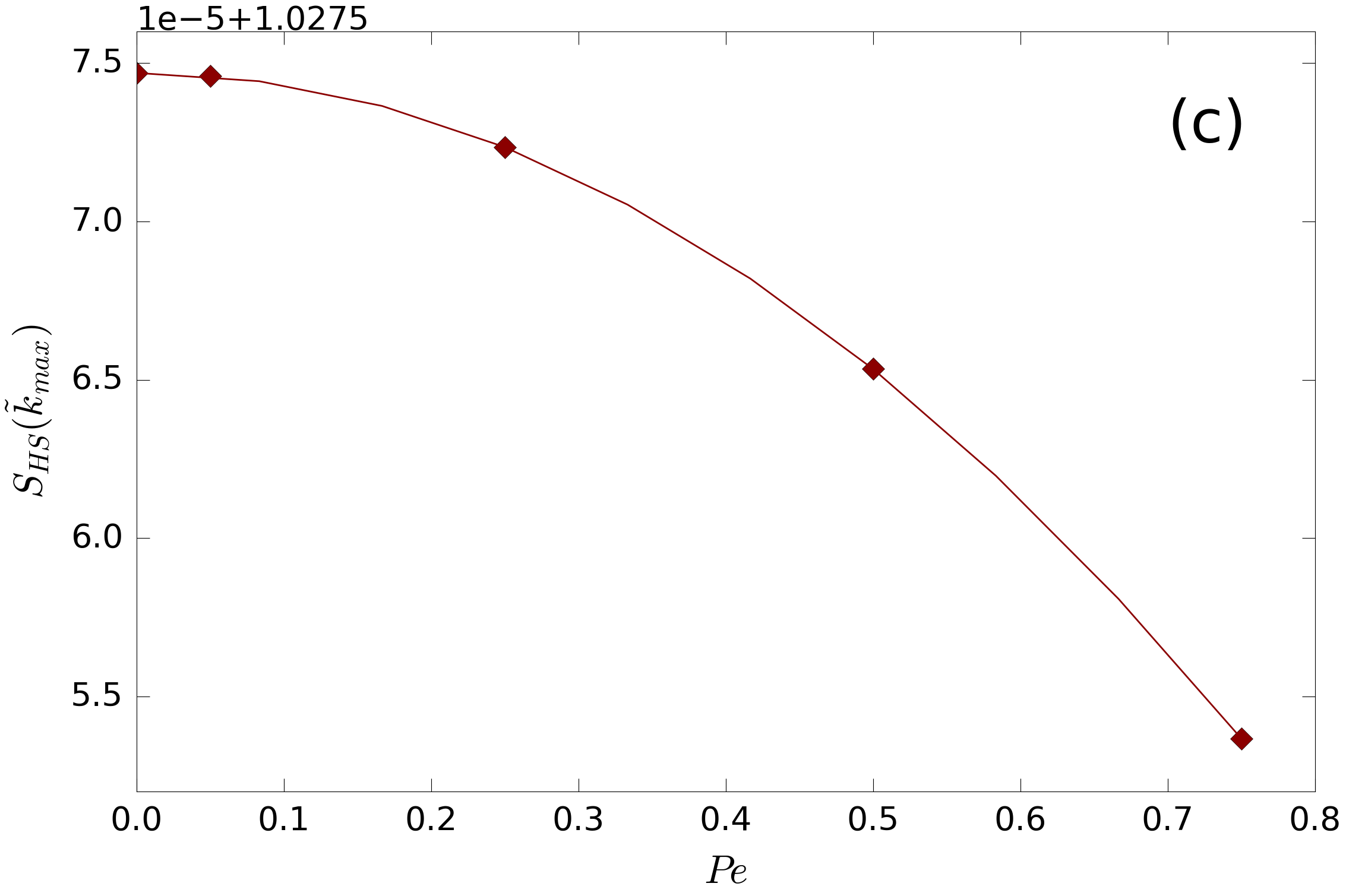}
    \caption{Maximum value of the structure factor first peak as a function of $Pe$. \textbf{Panel (a)} shows the maximum in the extensional quadrants, which decreases quadratically with quadratic coefficient $-2.00 \times 10^{-5}$. \textbf{Panel (b)} shows the maximum in the compressing quadrants, which decreases quadratically with a similar quadratic coefficient, $-1.74 \times 10^{-5}$. \textbf{Panel (c)} shows the total average on the solid angle, which decreases quadratically, but with a rather different quadratic coefficient, $-3.73 \times 10^{-5}$.}
\label{fig:broadening_S_max_vs_Pe_small_Pe_HS}
\end{figure}

The peak decreases again quadratically as a function of $Pe$, as illustrated in Fig. \ref{fig:broadening_S_max_vs_Pe_small_Pe_HS}. The quadratic coefficient is similar in extension (a), $-2.00 \times 10^{-5}$, and compression (b), $-1.74 \times 10^{-5}$, and is double on average (c), $-3.73 \times 10^{-5}$.

\begin{figure}[h!]
    \centering
    \includegraphics[width=0.45 \textwidth, keepaspectratio]{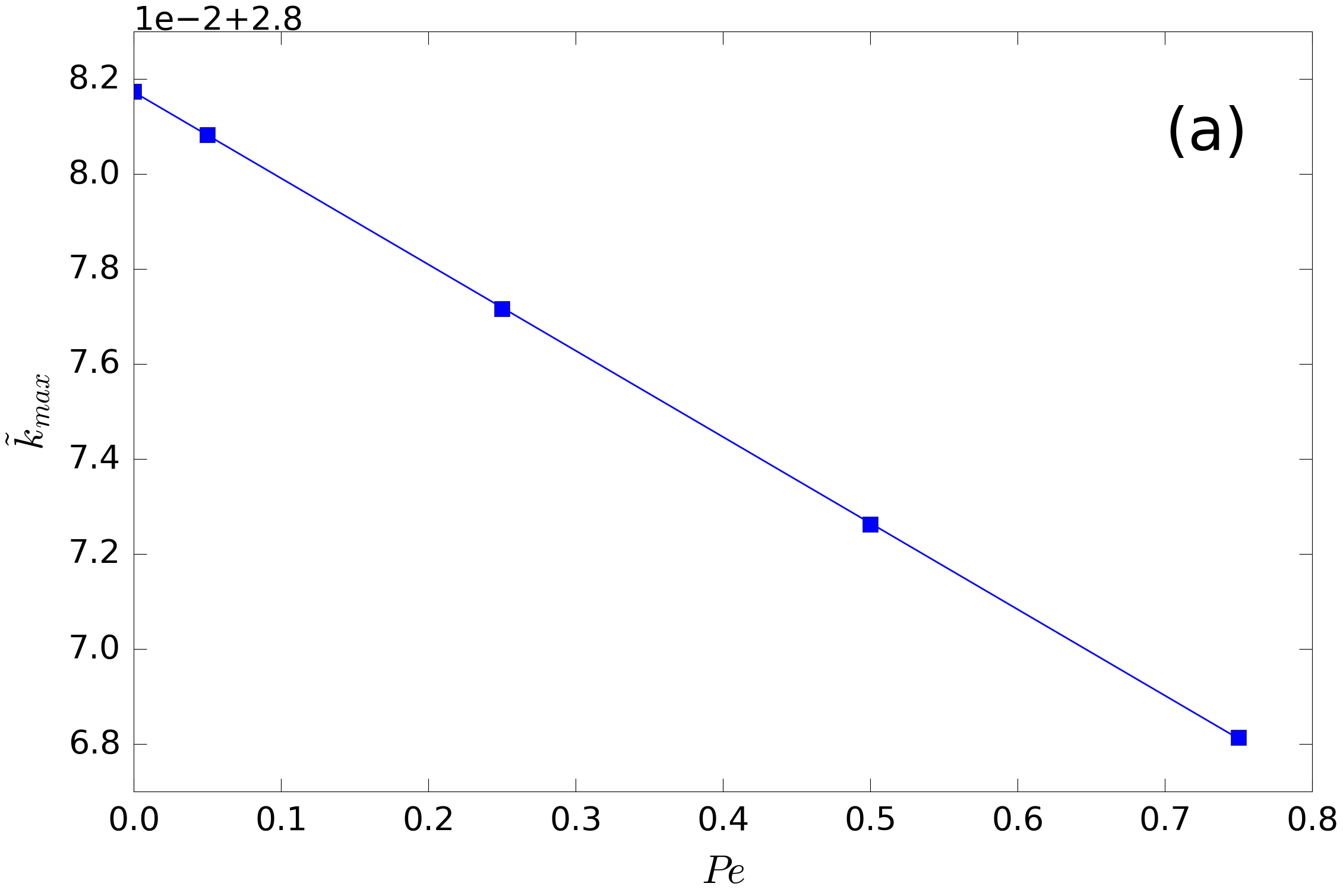}
    \\
    \vspace{0.4cm}
    \includegraphics[width=0.45 \textwidth, keepaspectratio]{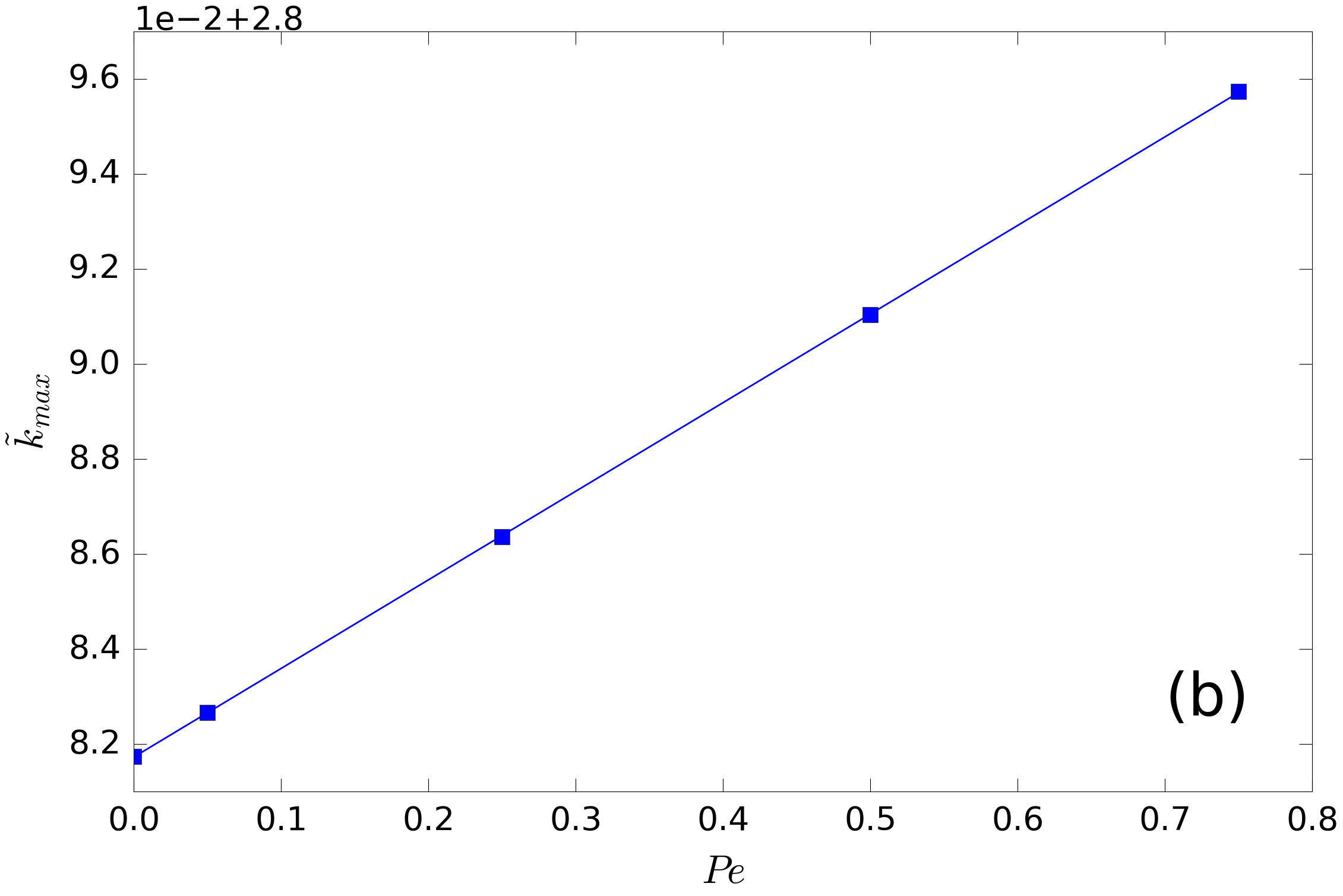}
    \\
    \vspace{0.4cm}
    \includegraphics[width=0.45 \textwidth, keepaspectratio]{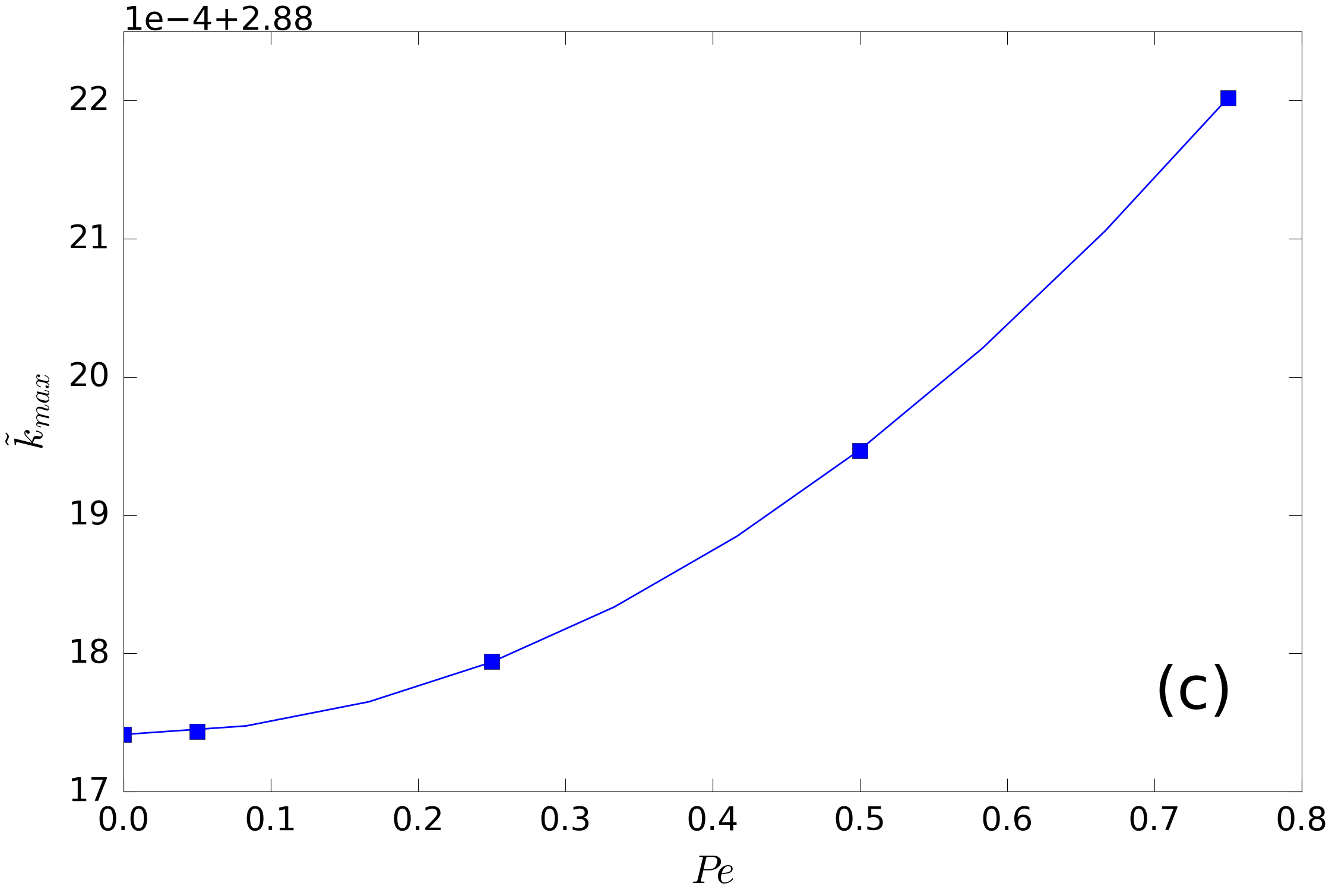}
    \caption{Position of the structure factor first peak as a function of $Pe$. \textbf{Panel (a)} shows the peak position in the extensional quadrants, which is shifted linearly to the left with linear coefficient $-0.01815$. \textbf{Panel (b)} shows the peak position in the compressing quadrants, which is shifted linearly to the right with linear coefficient $0.01865$. \textbf{Panel (c)} shows the total average on the solid angle, where the shift is quadratic with quadratic coefficient $0.00081$.}
\label{fig:broadening_k_max_vs_Pe_small_Pe_HS}
\end{figure}

From Fig. \ref{fig:broadening_k_max_vs_Pe_small_Pe_HS} we see that the peak is shifted towards lower and higher $\tilde{k}$ in the extensional (a) and compressing (b) quadrants, respectively. In this case the shift is linear with $Pe$, while on average (c) the shift is quadratic. All the coefficients are reported in the caption.

\subsection{Comparison with experiments of Ref.\cite{deKruif}}
Hard spheres have been investigated experimentally by de Kruif et al. \cite{deKruif}, who studied sterically-stabilized silica spheres by means of small-angle neutron scattering. They created an apparatus allowing them to measure the structure factor along the extensional and compressing axes, which form $45 ^{\circ}$ angles with the Cartesian axes and therefore correspond to the bisectors of what we have defined as extensional and compressing quadrants. A quantitative comparison of the results is not possible because we have averaged the structure factor over the quadrants, while the experimental one is measured along a specific direction. Even more important is the different concentration regime, since we have to work with dilute solutions ($\phi = 0.04$), while experiments are usually conducted at high concentration ($\phi=0.45$ in \cite{deKruif}).\\

Notwithstanding this, however, they observed some of the features that we have predicted here. In extension, they found that $S(\tilde{k})$ increases with $Pe$ at small $\tilde{k}$ and its peak decreases, as we have seen in Figs. \ref{fig:broadening_HS} and \ref{fig:broadening_S_max_vs_Pe_small_Pe_HS}. As opposed to our results though, their position of the maximum is shifted to higher $\tilde{k}$; however, they report that this position is particularly difficult to measure because of the flattening of the peak and affirm that this result is not completely certain. In our opinion, the shift to lower $k$ is a reasonable effect, since we expect the second and third nearest neighbour shells in the pair correlation function to move farther from the reference particle and thus the structure factor peak, which is related to their position via a Fourier transform, to move towards lower $k$. As for the Yukawa potential, this behaviour can be attributed to a local decrease in the particles concentration along this direction.

In the compressing sector, they observed that $S(\tilde{k})$ decreases as a function of $Pe$ at small $\tilde{k}$ and its peak increases and shifts towards higher values of $\tilde{k}$. Here the only difference with respect to our prediction is the growth of the peak; the lowering that we obtain, instead, can be explained as follows in terms of the difference regime of particle concentrations. At high concentrations, compression causes the particles in all the shells, including the second and third nearest-neighbour, to become more ordered and precisely localized, thus increasing the structure factor peak. In dilute conditions, though, we only expect the peak corresponding to the first shell in the pair correlation function to increase, since further shells are weakly populated; this increase in the first shell population should cause the second and third shells' population to lower due to overall conservation of particles, thus the structure factor peak decreases as well. The decrease and right-shift of the peak can be attributed again to the formation of transient accumulation of particles, which might occur together with an increase in concentration causing the decrease at small $\tilde{k}$.

\section{DLVO potential and parameters} \label{app:DLVO}

The interaction potential introduced in Section \ref{results} represents the effective interaction between charge-stabilized colloids at constant and low surface potential. This is a known result of the DLVO theory, developed independently by Derjaguin and Landau \cite{Derjaguin_1941} and Verwey and Overbeek \cite{Verwey_1948}. In this appendix we will only spend few words about its origin to justify our choice of the parameters $\Psi_0$, $\Gamma$ and $\kappa$; we refer to \cite{Hansen_1976} for its application to colloidal physics and to \cite{Verwey_1948} for its complete derivation.

The potential is obtained by treating the microions dispersed in the colloidal suspension within the linear Debye-H{\"u}ckel approximation in the low-potential regime. For ions with one protonic charge at temperature $T=298K$ this approximation is consistent when the potential is lower than $25 mV$ all over the space. Since the highest potential, $\Psi_0$, is the one reached at the particles surface, in the present work we have chosen (for all the results in Section \ref{results}) $\Psi_0 = 10 mV$.

Solution of the known Debye-H{\"u}ckel equation in dilute conditions for one colloidal particle with charge $q$ gives the following surface potential \cite{Verwey_1948,Hayter_1981}:
\begin{equation} \label{eq:Psi_0}
    \Psi_0 = \frac{q}{4 \pi \epsilon_0 \epsilon_r a (1+\kappa a)} .
\end{equation}
We operate under the assumption of constant surface potential, since we want to maintain the thermodynamic equilibrium of the colloid particles. The motion of such particles, indeed, is extremely slow compared to the time scale of the microions and the solvent particles and ions can rapidly rearrange or temporarily diffuse from the colloid surface to the solution so that the equilibrium is preserved.

Under this hypothesis \cite{Verwey_1948} derived the following expression for the interaction energy (in dimensionless units):
\begin{equation} \label{eq:VR_1}
    \tilde{V}(\tilde{r}) = \beta 4 \pi \epsilon_0 \epsilon_r a \Psi_0^2 \frac{e^{-\tilde{\kappa}(\tilde{r} - 2)}}{\tilde{r}} .
\end{equation}

By substituting Eq. (\ref{eq:Psi_0}) into Eq. (\ref{eq:VR_1}) it is possible to recognize Eq. (\ref{eq:Yuk_potential}), with the identification \cite{Hayter_1981,Hansen_1982}
\begin{equation} \label{eq:Gamma}
    \Gamma = \beta e^{2\tilde{\kappa}} 4 \pi \epsilon_0 \epsilon_r a \Psi_0^2 ,
\end{equation}
and $q=Z^* e$.


Eq. (\ref{eq:Gamma}) expresses the relation between the free parameters in the potential, $\Gamma$ and $\tilde{\kappa}$, and the surface electric potential $\Psi_0$, for which we have chosen the value $\Psi_0=10mV$. Verwey and Overbeek stated that their procedure leading to Eq. (\ref{eq:VR_1}) is valid for not-too-high values of the Debye screening parameter and quantified the upper limit as $\tilde{\kappa} \leq 3$ \cite{Verwey_1948}. Furthermore, our approximation for the dilute equilibrium solution, Eq. (\ref{eq:Se}), only gives reliable results when the pairwise potential $V(r)$ is not excessively repulsive. If the total integral of $V(r)$ is too large, we obtain unphysical negative values of $S_{eq}$ around $k=0$. To avoid this, we also restrict ourselves to $\tilde{\kappa} \geq 1.5$. To derive the results in Section \ref{results} we have chosen the value $\tilde{\kappa}=2$, which is in between these lower and upper values.

To avoid the restriction of low potential, which is necessary in the Debye-H{\"u}ckel linear theory, one could use the approximation for the interaction energy of spherical particles derived by Sader, Carnie and Chan \cite{Sader}. This is suitable for moderate-to-high potentials, up to $100 mV$, and is valid at large $\tilde{\kappa}$ for all inter-particle separations, as it is obtained from their ``modified HHF'' formula, which satisfies the corrects limits at small and large separation. Moreover, its form is as simple as the ``HHF'' formula from the Derjaguin approximation, despite being more accurate. Since the method illustrated in the present paper is completely general and independent of the choice of potential, it can be easily implemented for this effective interaction in the high-potential repulsion regime in future work.

\bibliographystyle{apsrev4-1}

\bibliography{biblio}

\end{document}